\documentclass[floats,prb,twocolumn]{revtex4}
\usepackage{epsfig}
\usepackage{graphicx}
\usepackage{docs}

\def\be{\begin{equation}}
\def\ee{\end{equation}}
\def\ba#1{\begin{array}{#1}}
\def\ea{\end{array}}
\def\bn{\begin{enumerate}}
\def\en{\end{enumerate}}

\def\rr{\right}
\def\l{\left}

\def\ket#1{\l|#1\rr\rangle}

\def\beq{\begin{equation}}
\def\eeq{\end{equation}}

\begin{document}

\title{A Mott Glass to Superfluid Transition for Random Bosons in Two Dimensions}
\author{S. Iyer}
\author{D. Pekker}
\author{G. Refael}
\affiliation{Department of Physics, California Institute of
  Technology, MC 149-33, 1200 E.\ California Blvd., Pasadena, CA 91125}
\date{\today}
\begin{abstract}

We study the zero temperature superfluid-insulator transition for a two-dimensional model of interacting, lattice bosons
in the presence of quenched disorder and particle-hole symmetry.  We follow the approach of a recent series of papers by
Altman, Kafri, Polkovnikov, and Refael, in which the strong disorder renormalization group is 
used to study disordered bosons in one dimension.  Adapting this method to two dimensions, we
study several different species of disorder and uncover universal features of the superfluid-insulator
transition.  In particular, we locate an unstable finite disorder fixed point that governs the transition 
between the superfluid and a gapless, glassy insulator.  We present numerical evidence that this glassy
phase is the incompressible Mott glass and that the transition from this phase to the superfluid is driven by a  
percolation-type process.  Finally, we provide estimates of the critical exponents governing this transition.

\end{abstract}

\maketitle

\section{Introduction} 

\indent In the 1980s, seminal experiments on helium adsorbed in Vycor first attracted the attention of theorists to
the random boson problem~\cite{crooker1983superfluidity,reppy19844He}.  The onset of superfluidity in this disordered system showed, in some respects, 
striking similarities to ideal Bose gas behavior.  Thus, in the decade before the realization of Bose-Einstein
condensation in cold atomic gases, disordered bosonic systems were actually proposed as possible 
realizations of this elusive phenomenon.  While studies of disordered bosons did not ultimately lead to the observation of Bose-Einstein
condensation, the random boson problem continued to stimulate theoretical activity because of its 
considerable richness. The interplay of interactions and disorder leads to a variety of phases in bosonic systems; the description of these phases and the transitions between them is an ongoing
challenge~\cite{weichman2008dirty,fisher1989boson}.

\indent The theoretical investigation of random bosons in one dimension was pioneered by Giamarchi
and Schulz, who described the transition to superfluidity in the presence of perturbatively weak disorder\cite{giamarchi1988anderson}.  
Subsequently, Fisher, Weichman, Grinstein, and Fisher expanded upon the
work of Giamarchi and Schulz by establishing the now canonical zero temperature phase diagram of the Bose-Hubbard model with chemical 
potential disorder.  The superfluid and Mott insulating phases of the clean model are separated by another
insulating phase, the Bose glass.  In this glassy phase, there exist rare-regions in which the energetic gap for adding
another boson to the system vanishes.  However, any additional bosons are localized by the disordered environment,
rendering the phase a gapless, compressible insulator.  Fisher et al.\ argued that there should be no direct superfluid-Mott insulator transition in 
the presence of quenched, uncorrelated disorder, or in other words, that the Bose glass \textit{always} intervenes
between the phases of the clean model\cite{fisher1989boson}.  The general picture formulated by these authors 
has been vindicated by over two decades of subsequent theoretical and numerical work~\cite{pollet2009absence}.
On the experimental front, following the observation of the Mott insulator to superfluid transition in cold atomic gases by Greiner et al.\
\cite{greiner2002quantum}, there has been progress towards the realization of a Bose glass,
including suggestive but inconclusive evidence that the phase has already been observed~\cite{schulte2005routes,
fallani2007ultracold,billy2008direct,white2009strongly}.

\indent Meanwhile, evidence has accumulated that disordered bosonic systems may exhibit more exotic glassy 
phases in the presence of particle-hole symmetry.  One such phase, the so-called Mott glass, differs from the Bose glass in that it is incompressible.
This phase was originally proposed for systems of disordered fermions by Giamarchi, Le Doussal, and Orignac,
but these authors predicted that it can exist in bosonic systems as well\cite{giamarchi2001competition}.
Subsequently, Altman, Kafri, Polkovnikov, and Refael studied a variant of the Bose-Hubbard model in which 
chemical potential disorder is omitted in favor of strong disorder in the on-site interaction and hopping.  In the limit of large, commensurate boson filling, 
this model is equivalent to a chain of quantum rotors that can describe a Josephson junction array.  
Relative to the large filling, this rotor model exhibits an exact particle-hole symmetry which has important consequences for the phase diagram.
Through a strong disorder renormalization group analysis, Altman et al.\ found that this symmetry results in the appearance of a Mott glass phase in this model\cite{altman2004pha}.  
The same authors later considered a generalization of this rotor model which allows for random offsets in the filling, effectively reintroducing a chemical potential.  
Generic disorder in the random offsets violates the particle-hole symmetry, and in this case, the rotor model exhibits the usual Bose glass phase~\cite{altman2010superfluid,weichman2008particle}. 
This breakdown of the Mott glass demonstrates the link between exotic phases and symmetries of the disordered Hamiltonian.
Irrespective of the identity of the glassy phase, Altman et al.\ also found that the superfluid-insulator transition at strong disorder lives in
a different universality class from the weak disorder transition of Giamarchi and Schulz~\cite{altman2004pha,altman2010superfluid}.

\indent In this paper, we study the two-dimensional analogue of the rotor model considered by Altman et al. in order
to investigate the particle-hole symmetric random boson problem in $d > 1$.  One of our goals is to look for the Mott glass, but
more generally, we aim to characterize the phases and the superfluid-insulator transition of our model.  Like Altman et al., 
the tool we use to do this is the strong disorder real space renormalization group (RG), first formulated by Ma and Dasgupta\cite{dasgupta1980low} 
to study the 1D Heisenberg antiferromagnet about thirty years ago. 
A numerical application of the RG by Bhatt and Lee followed shortly thereafter\cite{bhatt1982scaling}, and the method was later expanded upon and applied to 
more general spin models by Daniel Fisher\cite{fisher1994random}.  Strong disorder renormalization has proven to be a powerful tool in the analysis of 
several disordered systems, especially in one dimension.  In higher dimensions, application of the RG has been rarer, because
in addition to the generic intractability of analytical approaches \cite{fisher1999phase}, 
there are few known transitions that exhibit so-called \textit{infinite randomness}, a property that guarantees that the RG becomes 
asymptotically exact near criticality.  The random boson model is not expected to exhibit infinite randomness, and indeed, 
the numerical data we present in this paper is consistent with this expectation.  Hence, in addition to physical questions about
the phases and phase transitions of the model, our work also aims to address a methodological question: might the
strong disorder RG give useful information about a model, even when confronted with the twin difficulties of higher dimensionality
and the absence of infinite randomness?

\subsubsection{Summary of the Results}

\begin{figure}
\centering
\includegraphics[width=8cm]{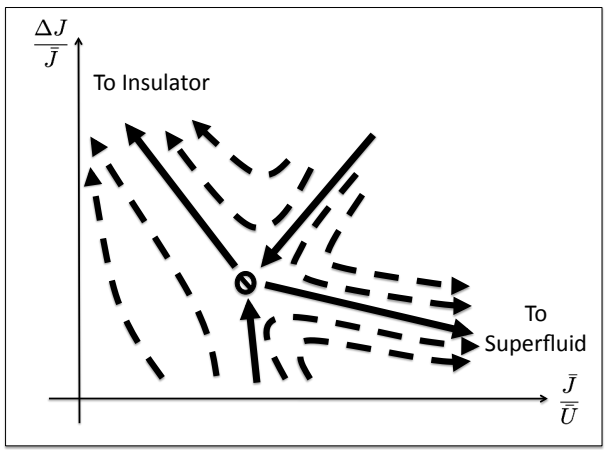}
\caption{A schematization of the universal features of the proposed flow diagram.  The $x$-axis gives the ratio of the mean of the renormalized Josephson coupling distribution to the
mean of the renormalized charging energy distribution.  The $y$-axis gives the ratio of the standard deviation of the Josephson coupling distribution to its mean.  In this context,
the Josephson coupling distribution only includes the dominant $2\tilde{N}$ couplings in the renormalized $J$ distribution, where $\tilde{N}$ is the number of
sites remaining in an effective, renormalized lattice.  See the text of Section \ref{sec:RG} for the reasoning behind the exclusion of weaker Josephson couplings from statistics.}
\label{flows}
\end{figure}

\indent Our main results are as follows:  we present numerical evidence for the existence of an unstable 
finite disorder fixed point of the RG flow, near which the distributions of Josephson couplings $J_{jk}$ and
charging energies $U_j$ in the rotor model flow to universal forms.  A schematic picture of this
unstable fixed point and the flows in its vicinity is given in Figure \ref{flows}.  

\indent To the left of the diagram, flows propagate towards a regime in which the ratio of $\bar{J}$, the mean of the Josephson
couplings, to $\bar{U}$, the mean of the charging energies, vanishes; meanwhile, the ratio of $\Delta J$, the
width of the Josephson coupling distribution, to $\bar{J}$ grows very large.  These flows terminate in one of 
two insulating phases.  The first is a conventional Mott insulator, in which it is energetically unfavorable for the 
particle number to fluctuate from the large filling at any site.  The other is a glassy phase, in which there 
exist rare-regions of superfluid ordering.  As the thermodynamic limit is approached, arbitrarily large 
rare-regions appear, driving the gap for charging the system to zero.  However, the density of the 
largest clusters decays exponentially in their size, and the size of the largest cluster in a typical sample
does not scale extensively in the size of the system.  Moreover, the largest clusters are so rare that they cannot
generate a finite compressibility.  Thus, the phase is a Mott glass.

\indent This insulating phase gives way to global superfluidity when the
rare-regions of superfluid ordering percolate, producing a macroscopic cluster of superfluid ordering. 
The appearance of  the macroscopic cluster is associated with flows that propagate towards the lower right of Figure \ref{flows}, indicating that the unstable fixed point
governs the glass-superfluid transition.  Our numerical implementation of the strong disorder RG allows 
us to extract estimates for the critical exponents that characterize this transition.  We are thus able to construct a compelling picture of the 
superfluid-insulator transition: a picture that must, however, be checked by other methods because of the perils of 
employing the strong disorder RG method in the vicinity of a finite disorder fixed point.

\subsubsection{Organization of the Paper}

\indent We begin our analysis of the disordered rotor model in Section~\ref{sec:model} by introducing the model, noting its relationship to the standard disordered 
Bose-Hubbard model and its special symmetries.  We also discuss the clean limit and the disordered
problem in one dimension.  Section~\ref{sec:RG} is devoted to a description of the strong disorder renormalization group, as applied to the disordered rotor model.  
In discussing the method, we emphasize the adaptations needed to use the technique in dimensions greater than one.  
We then present data collected from our numerical implementation of the strong disorder renormalization procedure in Section~\ref{sec:data} and subsequently explore what the data can tell us
about the zero temperature phases and quantum phase transitions of our random boson model in Section~\ref{sec:phases}. 
In Section~\ref{sec:conclusions}, we summarize the results, make connections to experiments, and give an outlook.

\section{The Model}
\label{sec:model}

\indent In motivating the model that we study in this paper, we begin by writing down a disordered 
Bose-Hubbard Hamiltonian that includes randomness in the interaction and hopping along with
the usual chemical potential disorder:
\begin{eqnarray}
\hat{H}_{\text{bh}} & = & - \sum_{\langle jk \rangle} t_{jk} (\hat{a}^\dagger_j \hat{a}_k + \hat{a}^\dagger_k \hat{a}_j)
                           + \sum_{j} U_j \hat{a}^\dagger_j \hat{a}_j(\hat{a}^\dagger_j \hat{a}_j-1) \nonumber \\
                       & \quad &    - \sum_{j} \mu_j \hat{a}^\dagger_j \hat{a}_j 
\label{Hbh}
\end{eqnarray}
Here, the creation and annihilation operators satisfy bosonic commutation relations:
\begin{equation}
\label{comm}
[\hat{a}_j,\hat{a}^\dagger_k] = \delta_{jk}
\end{equation}
and the hopping is between all nearest-neighbor sites on an $L \times L$ square lattice with periodic boundary
conditions.  
An alternative representation of this model is given by the number and phase operators:
\begin{equation}
\label{numberphase}
\hat{a}_j = e^{i \hat{\phi}_j} \sqrt{\hat{N}_j} 
\end{equation}
\begin{equation}
\label{numberphasecomm}
[\hat{\phi}_j,\hat{N}_k]  =   i \delta_{jk}
\end{equation}
In terms of these operators:
\begin{eqnarray}
\label{Hbhnp}
\hat{H}_{\text{bh}} & = & - \sum_{\langle jk \rangle} t_{jk} \left(\sqrt{\hat{N}_j}e^{-i \hat{\phi}_j}e^{i \hat{\phi}_k} \sqrt{\hat{N}_k} \right. \nonumber \\
                       & \quad &   \quad \quad \left.+\sqrt{\hat{N}_k}e^{-i \hat{\phi}_k}e^{i \hat{\phi}_j}\sqrt{\hat{N}_j}\right) \nonumber \\
                       & \quad &    + \sum_{j} U_j (\hat{N}_j-1)\hat{N}_j - \sum_{j} \mu_j \hat{N}_j
\end{eqnarray}
To obtain a large, commensurate boson filling $N_0$, the chemical potential is tuned such that
the on-site interaction and chemical potential terms are minimized for $\hat{N}_j = N_0$.  This consideration fixes $\mu_j = (2N_0-1)U_j$.  
Then, if we expand the number operators around this large filling as $\hat{N}_j = N_0 + \hat{n}_j$, the Hamiltonian becomes:
\begin{eqnarray}
\hat{H}_{\text{bh}} & = & - \sum_{\langle jk \rangle} t_{jk}N_0 \left(\sqrt{1+\frac{\hat{n}_j}{N_0}}e^{-i \hat{\phi}_j}e^{i \hat{\phi}_k} \sqrt{1+\frac{\hat{n}_k}{N_0}} \right. \nonumber\\
                       & \quad &   \quad \quad \left.+\sqrt{1+\frac{\hat{n}_k}{N_0}}e^{-i \hat{\phi}_k}e^{i \hat{\phi}_j}\sqrt{1+\frac{\hat{n}_j}{N_0}}\right) \nonumber \\
                       & \quad &    + \sum_{j} U_j \hat{n}^2_j + (\text{const.})
\end{eqnarray}
The operators $\hat{n}_j$ now correspond to the particle number deviations from the large filling $N_0$.  As such, $n_j$ can take on any integer
value from $-N_0$ to $\infty$, but we assume that $N_0$ is so large that we can let $n_j$ run from $-\infty$ to $\infty$.
The same approximation allows us to drop subleading (in $\frac{1}{N_0}$) terms in the hopping.  We finally define the couplings 
$J_{jk} = 2t_{jk}N_0$ to arrive at the quantum rotor Hamiltonian:
\begin{equation}
\label{rot}
\hat{H} = - \sum_{\langle jk \rangle} J_{jk} \cos(\hat{\phi}_j-\hat{\phi}_k) + \sum_{j} U_j \hat{n}^2_j
\end{equation}
This model, constructed as the large filling limit of a Bose-Hubbard Hamiltonian, can also be viewed as a description of a two-dimensional array 
of superconducting islands connected by Josephson junctions~\cite{altman2004pha,altman2010superfluid}.  Moreover, Vosk and Altman
have recently demonstrated that the one-dimensional model is relevant to cold atomic gases of
rubidium-87\cite{vosk2011superfluid}. 

\indent When the Josephson couplings $J_{jk}$ and charging energies $U_j$ are uniform,
the rotor model (\ref{rot}) exhibits a quantum phase transition between superfluid and Mott insulating phases
at zero temperature.  This transition is in the universality class of the three-dimensional classical 
XY model~\cite{weichman2008dirty,fisher1989boson,bruder2005bose}, and one recent study determines that the transition 
occurs at $\frac{J}{U} \approx 0.345$\cite{teichmann2009bose}.  
The critical exponent governing the divergence of the correlation length at the clean transition is $\nu \approx 0.663$\cite{gottlob1993critical}.
This exponent violates the Harris criterion:
\begin{equation}
\label{harris}
\nu d \geq 2
\end{equation}
when $d = 2$.  Violation of the Harris criterion generically indicates that disorder will either change the universality class of the clean transition
or completely smear it away.  In the former case, one or more Griffiths phases will separate the phases of the clean system~\cite{harris1974effect, vojta2010quantum}.  
The nature of this intervening region depends upon the specific model in question.  
In 1D at $T = 0$, Altman et al.\ found an incompressible Mott glass phase and a quantum phase transition of Kosterlitz-Thouless type between this 
glassy phase and the superfluid\cite{altman2004pha}.  The RG fixed point that controls this transition
actually occurs at a point in the flow diagram where all $U_j = 0$, meaning that the transition can be tuned by only 
varying the disorder in the Josephson couplings at arbitrarily weak interaction strength.

\indent In our work, we introduce disorder by choosing the initial distributions of charging energies and Josephson
couplings, $P_i(U)$ and $P_i(J)$ to have one of the following forms:
\begin{enumerate}
\item Gaussian distributions truncated at three standard deviations:
\begin{equation}
\label{Pgaussian}
P_i(x) \propto \exp{\left[-\frac{(x-x_0)^2}{2\sigma^2_x}\right]}
\end{equation}
for $x \in (x_0-3\sigma_x, x_0+3\sigma_x)$.
\item Power law distributions with upper and lower cutoffs:
\begin{equation}
\label{Ppl}
P_i(x) = \frac{\eta+1}{x_{\text{max}}^{\eta+1}-x_{\text{min}}^{\eta+1}}x^\eta
\end{equation}
for $x \in (x_{\text{min}},x_{\text{max}})$.
\item Flat distributions with upper and lower cutoffs:
\begin{equation}
\label{Pflat}
P_i(x) = \frac{1}{x_{\text{max}}-x_{\text{min}}}
\end{equation}
for $x \in (x_{\text{min}},x_{\text{max}})$.  Of course, this is just a power law with exponent $\eta = 0$.
\item ``Bimodal" distributions consisting of two flat peaks centered at $x_\ell$ and $x_h$:
\begin{equation}
\label{Pbimodal} 
P_i(x) = \frac{1}{2\delta x}
\end{equation}
for $x \in (x_\ell-\frac{\delta x}{2},x_\ell+\frac{\delta x}{2})$ and $x \in (x_h-\frac{\delta x}{2},x_h+\frac{\delta x}{2})$.
\end{enumerate}
All of these distributions have positive lower and upper cutoffs ($x_{\text{min}}$ and $x_{\text{max}}$ respectively) and have zero weight for 
$x$ outside of these bounds.
This restriction avoids the complications of frustration in the phase degrees of freedom and the pathologies of the 
particle sinks that result from on-site charging spectrums that are unbounded from below. 

\indent Even in the presence of disorder, the Hamiltonian (\ref{rot}) respects two important symmetries.  
First, there is the global U(1) phase rotation symmetry:
\begin{equation}
\label{symmU1rot}
\hat{\phi}_j \rightarrow \hat{\phi}_j + \varphi
\end{equation}
This means that the Hamiltonian conserves total particle number:
\begin{equation}
\label{particlenumberconservation}
\hat{n}_{\text{tot}} = \sum_j \hat{n}_j
\end{equation}
The model is also globally particle-hole symmetric: 
\begin{eqnarray}
\hat{n}_j & \rightarrow & -\hat{n}_j \nonumber \\
\hat{\phi}_j & \rightarrow & -\hat{\phi}_j
\label{symmph}
\end{eqnarray}
The particle-hole symmetry exists because the chemical potential coupling to the true particle number $\hat{N}_j$
has been tuned precisely to the value that enforces the large, commensurate filling.  If this chemical potential
is allowed to deviate from this value, then it would manifest in a chemical potential coupling to the particle number
fluctuation $\hat{n}_j$, or equivalently, in offsets to the large filling.  Such terms are absent from our rotor Hamiltonian (\ref{rot}),
but let us momentarily consider the on-site charging spectrum (as a function of $n_j$) in the general situation 
where a filling offset $\delta n_j$ may be present:
\begin{equation}
\label{chargingspectrum}
E_j(n_j) = U_j(n_j - \delta n_j)^2
\end{equation}
The integer value of $n_j$ that minimizes this energy changes at half-integer $\delta n_j$.
For sufficiently strong disorder in $\delta n_j$, the introduction of an arbitrarily small global chemical potential shift
will bring a finite fraction of sites $j$ arbitrarily close to these density changing points.
Thus, a finite density of particles will be added to the system, making it compressible.  Nevertheless, these particles will be localized
by the disordered environment, leaving the system globally insulating.  This is the mechanism behind the formation of the Bose glass\cite{fisher1989boson}.  
The situation changes at two special particle-hole symmetric points.  One occurs when $\delta n_j$ is restricted to be
integer or half-integer.  Then, at the half-integer sites, there is a degeneracy in the 
charging spectrum:
\begin{equation}
\label{degeneratespectrum}
E_j\left(\delta n_j -\frac{1}{2}\right) = E_j\left(\delta n_j + \frac{1}{2}\right)
\end{equation}
In the 1D model, Altman et al.\ showed that this 
degeneracy gives rise to a random singlet glass\cite{altman2010superfluid}, but we will not explore this situation
in this paper.  Instead, we will focus on the other particle-hole symmetric point where $\delta n_j = 0$ for all sites $j$.
Now, if the $U_j$ are distributed such that $U_{\text{min}} > 0$, the on-site charging spectrum always
has a unique minimum that is protected by a gap.  The usual mechanism for Bose glass formation
is evaded, and this allows for the possibility of realizing more exotic glassy phases.  Thus, the particle-hole symmetry 
is a crucial feature of our model that can influence the nature of the intervening glassy phase.

\section{Strong Disorder Renormalization of the 2D Disordered Rotor Model}
\label{sec:RG}

\indent As mentioned previously, the tool that we use to study the disordered
rotor model is the strong disorder real space renormalization method.  
Here we briefly review the method and discuss its application to the model at hand.

\indent At first glance, problems involving strong quenched disorder may appear to be substantially
more complicated than their clean counterparts.  However, one way to motivate the strong disorder renormalization
procedure is to consider that, in some cases, strong disorder can actually serve
as an advantage.  In particular, disorder can make the problem of finding the quantum ground state of a model
more local.  Having identified the strongest of all the disordered couplings in the Hamiltonian, we can then use the 
assumption of strong disorder to argue that other couplings
in the vicinity (in real space) of this dominant coupling are likely to be much weaker.  This means
that the ground state can locally be approximated by satisfying the dominant coupling.  Other 
terms in the Hamiltonian can then be incorporated as corrections.  Quite often, these other terms
are treated by means of perturbation theory, but this need not always be the case.  These
corrections manifest as new couplings in the model, and thus, the procedure yields a new effective
Hamiltonian.  Since part of the ground state is specified in this step, some degrees
of freedom of the system are decimated away.  By repeating the procedure, we can iteratively specify the entire
ground state~\cite{dasgupta1980low,bhatt1982scaling}.  Moreover, we can examine the way in which the probability distributions of the
disordered couplings flow as the renormalization proceeds.  One possibility is that the the model looks
more and more disordered at larger length scales near criticality.  This is the case for random transverse 
field Ising models in one and two dimensions~\cite{fisher1999phase,kovacs2010renormalization,motrunich2000infinite}.  
In such cases, the model flows towards infinite randomness, and the strong disorder renormalization group becomes
asymptotically exact near criticality\cite{fisher1994random}. Of course, it is also possible to flow towards finite or weak disorder,
and in these cases, the reliability of the RG is less clear.  We discuss this issue extensively, as it pertains to
the disordered rotor model, in Appendix D.

\subsection{The Basic RG Steps}

\indent We now concretize these ideas by application to the rotor Hamiltonian (\ref{rot}).  In
our model of random bosons in 2D, there are two types of disordered
couplings, namely charging energies and Josephson couplings.  In each step of the
renormalization, we identify the maximum of all of these couplings, which defines the RG scale:
\begin{equation}
\label{RGscale}
\Omega = \text{max}\left[\{U_j\},\{J_{jk}\}\right]
\end{equation}
How we then proceed depends upon which type of coupling is dominant.  

\subsubsection{Site Decimation}

\begin{figure}
\centering
\includegraphics[width=8cm]{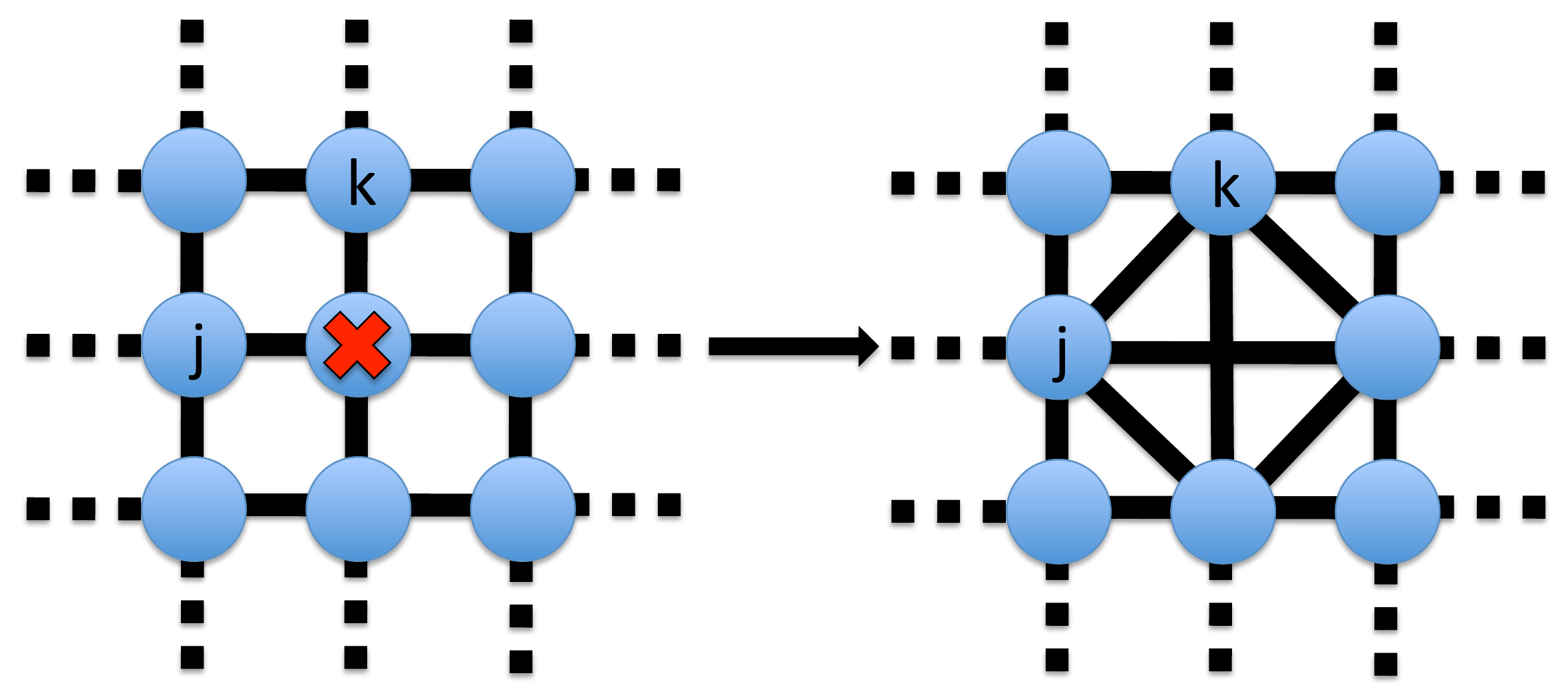}
\caption{The site decimation RG step: The site marked with the $X$ has the dominant charging energy and
is decimated away, generating bonds between neighboring sites $j$ and $k$ with the effective coupling
given in equation (\ref{sitedec}).  The new local structure of the lattice is shown to the right.}
\label{sitedecfig}
\end{figure}

\indent Consider the case where the charging energy on site $X$ is dominant.  We define a local Hamiltonian 
in which this charging energy term is chosen to be the unperturbed piece.  All Josephson couplings entering 
the corresponding site are considered to be perturbations:
\begin{equation}
\label{Hjsd}
\hat{H}_{X} = U_X \hat{n}^2_X - \sum_{k} J_{Xk} \cos{(\hat{\phi}_X-\hat{\phi}_k)}
\end{equation}
Satisfying the dominant coupling means setting $n_X = 0$ to leading order.  This defines a 
degenerate manifold of local ground states: $\ket{0,\{n_k\}}$.  In these kets, the first term
corresponds to zero number fluctuation on site $X$ and the second specifies the number
fluctuations on all sites connected to $X$ by a Josephson coupling.  The degenerate space
is infinitely large, corresponding to all possible choices of $\{n_k\}$.  However, all matrix
elements of the perturbative Josephson couplings in this ground state manifold vanish.
The leading corrections then come from second order degenerate perturbation
theory, in which we calculate corrections coming from excited states:
\begin{eqnarray}
\ket{0,\{n_k\}}' & \approx & \ket{0,\{n_k\}}  \\
                          & \quad & + \sum_{m \in k} \frac{J_{Xm}}{2U_X}(\ket{1,n_m-1}  + \ket{-1,n_m+1}) \nonumber
\label{sdperturbed}
\end{eqnarray}
In the terms giving the perturbative corrections, we assume that the number fluctuations on
all neighboring sites except $m$ remain unmodified from their values in $\{n_k\}$.
We next consider the matrix elements of these states in $\hat{H}_{X}$.  Up to a constant term,
these matrix elements are identically those that would result from an effective Josephson 
coupling:
\begin{equation}
\label{sitedec}
\tilde{J}_{jk} = \frac{J_{jX}J_{Xk}}{U_X}
\end{equation}
between each two sites that were coupled to site $X$ before the decimation step\cite{altman2004pha}.  This
process of site decimation is illustrated in Figure \ref{sitedecfig}.

\subsubsection{Link Decimation}

\begin{figure}
\centering
\includegraphics[width=8cm]{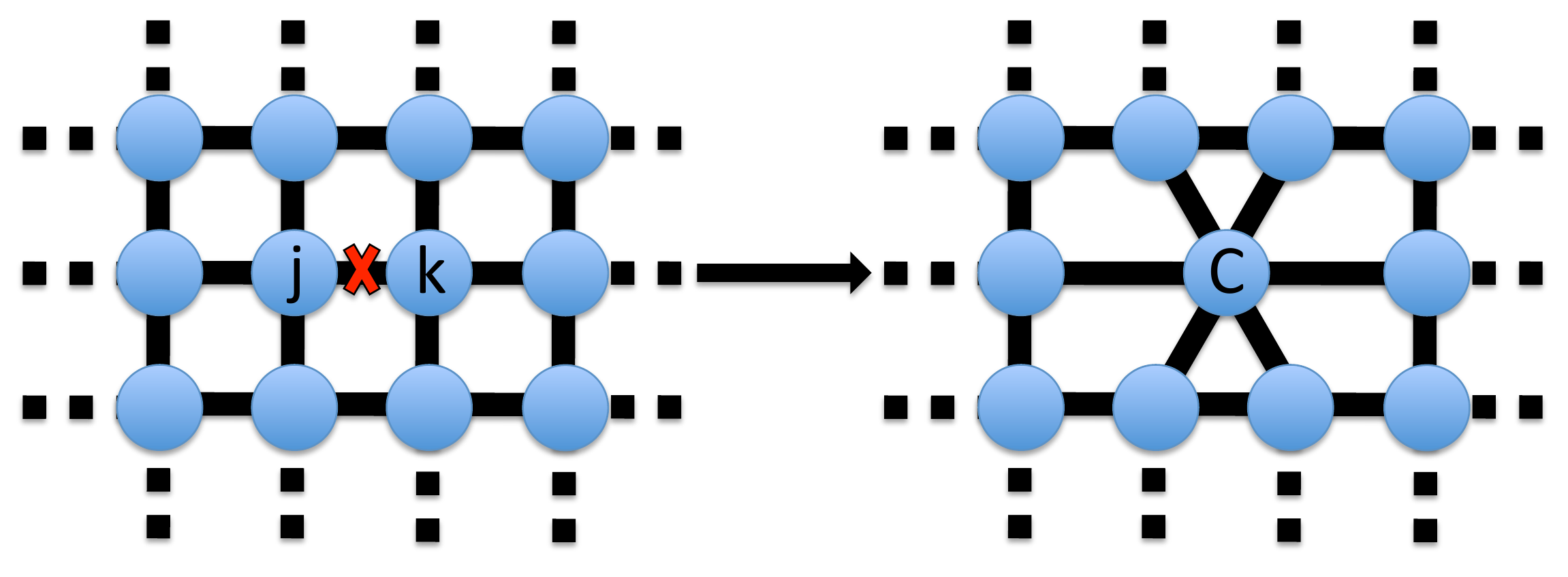}
\caption{The link decimation RG step: The crossed link has the dominant Josephson coupling.  The two sites
it joins are merged into one cluster, resulting the effective lattice structure shown to the right.  The cluster $C$ has
an effective charging energy given in equation (\ref{linkdec}).}
\label{linkdecfig}
\end{figure}

\indent Now, suppose that a Josephson coupling is the dominant energy scale.  In this case, the local Hamiltonian is:
\begin{equation}
\label{Hjkld}
\hat{H}_{jk} = U_j \hat{n}^2_j+U_k \hat{n}^2_k - J_{jk} \cos{(\hat{\phi}_j-\hat{\phi}_k)}
\end{equation}
The local approximation to be made here is that, to lowest order, the phases on these adjacent sites should be 
locked together.  In other words, the degree of freedom to be specified is the relative phase $\phi_j-\phi_k$.  This motivates
the introduction of new operators:
\[\hat{n}_C = \hat{n}_j + \hat{n}_k\]
\[\hat{\phi}_C = \frac{U_k \hat{\phi}_j + U_j \hat{\phi}_k}{U_j+U_k}\]
\[\hat{n}_R = \frac{U_j \hat{n}_j - U_k \hat{n}_k}{U_j+U_k}\]
\begin{equation}
\label{ldclustreloperators}
\hat{\phi}_R = \hat{\phi}_j - \hat{\phi}_k
\end{equation}
These operators satisfy the commutation relations:
\[[\hat{\phi}_C,\hat{n}_C] = i\]
\begin{equation}
\label{ldclustrelcommutators}
[\hat{\phi}_R,\hat{n}_R] = i
\end{equation}
with all other commutators vanishing.  Thus, the transformation preserves the algebra of number and phase
operators.  A subtlety arises for the relative coordinate operators $\hat{n}_R$ and $\hat{\phi}_R$ because, as defined above,
$n_R$ need not be an integer and $\phi_R \in [-2\pi,2\pi)$ as opposed to $\phi_R \in [0,2\pi)$.  To deal with this
difficulty, we may make the additional approximation of treating $\phi_R$ as a noncompact variable.  This makes
$n_R$ continuous instead of discrete.   
Then, in terms of the new cluster and relative coordinate operators, the local Hamiltonian (\ref{Hjkld}) reads:
\begin{equation}
\label{Hjkld2}
\hat{H}_{jk} = \frac{U_j U_k}{U_j+U_k} \hat{n}^2_C+(U_j+U_k) \hat{n}^2_R - J_{jk} \cos{(\hat{\phi}_R)}
\end{equation}
To lowest order, we set $\phi_R = 0$.  This decimation of the relative phase leaves the cluster phase $\phi_C$
unspecified, so two phase degrees of freedom have been reduced to one.  The first term in $\hat{H}_{jk}$ shows that the 
inverse charging energies add like the capacitances of capacitors in parallel to give the charging energy for
the cluster:
\begin{equation}
\label{linkdec}
\tilde{U}_C = \frac{1}{\frac{1}{U_j}+\frac{1}{U_k}} = \frac{U_j U_k}{U_j+U_k}
\end{equation}
Figure \ref{linkdecfig} depicts this process of link decimation\cite{altman2004pha}.

\indent Higher order corrections to this picture, arising from harmonic vibrations of the
phases that make up the cluster, can be obtained by considering the part of the local Hamiltonian involving
the relative coordinate.  These terms act approximately like a simple harmonic oscillator Hamiltonian on the basis of $\hat{n}_R$
eigenstates.  Thus, the ground state for the relative coordinate can be approximated by a simple harmonic
oscillator ground state:
\begin{equation}
\label{lhogs}
\ket{\psi_R} \approx \frac{\gamma^\frac{1}{8}}{\pi^\frac{1}{4}} \int_{-\infty}^{\infty} dn_R \exp{\left[-\frac{\gamma^\frac{1}{2}}{2}n^2_R\right]}\ket{n_R}
\end{equation}
with:
\begin{equation}
\label{ldgamma}
\gamma = \frac{2(U_j+U_k)}{J_{jk}}
\end{equation}

\indent This approximation is used in the numerics to compute \textit{Debye-Waller factors}
that modify Josephson couplings entering the newly formed cluster.  Quantum fluctuations of $\phi_R$
weaken the phase coherence of the cluster, and consequently, suppress these Josephson couplings.
Mathematically, the Debye-Waller factors arise because, in writing down the local, two site Hamiltonian 
(\ref{Hjkld}), we have neglected that $\hat{\phi}_R$ also appears in the other links penetrating the two sites $j$
and $k$.   Consider a Josephson coupling from a third site $m$ to the site $j$.  This corresponds to 
a term in the full Hamiltonian (\ref{rot}):
\begin{eqnarray}
\cos{(\hat{\phi}_m-\hat{\phi}_j)} & = & \cos{\left(\hat{\phi}_m-\hat{\phi}_C-\mu_1\hat{\phi}_R\right)} \nonumber \\
                                                        & = & \cos{\left(\hat{\phi}_m-\hat{\phi}_C\right)}\cos{\left(\mu_1\hat{\phi}_R\right) } \nonumber \\
                                                        & \quad &  + \sin{\left(\hat{\phi}_m-\hat{\phi}_C\right)}\sin{\left(\mu_1\hat{\phi}_R\right) } \nonumber \\
                                                        & \approx & \cos{\left(\hat{\phi}_m-\hat{\phi}_C\right)}\langle\cos{\left(\mu_1\hat{\phi}_R\right)}\rangle + \nonumber \\
                                                        & \quad & \sin{\left(\hat{\phi}_m-\hat{\phi}_C\right)}\langle\sin{\left(\mu_1\hat{\phi}_R\right)}\rangle
\label{lddwcalculation}                                                      
\end{eqnarray}
with:
\begin{equation}
\label{lddwmu}
\mu_j = \frac{U_j}{U_j+U_k}
\end{equation}
The angle brackets in the final line of equation (\ref{lddwcalculation}) refer to averages taken in the relative coordinate ground state (\ref{lhogs}).
The expectation value of the sine vanishes, and the expectation value of the cosine yields the Debye-Waller
factor:
\begin{equation}
\label{dwj}
c_{\text{DW},j} \approx \frac{\sin^2(\pi \mu_j)}{\pi^2} \sum_{q = -\infty}^\infty \frac{(q^2+\mu^2_j)}{(q^2-\mu^2_j)^2}\exp{\left(-\frac{\gamma^\frac{1}{2}}{4}q^2\right)}
\end{equation}
In the numerics, we truncate the calculation of this sum at a specified order, $|q_{\text{max}}| = 20$, and multiply the
Josephson coupling $J_{mj}$ by the result to find the new Josephson coupling $\tilde{J}_{mC}$ penetrating
the cluster.  Note that the Debye-Waller factor for links penetrating the site $k$ is, in general, not equal to $c_{\text{DW},j}$,
but its calculation is completely analogous.

\subsection{Adaptations for 2D}

\subsubsection{Sum Rule}

\begin{figure}
\centering
\includegraphics[width=8cm]{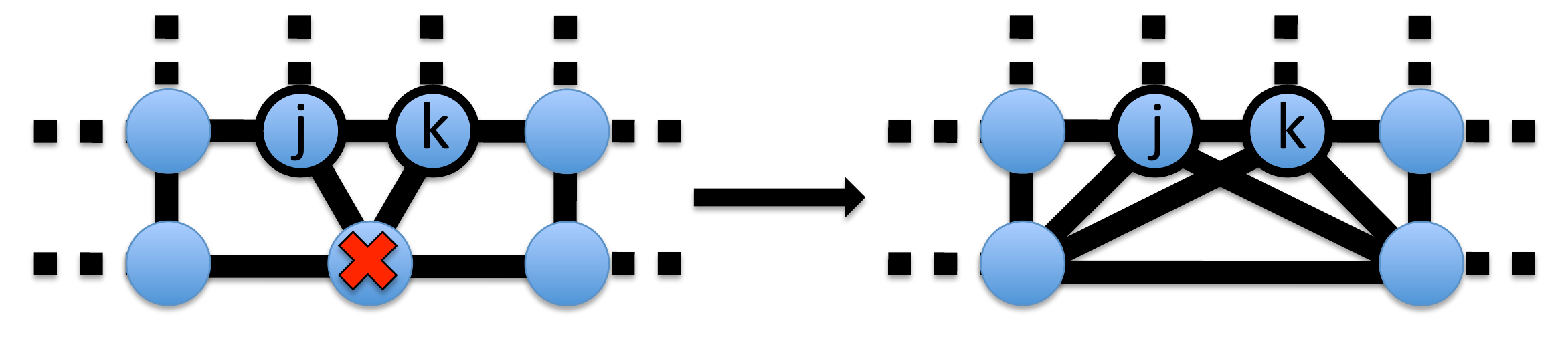}
\caption{Site decimation with the sum rule: The site with the dominant charging energy (marked with an $X$) 
is coupled to two sites ($j$ and $k$) that are already coupled to one another.   
After site decimation, the effective Josephson coupling between sites $j$ and $k$ is the sum of the old coupling 
and the effective coupling generated through decimation of site $X$ (see equation (\ref{specsitedec})).}
\label{spsitedecfig}
\end{figure}

\begin{figure}
\centering
\includegraphics[width=8cm]{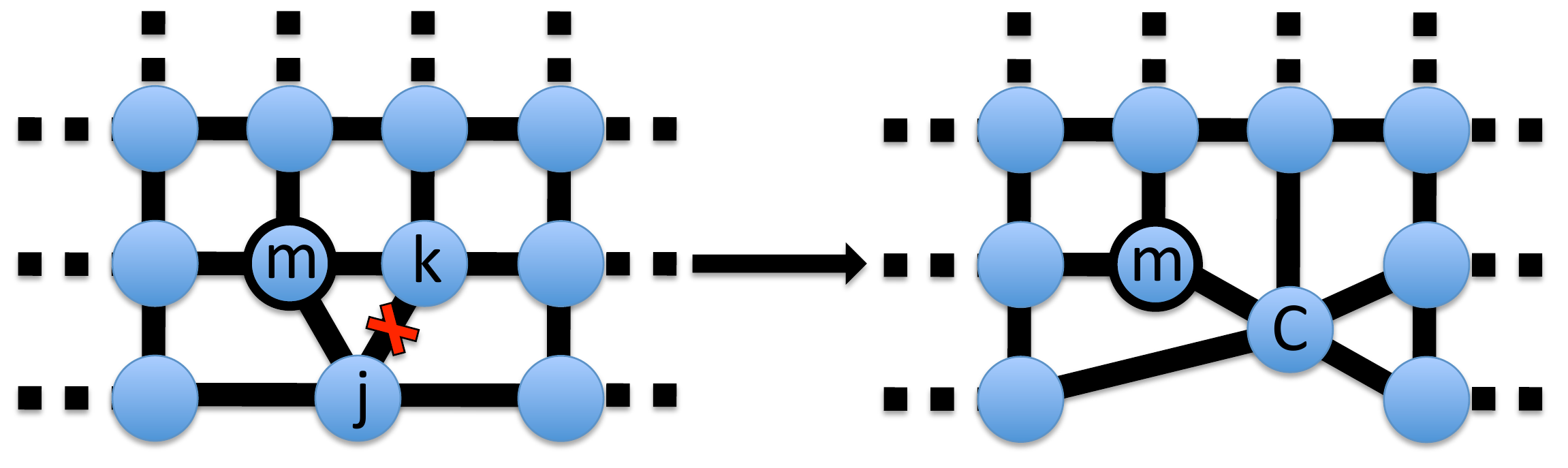}
\caption{Link decimation with the sum rule: The two sites connected by the dominant Josephson coupling (sites $j$ and $k$) 
are both coupled to a third site $m$.  Following link decimation, the effective Josephson coupling between site $m$ and the cluster $C$ 
is the sum of the two preexisting couplings between site $m$ and sites $j$ and $k$ (see equation (\ref{speclinkdec})), up to corrections coming 
rom Debye-Waller factors.}
\label{splinkdecfig}
\end{figure}

\indent As shown by Altman, Kafri, Polkovnikov, and Refael, the two renormalization steps outlined above 
form the basis of the strong disorder renormalization group for the disordered rotor model in 1D\cite{altman2004pha}.  
In higher dimensions, the geometry of the lattice changes as the renormalization proceeds, and this presents complications.
For example, in Figure \ref{spsitedecfig}, decimation of a site $X$ produces an effective Josephson coupling between two sites $j$ and $k$ 
that are already linked to one another by a preexisting Josephson coupling.  In our numerics, we sum the preexisting and new coupling to form 
the effective coupling between the remaining sites:
\begin{equation}
\label{specsitedec}
\tilde{J}_{jk} = J_{jk}+\frac{J_{jX}J_{Xk}}{U_X}
\end{equation} 
A similar situation can arise during link decimation.  In Figure \ref{splinkdecfig}, a cluster is formed by two sites $j$ and $k$, each of which is connected to a
third site $m$.  Up to corrections coming from Debye-Waller factors, the effective Josephson coupling joining site $m$ to the cluster is:
\begin{equation}
\label{speclinkdec}
\tilde{J}_{mC} = J_{mj}+J_{mk}
\end{equation} 

\indent Some authors replace the \textit{sum rule} in equations (\ref{specsitedec}) and (\ref{speclinkdec}) with a \textit{maximum rule} 
~\cite{motrunich2000infinite,kovacs2010renormalization,fisher1999phase}.  The motivation behind the maximum rule is that, in the strong disorder limit:
\begin{equation}
\label{maxrule}
\text{max}\left[J_{jk},\frac{J_{jX}J_{Xk}}{U_X}\right] \approx J_{jk}+\frac{J_{jX}J_{Xk}}{U_X}
\end{equation}
This should be a good approximation in an infinite disorder context.  For our model however, we find that the sum rule increases the
class of distributions which find the unstable fixed point depicted schematically in Figure \ref{flows}.  For further discussion of 
the difference between the sum and maximum rules, please consult Appendix A. 

\subsubsection{Thresholding}

\indent In dimensions greater than one, there is a tendency for the numerics to slow down considerably if the renormalization procedure involves a lot of site decimation.  
Again, the source of the problem is the evolution of the lattice under the RG.  If a site $X$ is decimated, then effective
links are generated between each pair of sites that were previously coupled to $X$.  Thus, site decimation generates many new couplings, increasing
the coordination number of the effective lattice.  At the same time, the site decimation step takes computer time quadratic in the 
coordination number of the site being decimated.  To apply the procedure to large lattices, it is necessary to find a way to circumvent
this difficulty.

\indent At the beginning of the RG, we specify a thresholding parameter, which we call $\alpha$.  During a site decimation, 
if a new Josephson coupling is created between sites $j$ and $k$ such that:
\begin{equation}
\label{threshold}
\tilde{J}_{jk} = \frac{J_{jX}J_{Xk}}{U_X} < \alpha U_X = \alpha \Omega
\end{equation} 
then the coupling is thrown away.  For convenience in implementation, the new bond is ignored only if it does not sum with a preexisting
Josephson coupling.  If $\alpha$ is chosen to be very small, then ignoring the coupling will hopefully not 
affect the future course of RG.  However, to be more careful, it is better to perform an extrapolation in the threshold $\alpha$ 
to see if the numerics converge.  Using this thresholding procedure, we are able to reach lattices up to size $300\times300$, 
if we additionally require averaging over a reasonably large number of disorder samples.  
In this paper, unless otherwise stated, we always use $10^3$ samples for any given choice of distributions.

\subsubsection{Distribution Flows}

\indent Typically, in an application of the strong disorder renormalization method, it is 
interesting to monitor the flow of the distributions of the various couplings as the RG proceeds. 
This is straightforward for a 1D chain, but in higher dimensions, there is yet again a 
complication from the evolving lattice structure.  As the renormalization proceeds, it is possible
to generate very highly connected lattices.  Many of the effective Josephson couplings will, however,
be exceedingly small. Incorporating these anomalously small couplings into statistics can be misleading.
Despite the large number of weak bonds, there may exist a number of strong bonds sufficient to produce superfluid clusters. 
In fact, including the weak bonds in statistics is analogous to polluting the statistics with the inactive next-nearest 
neighbor Josephson couplings in the original lattice.  It is more appropriate to follow Motrunich et al.\ and focus
on the largest $O(\tilde{N})$ Josephson couplings, where $\tilde{N}$ is the number of sites remaining in the 
effective lattice\cite{motrunich2000infinite}.  In the remainder of the paper, the ``Josephson 
coupling distribution" will therefore refer solely to the dominant $2\tilde{N}$ effective Josephson couplings
at any stage in the RG, and all statistics will be done only on these $2\tilde{N}$ couplings.

\section{Numerical Application of the Strong Disorder RG}
\label{sec:data}

\indent In this section, we present numerical data collected from our implementation
of the strong disorder renormalization group.  First, we explore the strong disorder RG flows of the
distributions of charging energies and Josephson couplings.  This investigation points to the existence
of an unstable fixed point of the RG flow.  We find that the presence of this fixed point is robust to many different
changes in the choices of the initial distributions.  Next, we examine the distributions generated by 
the RG near this fixed point and find that universal physics arises in its vicinity.  Subsequently,
we proceed away from the fixed point to study properties of the phases of the disordered rotor model.
We find phases that we tentatively identify as Mott insulating, glassy, and superfluid, and furthermore,
we find that the unstable fixed point governs the putative glass-superfluid transition.  We defer
detailed interpretation of the data and analysis of the transition to Section \ref{sec:phases}.

\subsection{Flow Diagrams and the Finite Disorder Fixed Point}

\begin{figure}
\centering
\includegraphics[width=8cm]{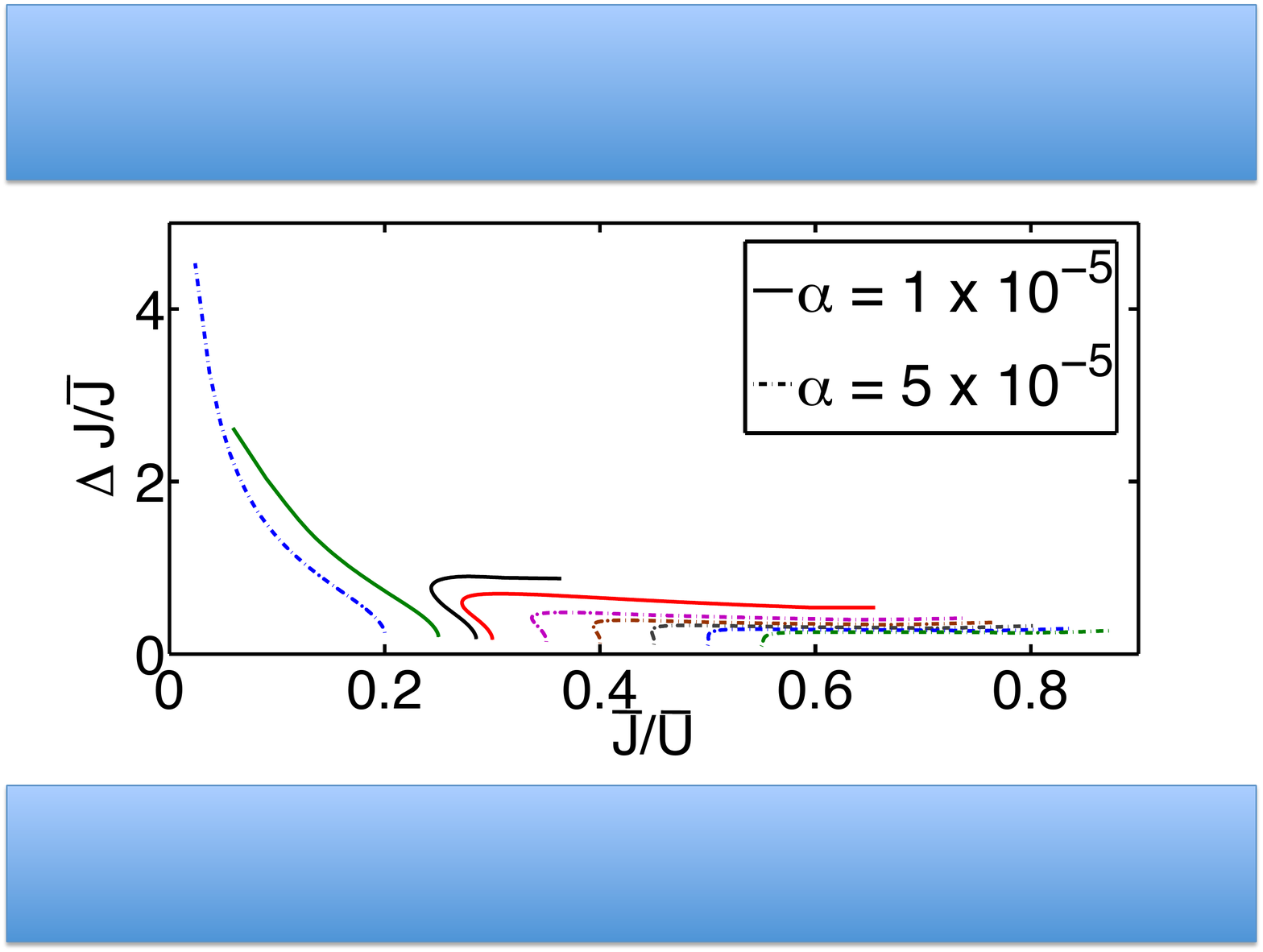}
\caption{The projection, in the $\Delta J/\bar{J}$ vs. $\bar{J}/\bar{U}$ plane, of the flows of the coupling distributions at different stages of the RG.  The initial distributions $P_i(U)$ and $P_i(J)$ are both truncated Gaussians, and $J_0$ (the center of the initial $J$ distribution) is used as the tuning parameter. Each flow corresponds to a different choice of the tuning parameter.  The flows start at the bottom of the figure and go up and to the left or up and to the right.  A smaller value of the thresholding parameter is used near criticality as indicated by the legend.  All runs were done on $L = 100$ lattices.}
\label{flowg}
\end{figure}

\begin{figure}
\centering
\includegraphics[width=8cm]{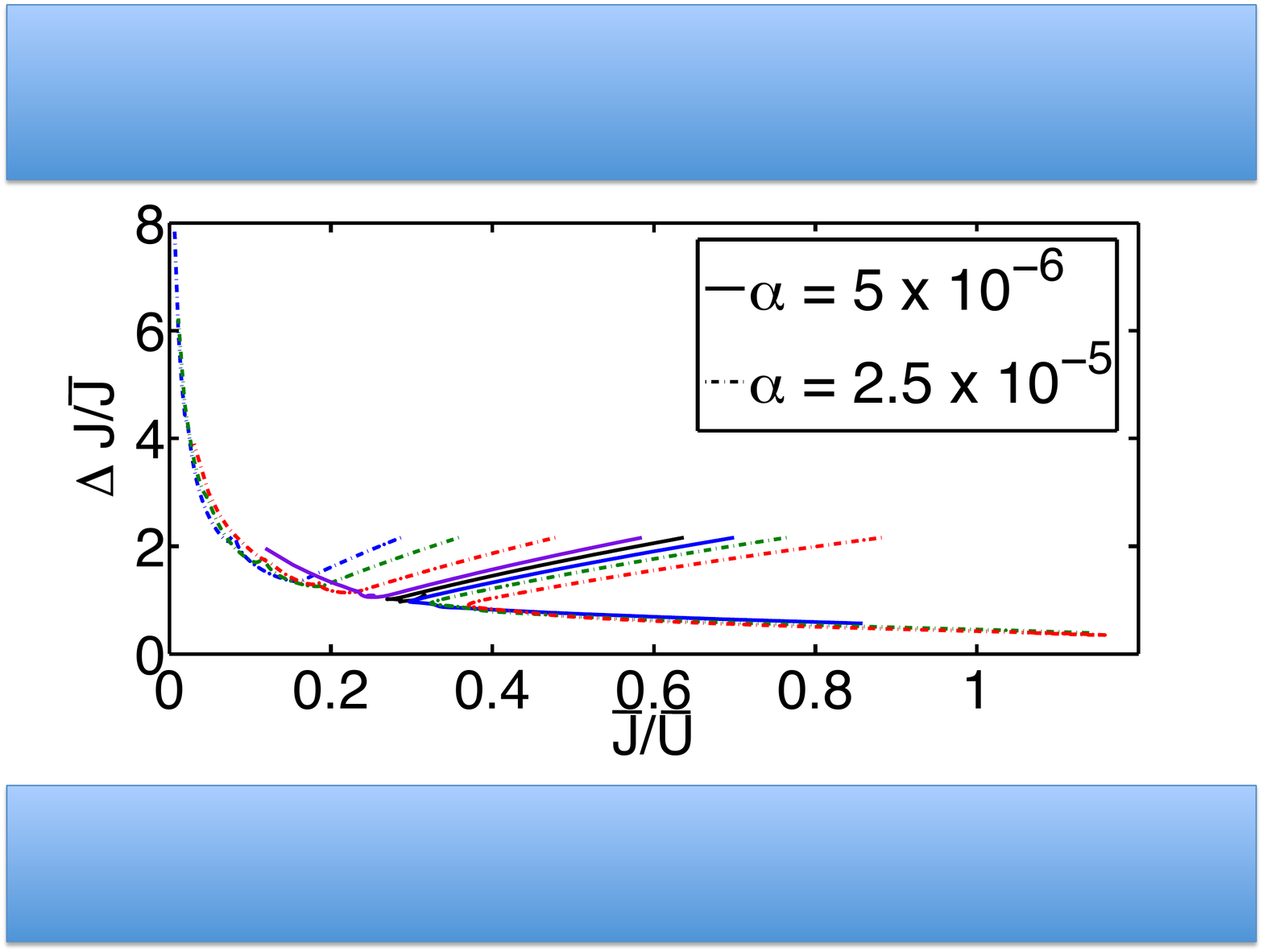}
\caption{Same as Figure ~\ref{flowg}, except $P_i(U)$ is Gaussian and $P_i(J) \propto J^{-1.6}$, with cutoffs chosen to make the latter distribution very broad.  The parameter $U_0$ is used to tune through the transition.  The flows begin near the center of the figure.  To the left of the figure, flows initially propagate towards the bottom left but eventually turn around and propagate towards the top left.  To the right of the figure, flows initially propagate towards the bottom left but eventually turn around and propagate towards the bottom right.  All runs were done on $L = 300$ lattices.}
\label{flowJplUg}
\end{figure}

\begin{figure}
\centering
\includegraphics[width=8cm]{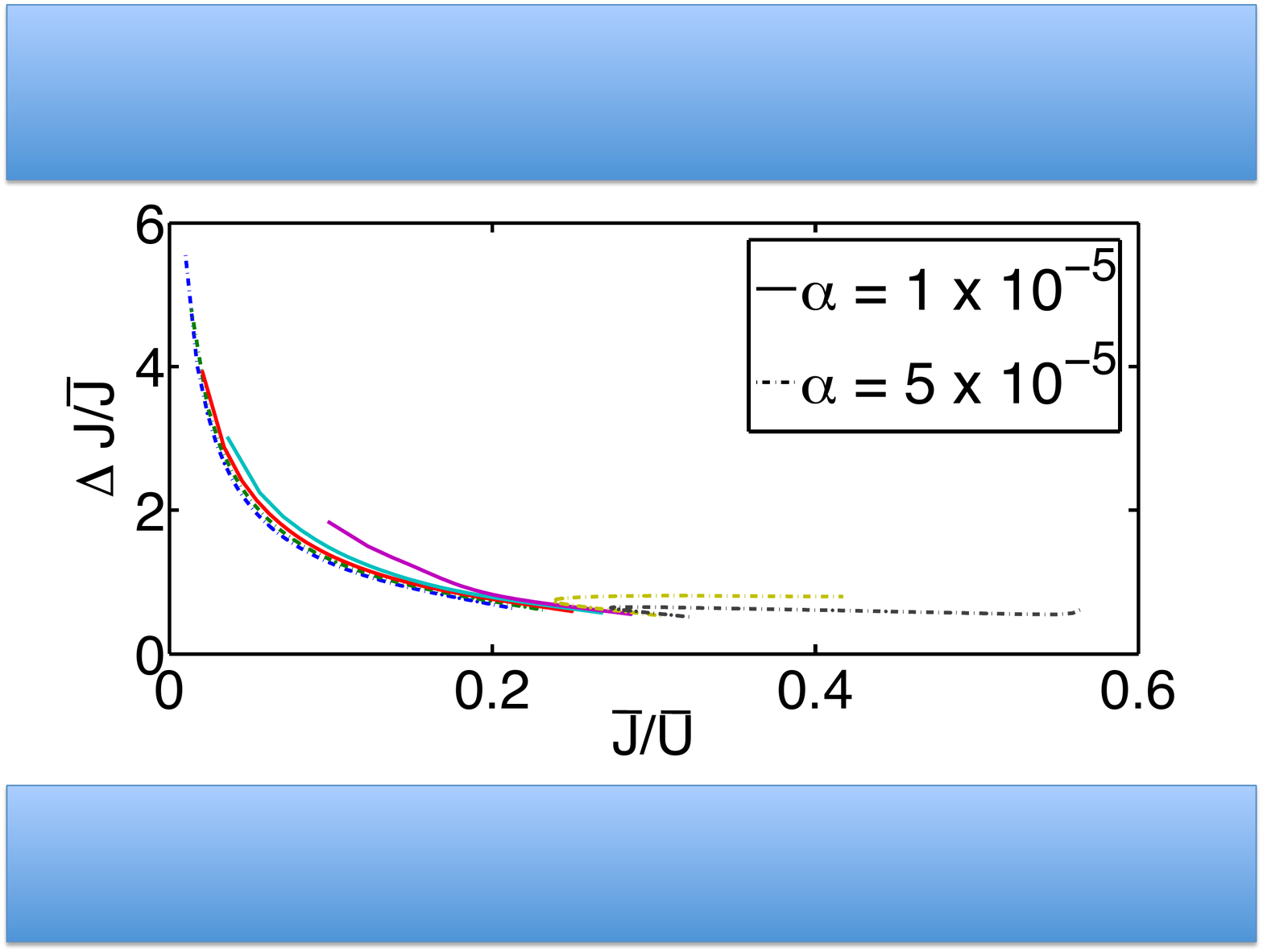}
\caption{Same as Figure ~\ref{flowg}, except $P_i(U) \propto U^5$ and $P_i(J) \propto J^{-3}$.  The tuning parameter is $J_{\text{min}}$, the lower cutoff of $P_i(J)$.  All runs were done on $L = 150$ lattices.}
\label{flowJ3U5}
\end{figure}

\indent In Figures \ref{flowg}-\ref{flowJ3U5}, we plot 
flows of quantities that characterize the Josephson coupling and charging energy distributions.  
We emphasize again that, in the context of our study of distribution flows, the ``Josephson coupling distribution" actually only includes the greatest $2\tilde{N}$ Josephson couplings,
where $\tilde{N}$ is the number of sites remaining in the effective lattice.  
After $M$ decimation steps of the RG, we stop the procedure and look at the remaining charging energies and these dominant Josephson couplings.  
We then use these values to update estimates for the mean and standard deviation corresponding to that step in the RG.  
For each realization of the randomness (i.e.~each sample), we do this for many different choices of $M$, and we repeat the process for $10^3$ 
realizations of the randomness.  Ultimately, this procedure gives a ``flow" that characterizes the disorder-averaged evolution of the 
distributions at different stages of the renormalization.

\indent The x-axes of Figures \ref{flowg}-\ref{flowJ3U5} give the ratio of the means of the two distributions.  Meanwhile, the y-axes 
give the ratio of the standard deviation of the Josephson coupling distribution to the mean of the distribution.  The plots actually show 2D 
projections of flows that occur in the space of all possible distributions.  At the very least, these plots imply the existence of a third axis, namely
$\frac{\Delta U}{\bar{U}}$, which may carry important information.  Nevertheless, these highly simplified 2D pictures are surprisingly effective
in describing the fate of the model with different choices of parameters.  In interpreting Figures \ref{flowg}-\ref{flowJ3U5}, the reader will likely
find it helpful to glance back to Figure \ref{flows}, which shows a schematization of the flows.

\indent Figure \ref{flowg} specifically corresponds to flows for initial distributions $P_i(U)$ and $P_i(J)$ that are Gaussian.  
The center of the Josephson coupling distribution is used as the tuning parameter.  To the left of the plot are two flows that propagate to the top left of the diagram, towards small $\frac{\bar{J}}{\bar{U}}$ and large
$\frac{\Delta J}{\bar{J}}$.  Since these flows propagate towards high $\bar{U}$, it is tempting to identify them as flowing towards an insulating regime.  
Meanwhile, to the right of the plot, there are seven flows that propagate towards high $\bar{J}$, and it is tempting to identify these as 
propagating towards a superfluid regime.  At the interface between these two behaviors, the flows ``slow down" and travel a shorter distance in the plane.
This behavior is suggestive of a separatrix flow that terminates at an unstable fixed point, as shown in Figure \ref{flows}. 

\indent Our next goal will be to show that the behavior indicating the presence of this unstable fixed point is robust to changes in the choice of the initial distributions.  
In Figure \ref{flowJplUg}, $P_i(U)$ is a Gaussian, and $P_i(J) \propto J^{-1.6}$.  The center of the charging energy distribution, $U_0$, is used as the tuning parameter.  
The numerical choices place the flows initially above and to the right of  the location of the unstable fixed point in the previous figure. From the point of view of the
strong disorder RG procedure, this choice of initial distributions is advantageous, because the flows begin in a regime of high disorder in $J$, where the procedure
is more accurate.  Later in the paper, after presenting evidence of the universal physics that emerges in the disordered rotor model, we will focus on this choice of distributions exclusively.  Therefore, we have collected
additional details about these distributions in Appendix C.  Note that the leftmost flows in Figure \ref{flowJplUg} 
initially propagate towards the lower left hand corner of the figure; then, they turn upward, continuing onward to lower $\frac{\bar{J}}{\bar{U}}$ but now also towards high $\frac{\Delta J}{\bar{J}}$.  
Hence, they share the same qualitative fate as the leftmost flows in Figure \ref{flowg}.  To the right of Figure \ref{flowJplUg},  the flows initially also propagate towards
the lower left; however, these flows ultimately turn around and propagate towards high $\frac{\bar{J}}{\bar{U}}$.  The separatrix that divides these two classes
of flows appears to terminate in the same critical region that was seen in Figure \ref{flowg}.

\begin{figure}
\begin{minipage}[b]{0.4cm}
       {\bf (a)}
       
       \vspace{3.3cm}
\end{minipage}
\begin{minipage}[t]{7.9cm}
       \includegraphics[width=7.8cm]{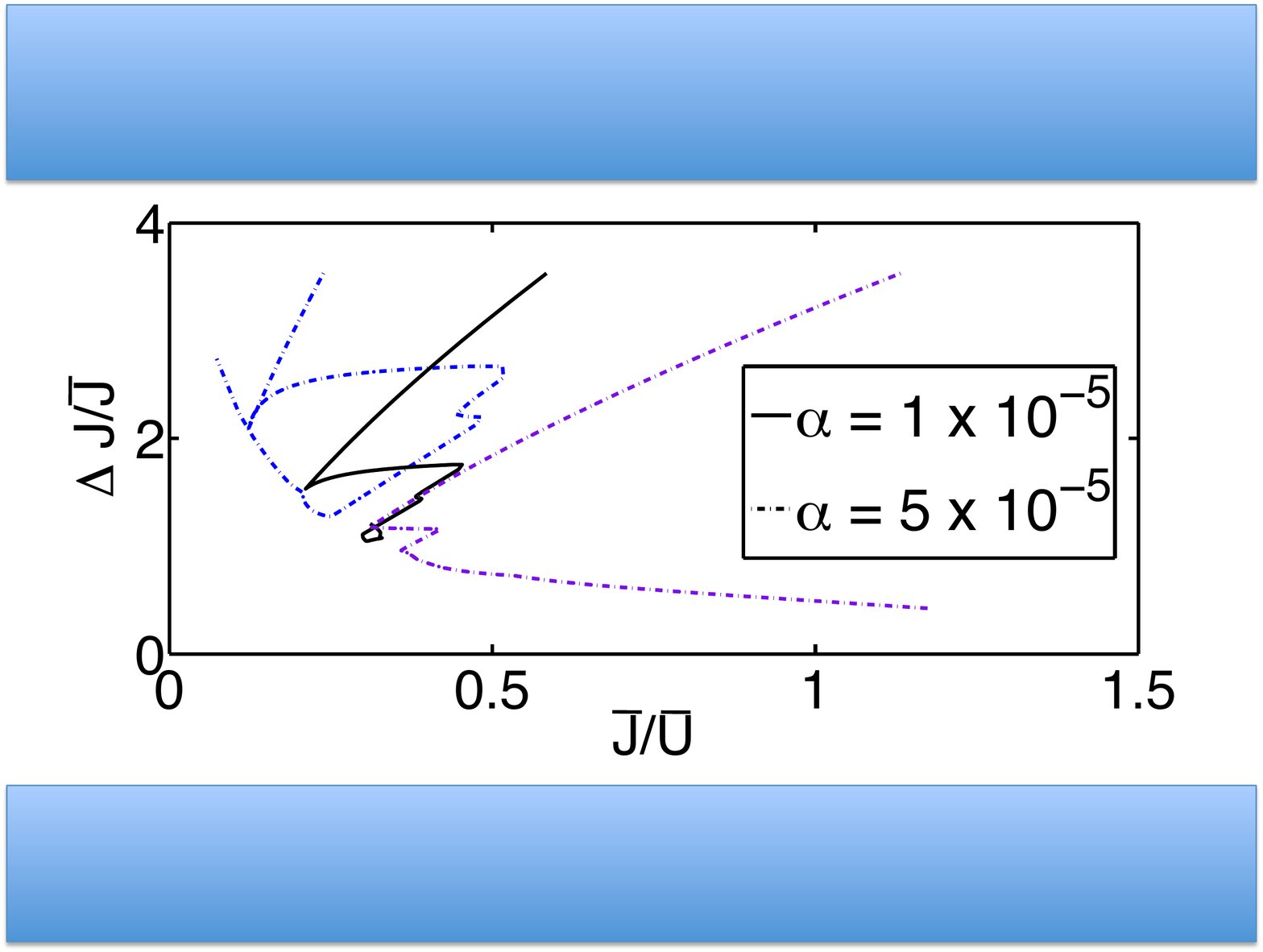}
\end{minipage}\\
\begin{minipage}[b]{0.4cm}
       {\bf (b)}

       \vspace{3.3cm}
\end{minipage}
\begin{minipage}[t]{7.9cm}
       \includegraphics[width=7.8cm]{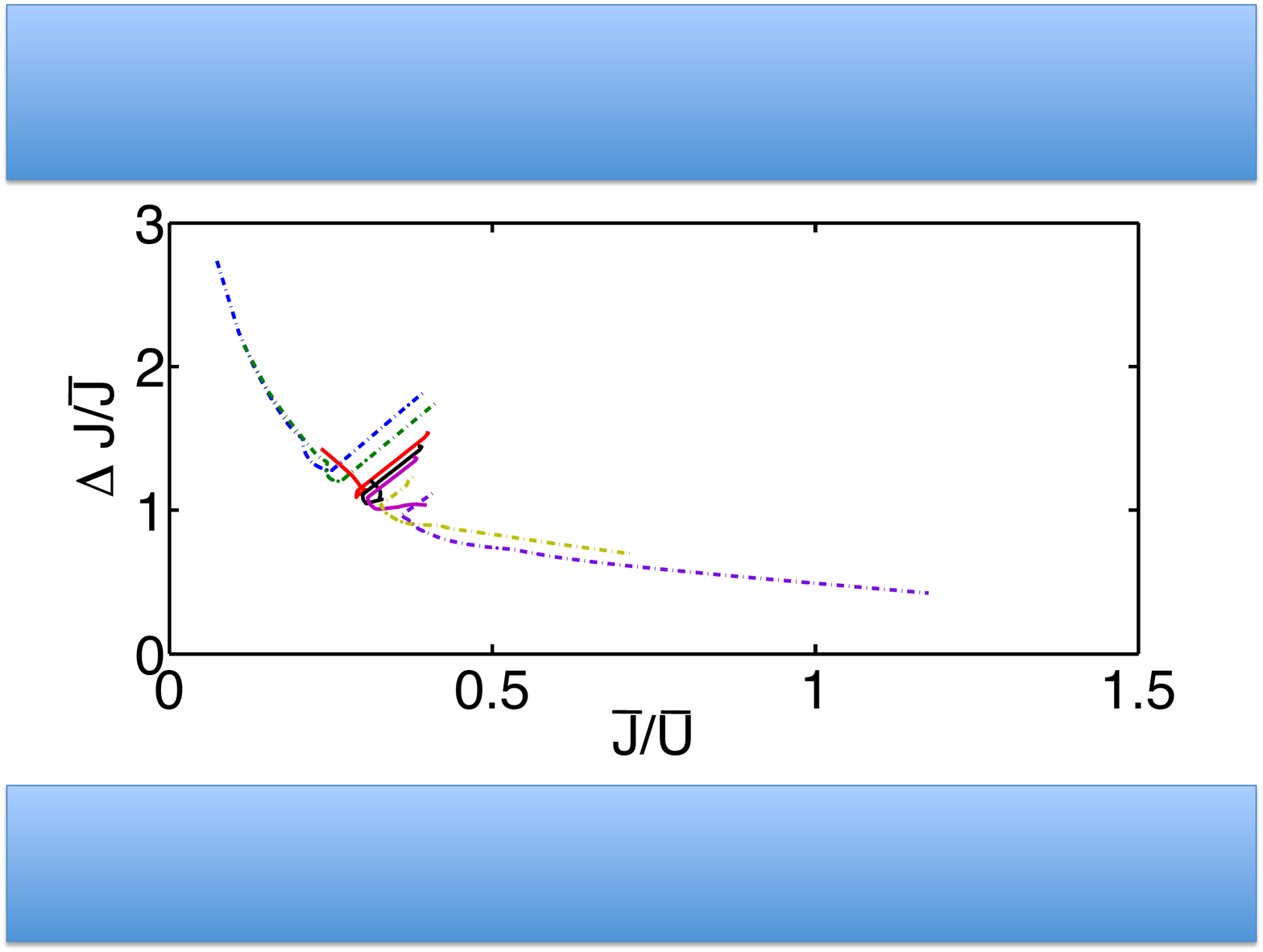}
\end{minipage}\\

\caption{In these numerical flow diagrams, $P_i(U)$ is bimodal and $P_i(J) \propto J^{-1.6}$, with cutoffs chosen to make the
latter distribution very broad.  In panel (a), we show a few sample flows that start near the top of the figure and initially propagate towards the lower left hand corner.  
The complex features of these flows reflect the structural details of the bimodal $U$ distribution.  
In panel (b), we exhibit the flows at late RG times, when the procedure has had an opportunity to renormalize away the microscopic details of the initial distributions.  Then, the flows are, at least
qualitatively, more similar to those seen in Figure \ref{flowJplUg}.  All runs were done on $L = 200$ lattices.}
\label{flowJplUb}
\end{figure}

\indent In Figure \ref{flowJ3U5}, we make yet another choice of initial distributions.  Now, $P_i(J) \propto J^{-3}$ and $P_i(U) \propto U^5$.  
The resulting flow diagram again suggests the presence of an unstable fixed point in the same critical region.  It would be misleading, however, 
to suggest that every flow diagram generated by the RG will have the nice properties of Figures \ref{flowg}-\ref{flowJ3U5}.  We provide a counterexample in Figure \ref{flowJplUb}, 
in which $P_i(U)$ is bimodal and $P_i(J) \propto J^{-1.6}$.   Panel (a) shows the extremely complicated behavior of some of the flows.  
These features are reflections of the structural details of the bimodal distribution.  We will see shortly that, at least in the vicinity of criticality, the RG works to wash away these details 
and construct universal distributions.  After these universal distributions are somewhat well approximated, the flows should be more well behaved, but in Figure \ref{flowJplUb}, 
we see a non-universal era of the flows, where the complexities of the initial distributions can manifest in complicated flows.  To bring out this point more clearly, we have
removed data for the early stages of the RG in panel (b).  Now the ``late RG-time" behavior of the flows falls more nicely in line with what is seen in Figure \ref{flowJplUg}.

\subsection{Universal Distributions}

\indent Near the unstable finite disorder fixed point of the RG flow, we expect universal physics
to emerge.  Certain aspects of the critical behavior should be independent of microscopic details, including
the structure of the initial distributions.  The universality of the fixed point should become evident in the forms of the \textit{renormalized}
distributions generated through the RG: whatever the initial distributions may be, they should evolve towards universal forms, 
provided that they put the system near criticality.  

\begin{figure}
\begin{minipage}[b]{0.4cm}
       {\bf (a)}
       
       \vspace{3.3cm}
\end{minipage}
\begin{minipage}[t]{7.9cm}
       \includegraphics[width=7.8cm]{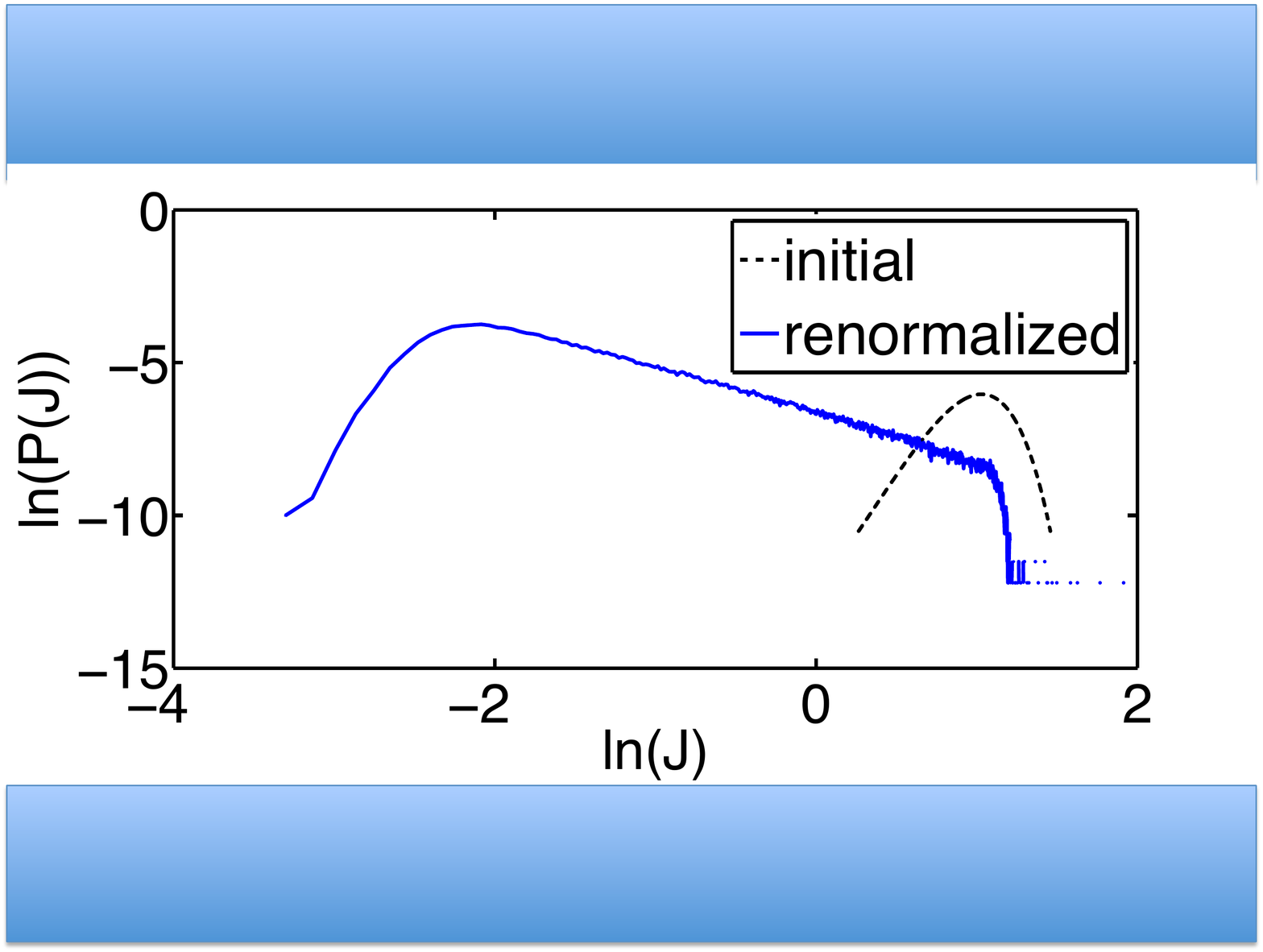}
\end{minipage}\\
\begin{minipage}[b]{0.4cm}
       {\bf (b)}

       \vspace{3.3cm}
\end{minipage}
\begin{minipage}[t]{7.9cm}
       \includegraphics[width=7.8cm]{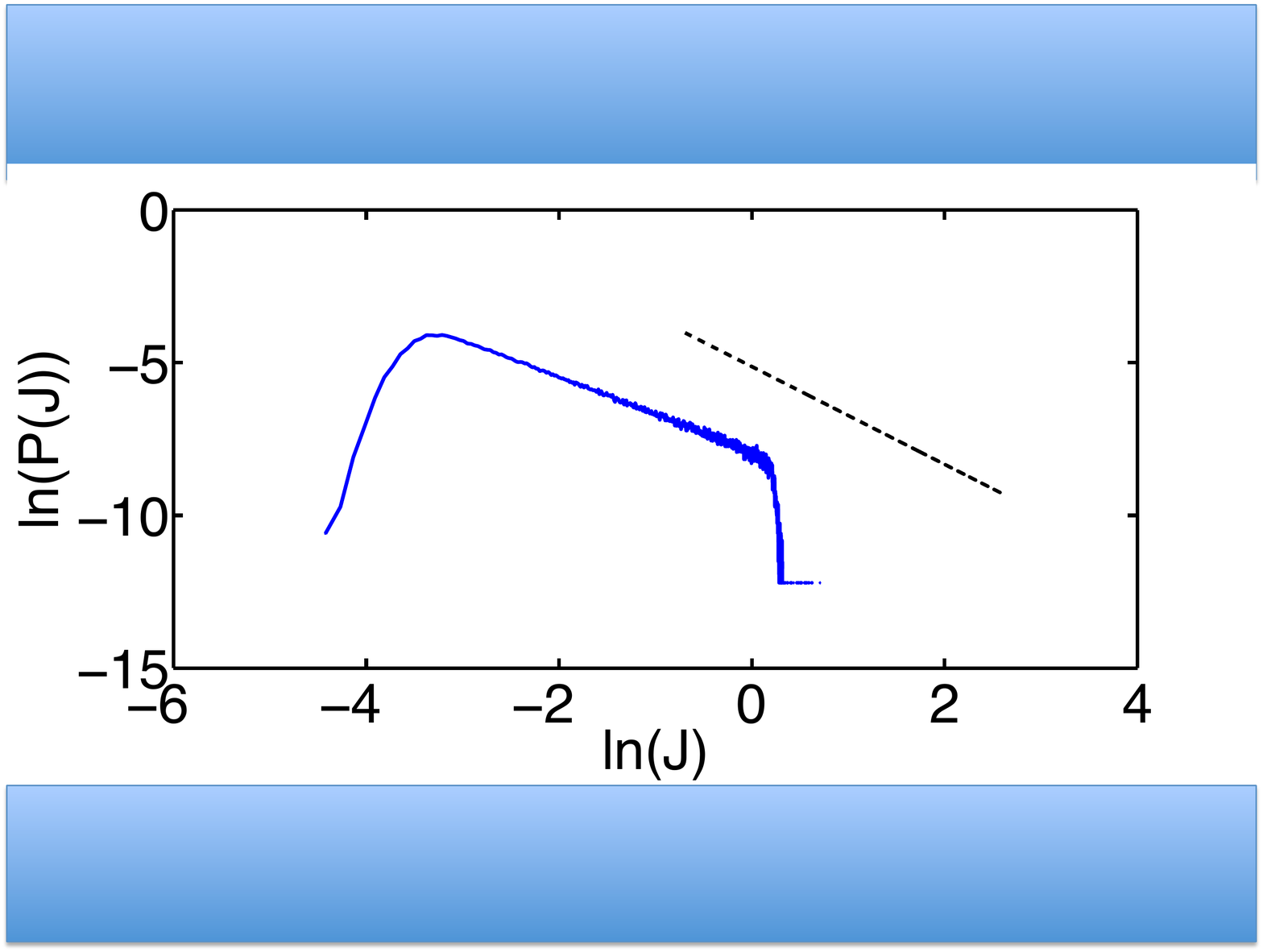}
\end{minipage}\\
\begin{minipage}[b]{0.4cm}
       {\bf (c)}

       \vspace{3.3cm}
\end{minipage}
\begin{minipage}[t]{7.9cm}
       \includegraphics[width=7.8cm]{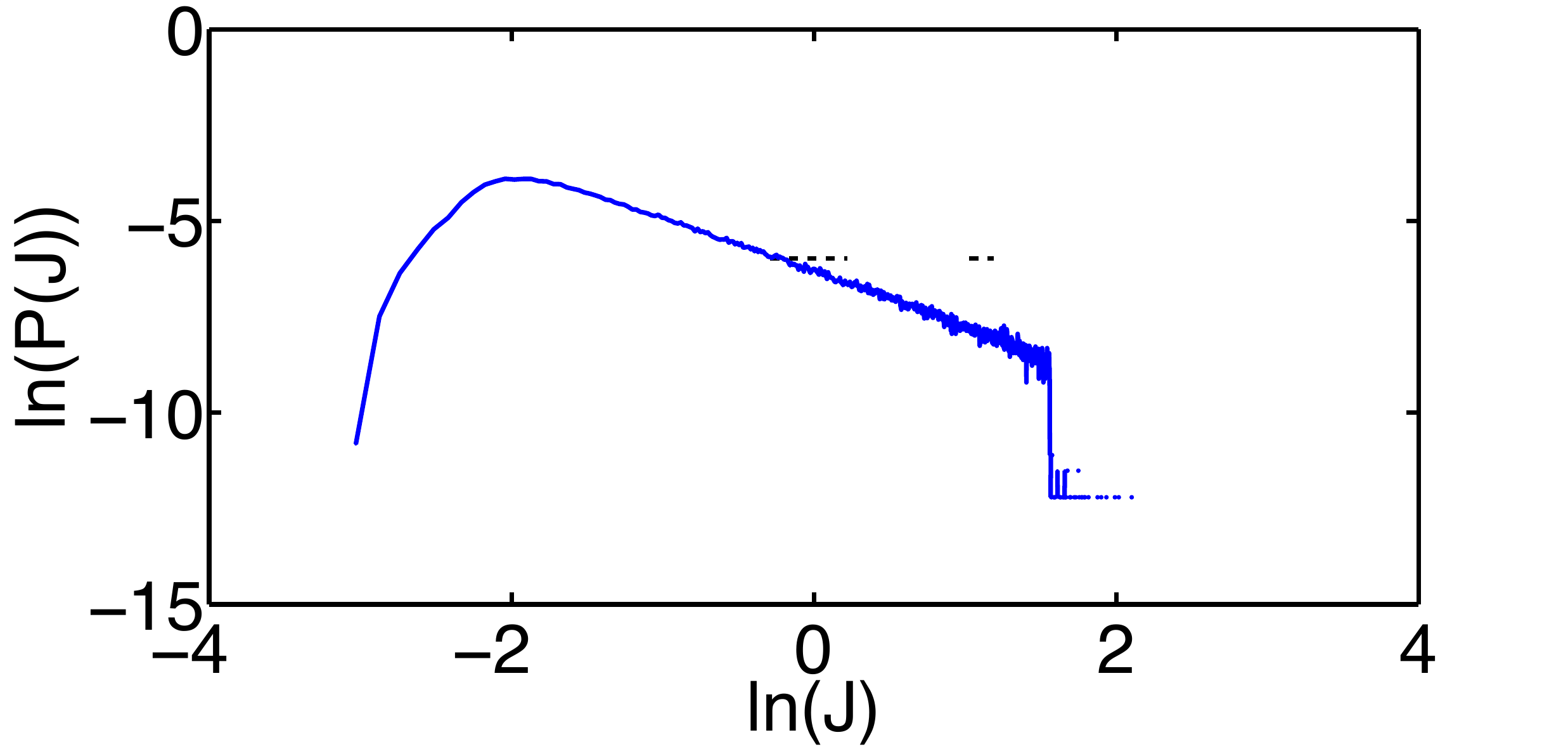}
\end{minipage}\\
\begin{minipage}[b]{0.4cm}
       {\bf (d)}

       \vspace{3.3cm}
\end{minipage}
\begin{minipage}[t]{7.9cm}
       \includegraphics[width=7.8cm]{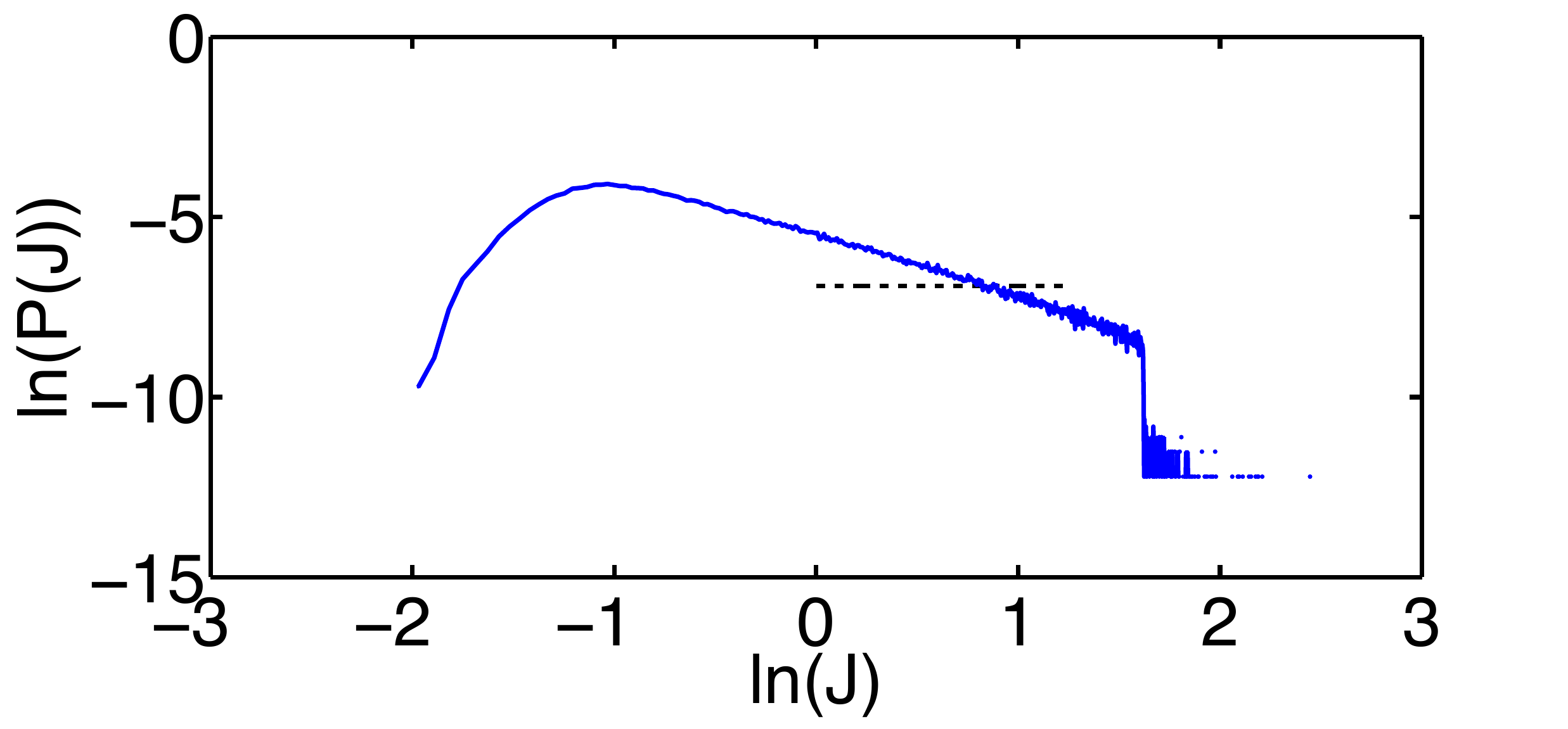}
\end{minipage}

\caption{Log-log plots of initial and renormalized Josephson coupling distributions for near-critical flows.  
All runs were done on $L = 100$ lattices with $\alpha = 5 \times 10^{-6}$.
Each plot shows the initial distribution and the distribution when the effective lattice has $1\%$ of the original number of sites.  
The initial distributions have four different forms, but the distributions after renormalization show a universal power law.
Note that the plots of initial distributions in these plots were not constructed from actual data (i.e.\ actual numerical
sampling of the distributions) but were instead constructed by hand.}
\label{universalJ}
\end{figure}

\begin{figure}
\centering
\includegraphics[width=8cm]{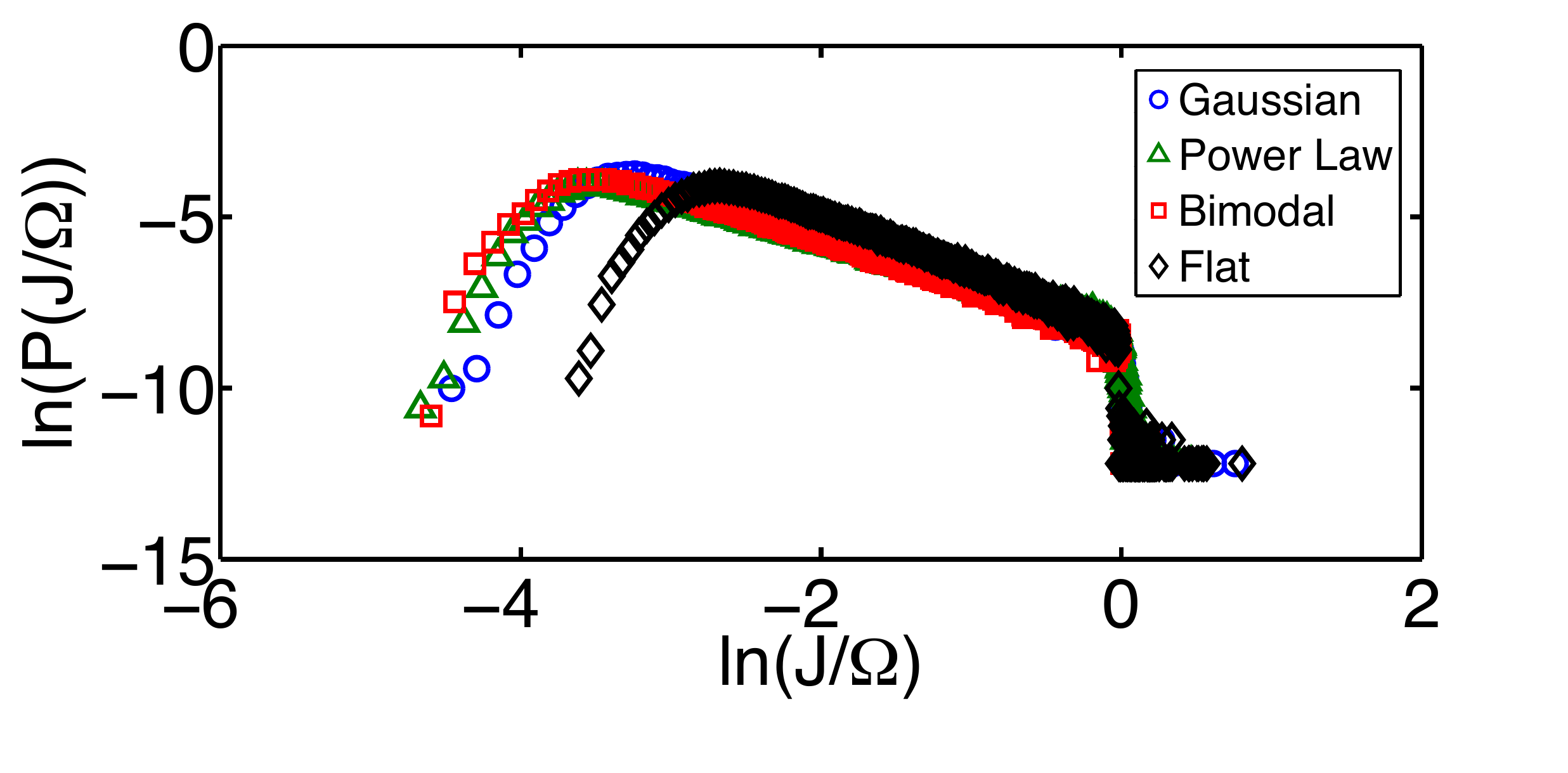}
\caption{The distributions from Figure \ref{universalJ}, with the Josephson coupling strength scaled by the mean RG scale $\Omega$ at the corresponding stage of the RG. 
The near collapse of the distributions reveals the universal power law form of the Josephson coupling distribution near the unstable fixed point.  A refined version
of this plot, showing data for larger lattices (but also different choices of initial distributions), can be found in panel (b) of Figure \ref{universalUscaled}.}
\label{universalJscaled}
\end{figure}

\indent We first focus on determining the universal form of the fixed point Josephson coupling distribution.  
Figure \ref{universalJ} shows data for the four different choices of the initial distributions that we explore 
in this paper.  In panels (a), (c), and (d), $P_i(U)$ and $P_i(J)$ have the same qualitative form, and in panel (b), 
$P_i(J) \propto J^{-1.6}$ and $P_i(U) \propto U^{1.6}$. 
We tune the parameters characterizing the distributions such that the flows propagate near the unstable fixed point, run the numerics on $100\times100$ lattices,
and plot the initial distributions alongside the renormalized distributions when $100$ sites remain in the effective lattice.  For the renormalized distributions, 
we again only include the dominant $2\tilde{N} = 200$ Josephson couplings for each sample.  The renormalized distributions suggest that the 
RG indeed washes away the details of the initial choices, leaving a power law in each case.  The universality of this power law is more striking in Figure \ref{universalJscaled}, 
where we plot the renormalized distributions for the four cases together.
In this plot, we scale $J$ for each of the four cases by the mean RG scale $\Omega$ when only $100$ sites remain.  
This scaling causes the distributions to nearly collapse onto one another, revealing the universal form:
\begin{equation}
\label{univdistJ}
P_{\text{univ}}\left(\frac{J}{\Omega}\right) \propto \left(\frac{J}{\Omega}\right)^{-\varphi}
\end{equation}
We will momentarily defer providing a numerical estimate of $\varphi$, in anticipation of presenting higher quality data, taken
from runs on larger lattices, below.

\indent We have not plotted the renormalized charging energy distributions for the four cases shown in Figure \ref{universalJ}.  Were we to do so, we would see that these
distributions, while showing hints of universality, are not as strikingly universal as the corresponding renormalized Josephson coupling distributions.  
The reason for this is the following:  in three out of the four cases, the initial distributions have $J_{\text{max}} < U_{\text{max}}$.
Several of the initial distributions we study in this paper satisfy this property, because in dimensions greater than one, interesting choices of
distributions typically have most bare charging energies greater than most bare Josephson couplings.  Otherwise, they almost certainly yield superfluid behavior.
Consequently, for three out of the four cases in Figure \ref{universalJ}, the RG begins with only site decimations.  These site decimations dramatically modify
the Josephson coupling distribution, but the charging energy distribution is, to a large extent, only truncated from above by the renormalization scale.  Later on in the procedure, 
after many sites have been decimated away, the RG enters a regime where the charging energy and Josephson coupling scales compete.  Only then do
link decimations begin to occur, and only then can the charging energy distribution begin to evolve in a nontrivial way.  However, by this point, there is far less 
RG time remaining for the fixed point distribution to emerge. 

\indent There are two ways to circumvent this difficulty.  One strategy is to note that this problem of insufficient RG time would not arise if we had access to 
arbitrarily large lattices.  We could follow the renormalization as long as necessary to construct the universal distributions.  Thus, we can try to explore larger lattices
up to the limits set by our computational capabilities.  On the other hand, another solution is to work with very wide initial distributions of Josephson couplings.  
These are distributions which have large $\frac{\Delta J}{\bar{J}}$.  As such, they correspond to flows that begin above the unstable fixed point in our 
$\frac{\Delta J}{\bar{J}}$ vs. $\frac{\bar{J}}{\bar{U}}$ flow diagrams.  Using such distributions, it is possible to engineer situations where most bare charging energies exceed most 
bare Josephson couplings but where, due to the presence of a small fraction of anomalously large Josephson couplings, $J_{\text{max}} > U_{\text{max}}$ at the beginning of the RG.  
If the parameters are chosen appropriately, the renormalization procedure will begin with a few link decimations, until the charging energy
and Josephson coupling scales meet.  After this point, the RG will feature an interplay of site and link decimations.  Thus, both the Josephson coupling
and charging energy distributions will evolve nontrivially.

\begin{figure}
\begin{minipage}[b]{0.4cm}
       {\bf (a)}
       
       \vspace{3.3cm}
\end{minipage}
\begin{minipage}[t]{7.9cm}
       \includegraphics[width=7.8cm]{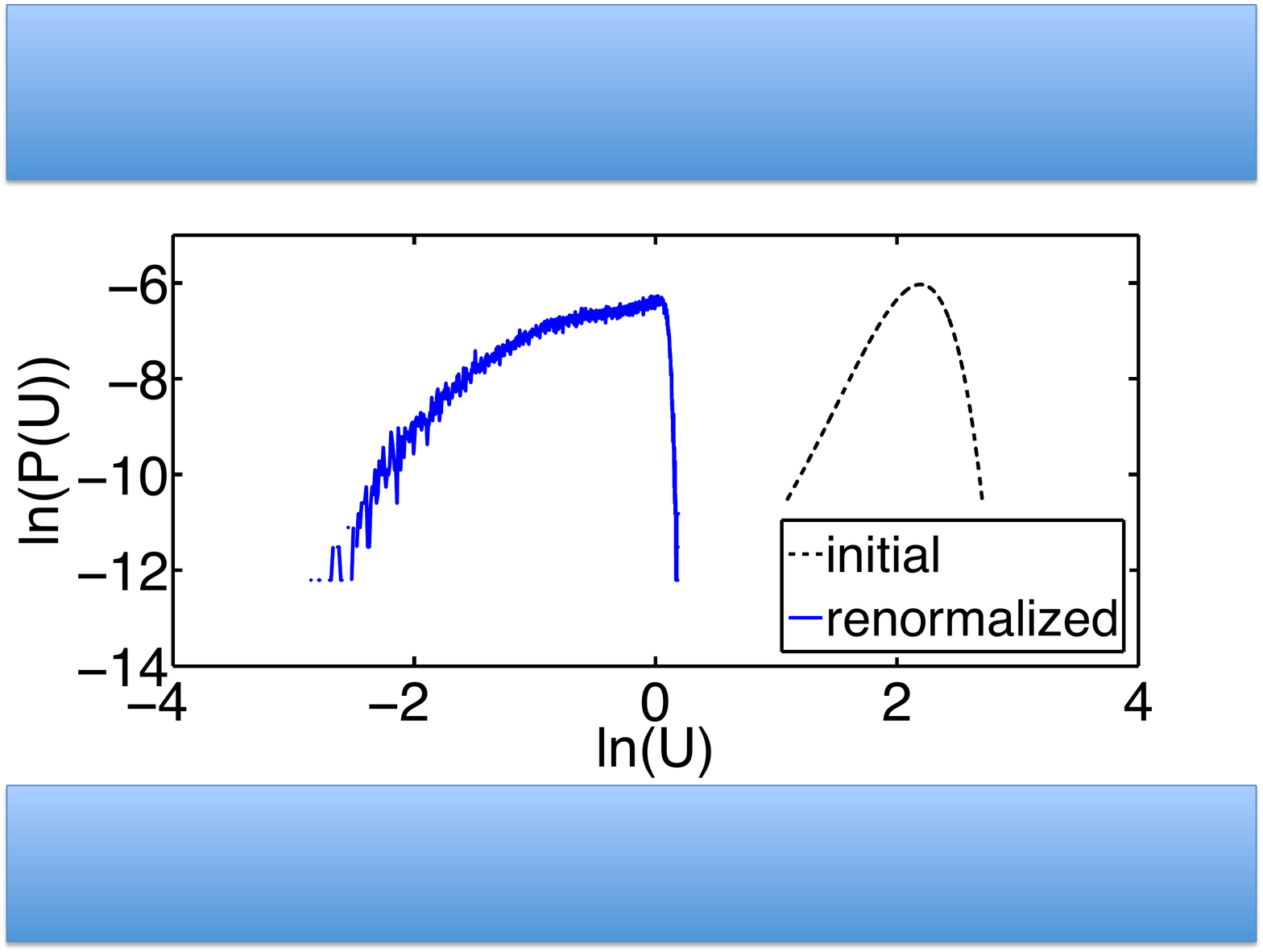}
\end{minipage}\\
\begin{minipage}[b]{0.4cm}
       {\bf (b)}

       \vspace{3.3cm}
\end{minipage}
\begin{minipage}[t]{7.9cm}
       \includegraphics[width=7.8cm]{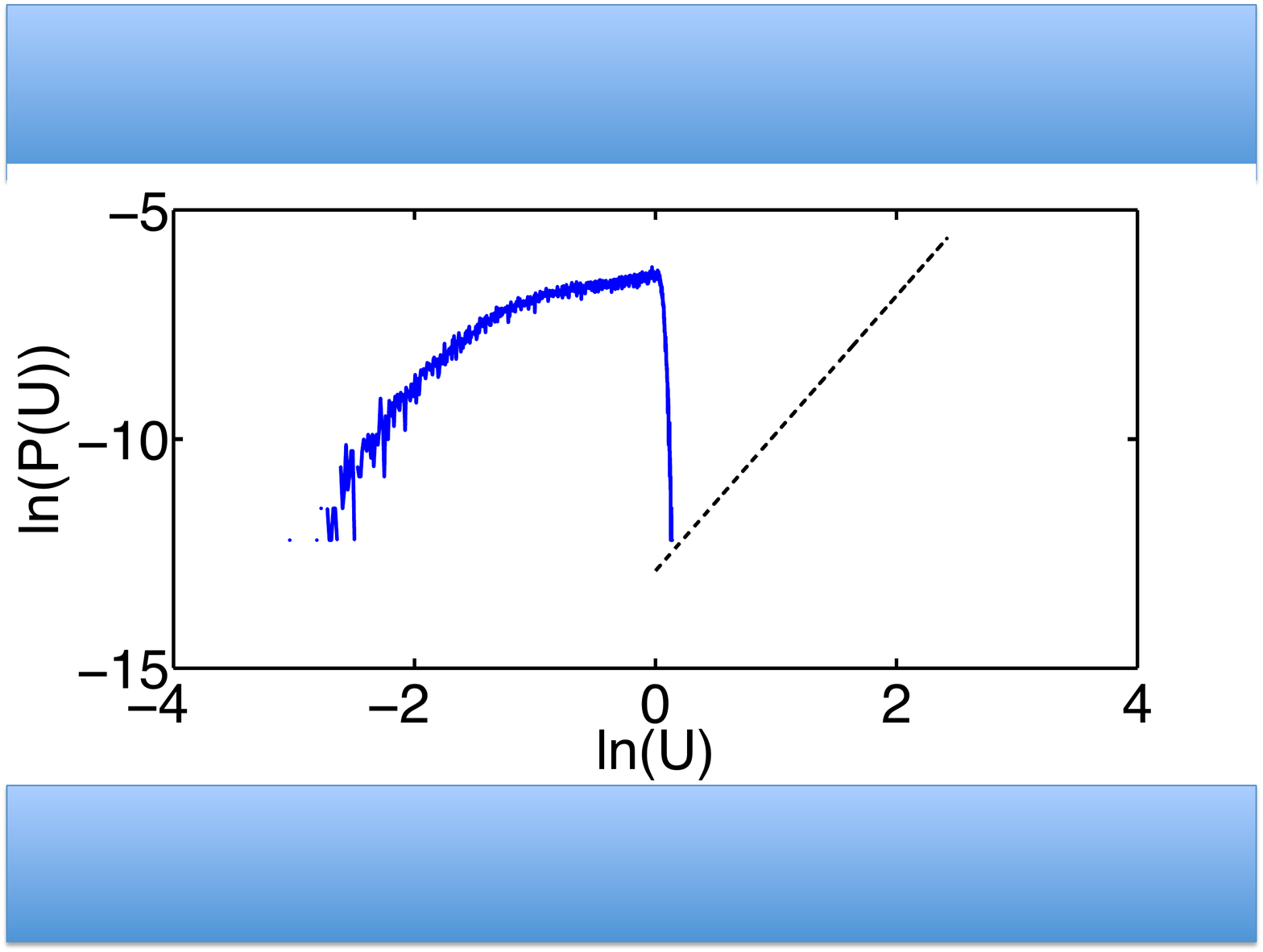}
\end{minipage}\\
\begin{minipage}[b]{0.4cm}
       {\bf (c)}

       \vspace{3.3cm}
\end{minipage}
\begin{minipage}[t]{7.9cm}
       \includegraphics[width=7.8cm]{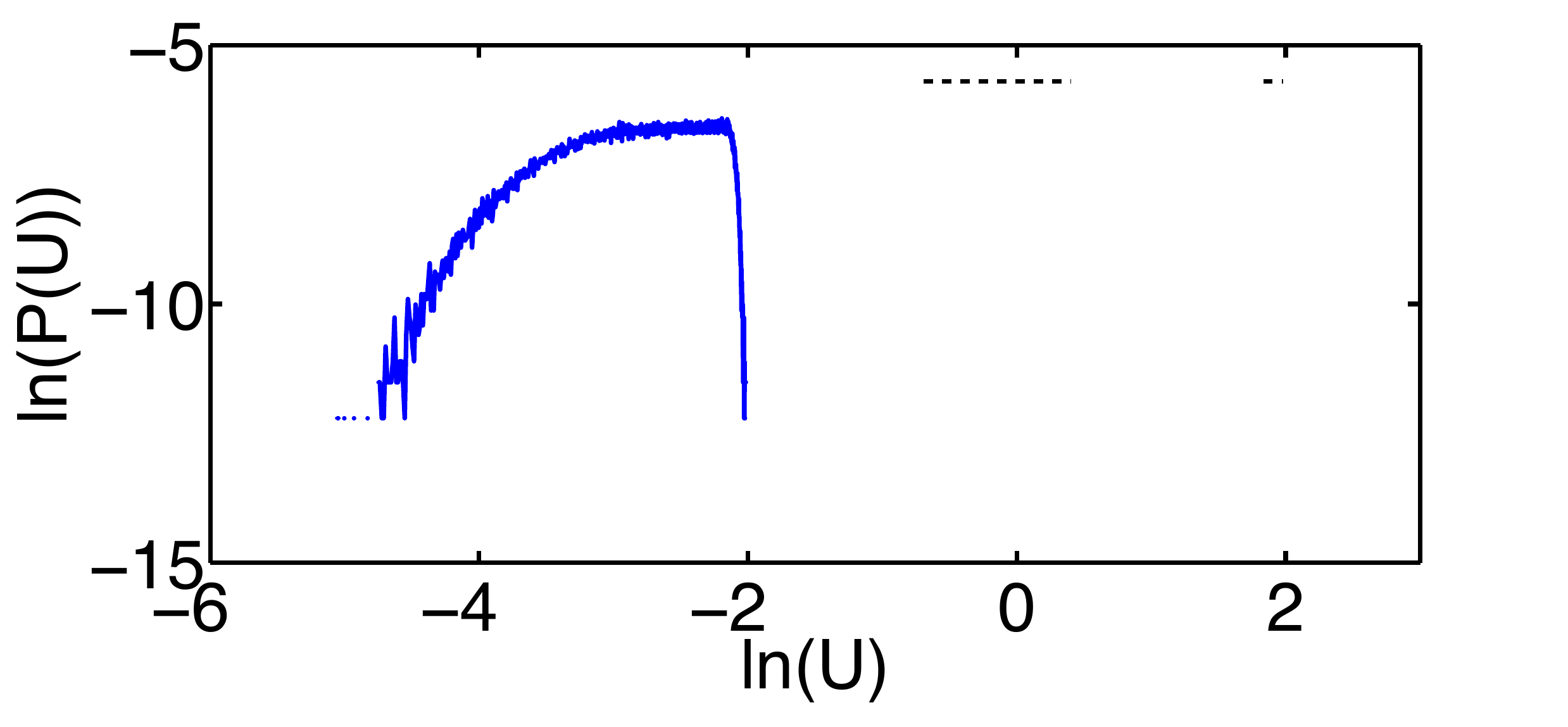}
\end{minipage}\\
\begin{minipage}[b]{0.4cm}
       {\bf (d)}

       \vspace{3.3cm}
\end{minipage}
\begin{minipage}[t]{7.9cm}
       \includegraphics[width=7.8cm]{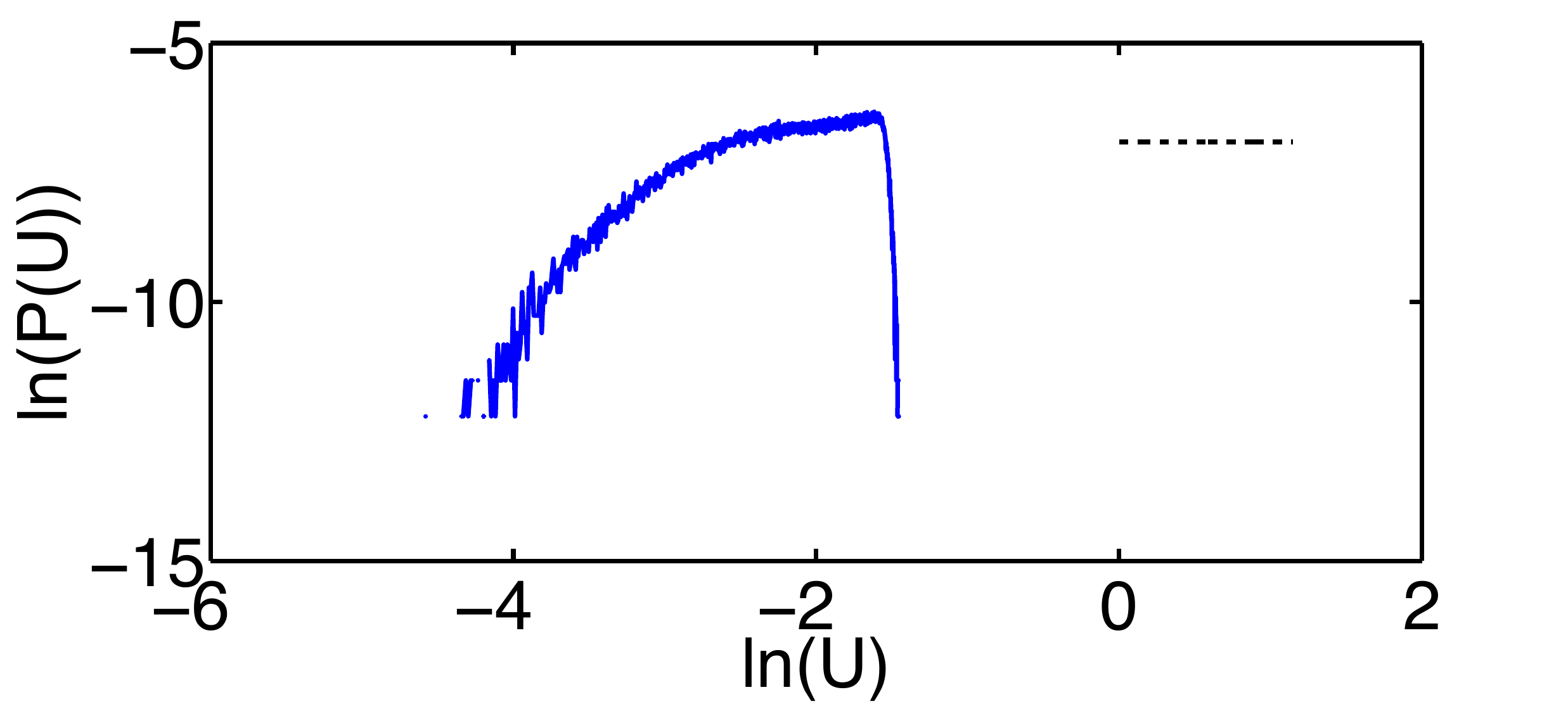}
\end{minipage}

\caption{Log-log plots of initial and renormalized charging energy distributions for near-critical flows.  
All runs were done on $L = 200$ lattices with $\alpha = 5 \times 10^{-6}$.
Each plot shows the initial distribution and the distribution when the effective lattice has $0.5\%$ of the original number of sites.  
The initial charging energy distributions have four different forms, but the distributions after renormalization show a universal form.
Note that the plots of initial distributions in these plots were not constructed from actual data (i.e.\ actual numerical
sampling of the distributions) but were instead constructed by hand.}
\label{universalU}
\end{figure}

\begin{figure}
\begin{minipage}[b]{0.4cm}
       {\bf (a)}
       
       \vspace{3.3cm}
\end{minipage}
\begin{minipage}[t]{7.9cm}
       \includegraphics[width=7.8cm]{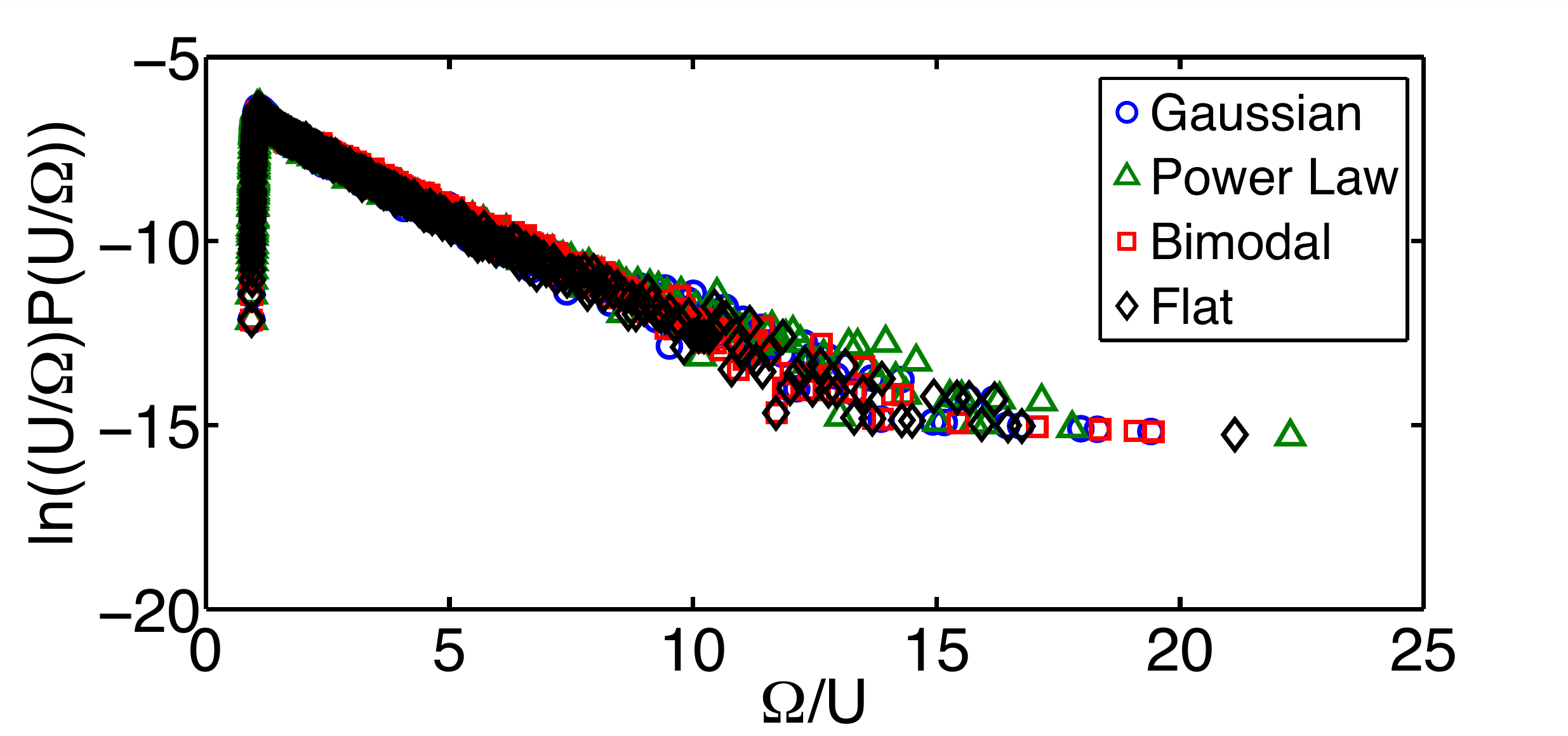}
\end{minipage}\\
\begin{minipage}[b]{0.4cm}
       {\bf (b)}

       \vspace{3.3cm}
\end{minipage}
\begin{minipage}[t]{7.9cm}
       \includegraphics[width=7.8cm]{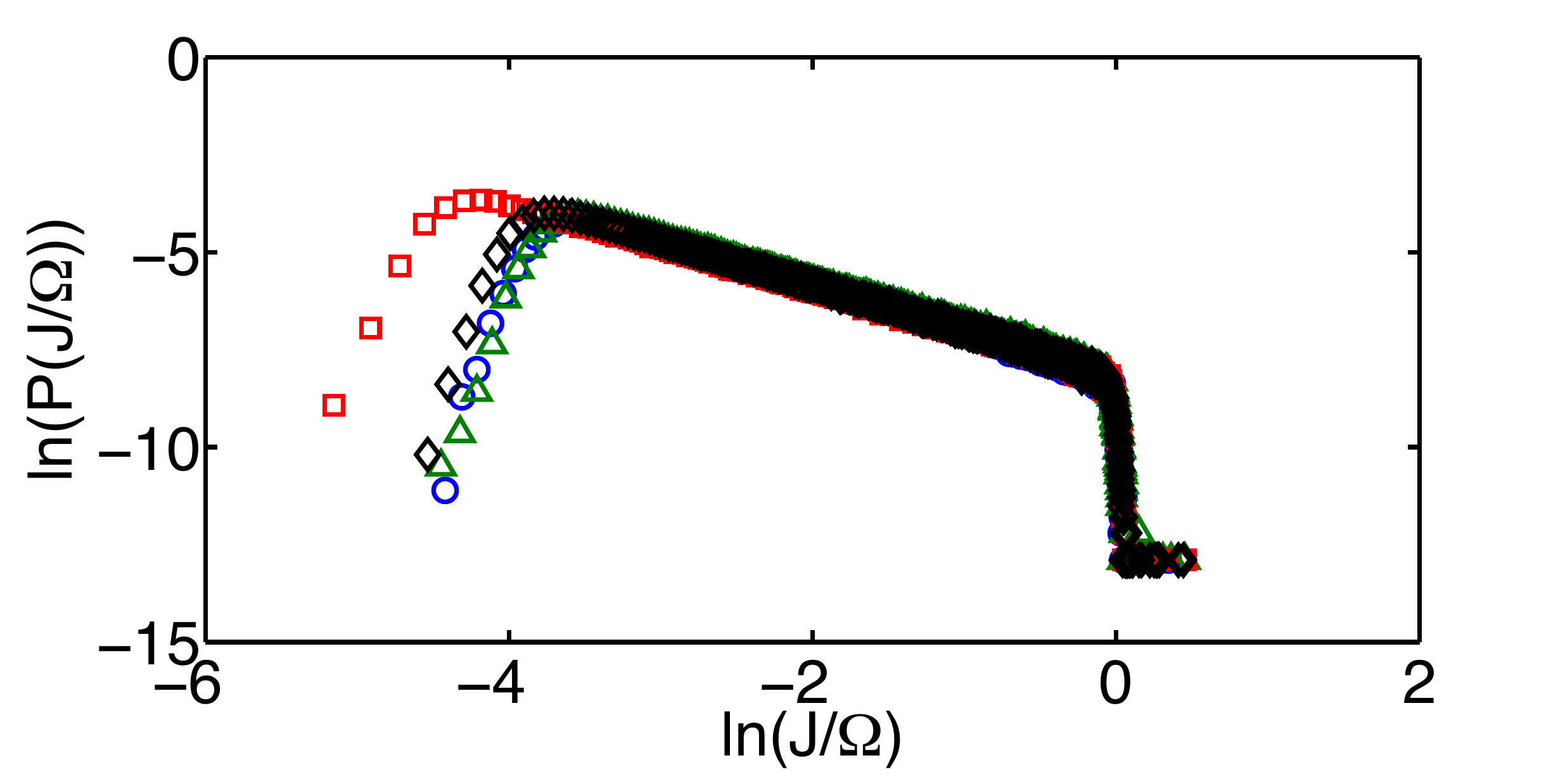}
\end{minipage}\\

\caption{In panel (a), the renormalized distributions from Figure \ref{universalU} are plotted together, with the charging energies $U$ scaled by the 
mean RG scale $\Omega$.  
In panel (b), we provide a similar plot for the renormalized Josephson coupling distributions produced by the runs in Figure \ref{universalU}.}
\label{universalUscaled}
\end{figure}

\indent To target the fixed point charging energy distribution, we apply both of the strategies.  We proceed to $200\times200$ lattices and compute renormalized
distributions when the effective lattice has $200$ sites remaining.  Additionally, we work with very wide initial Josephson coupling distributions.  
In order to achieve large initial $\frac{\Delta J}{\bar{J}}$, we restrict our attention to power law distributions of Josephson
couplings of the form $P_i(J) \propto J^{-1.6}$.  We vary the choice of the initial charging energy distribution and tune the parameters
near criticality.  The results are shown in Figure \ref{universalU}.  Now, the RG does generate a universal form for the charging energy distribution,
and in Figure \ref{universalUscaled}, we scale the renormalized distributions by the corresponding RG scales to expose the universality 
more clearly.  Figure \ref{universalUscaled} suggests that the functional form of the fixed point charging energy distribution may be:
\begin{equation}
\label{univdistU}
P_{\text{univ}}\left(\frac{U}{\Omega}\right) \propto \left(\frac{\Omega}{U}\right)^\beta \exp{\left[-\left(\frac{g_U}{\Omega}\right) \times \left(\frac{\Omega}{U}\right)\right]}
\end{equation}
where $g_U$ is an energy scale below which the charging energies are exponentially rare.
We have been unable to extract a good estimate of $\beta$.  Panel (a) of Figure \ref{universalUscaled} presupposes $\beta \approx 1$, and the linearity
of the plots suggests that this may be close to the actual value.  Taking $\beta = 1$ and focusing on the case where $P_i(U)$ is Gaussian (because that 
is the choice of initial distributions for which we have most accurately targeted criticality), we fit:  
\begin{equation}
\label{gUestimate}
\frac{g_U}{\Omega} \approx 0.66 \pm 0.02
\end{equation}
We should note that qualitatively similar charging energy distributions to those seen in Figure \ref{universalU} still emerge 
near criticality if we relax the restrictions of initially power law $J$ distributions and initially high $\frac{\Delta J}{\bar{J}}$.  This is true of the distributions
studied in Figure \ref{universalJ}, even in the flat and bimodal cases where $J_{\text{max}} < U_{\text{min}}$ initially and clusters can only form
due to the use of the sum rule.  As argued above, the additional restrictions we impose on $P_i(J)$ in Figure \ref{universalU}
simply allow the RG to construct the fixed point distributions more cleanly. 

\indent In the lower panel of Figure \ref{universalUscaled}, we additionally present data for the Josephson coupling distributions at the same stage of the RG.  
Again focusing on the data for the case where $P_i(U)$ is Gaussian, we can estimate:
\begin{equation}
\label{varphiestimate}
\varphi \approx 1.15 \pm 0.01
\end{equation}
Before proceeding, we should note that the form of the fixed point $J$ distribution allows us to construct an argument for the validity of the RG near criticality.  
We expand upon this argument greatly in Appendix D, but we will sketch the basic premise here.  Essentially, we should consider
the implications of the fixed point distributions for the reliability of each of the RG steps.
The validity of site decimation rests on the reliability of the perturbative treatment of the Josephson couplings entering the
site with the dominant charging energy.  The form of the critical Josephson coupling distribution immediately guarantees that
most Josephson couplings are much weaker than the RG scale.  For the Gaussian case in Figure \ref{universalUscaled}, the ratio
of the median $J$ to the RG scale is $\frac{J_{\text{typ}}}{\Omega} \approx 0.11 \pm 0.01$.  Hence, the site decimation is usually very safe.  In the case of link
decimation, a similar argument allows us to ignore, to leading order, other Josephson couplings penetrating the sites joined by the 
dominant coupling.  However, the structure of the critical \textit{charging energy} distribution, shown in Figure \ref{universalUscaled}, 
actually suggests that there can be a large number of charging energies of the same order as the RG scale; in particular, the ratio of the median $U$
to the RG scale is $\frac{U_{\text{typ}}}{\Omega} \approx 0.67 \pm 0.01$.
Consequently, the question of the reliability of link decimation reduces to the following:
in a single junction (or two-site) problem, how reliable is cluster formation when both charging energies are weaker than the Josephson
coupling but potentially of the same order-of-magnitude?  We address this question in Appendix D and find that the link decimation
step also seems to be reasonably safe.

\subsection{Physical Properties and Finite Size Scaling}

\indent To determine a preliminary classification of the phases of the model, we now measure four physical properties.  
First, we measure $s_{\text{max}}$, the size of the largest cluster formed by link decimations during the renormalization procedure.  
This corresponds to the largest domain of superfluid ordering in the system.  We also measure $s_2$, the size of the second
largest cluster.  We will see that the behavior of this quantity differs dramatically from that of $s_{\text{max}}$ in the superfluid phase, and
therefore, both are interesting quantities to measure.

\indent We also calculate the charging gap for the system, $\Delta_{\text{min}}$.  We can estimate this quantity by simply measuring the
charging energy of the final site remaining in the RG.

\indent Finally, we measure a susceptibility towards superfluid ordering.  Consider adding a perturbation of the 
following form to the rotor model Hamiltonian:
\begin{equation}
\label{Hchipert}
\hat{H}' = -h \sum_j \cos(\hat{\phi}_j)
\end{equation}
In the RG, we can evaluate the linear response of the system to this perturbation and measure the susceptibility:
\begin{equation}
\label{susc}
\chi = \frac{1}{L^2} \sum_j \left. \frac{\partial \langle \cos{\hat{\phi_j}}\rangle}{\partial h}\right|_{h = 0}
\end{equation}
The terms of this sum are computed during site decimation.  Perturbation theory gives a single site 
susceptibility of $\frac{1}{U_X}$, where $X$ is the site being decimated.  Neglecting corrections
from harmonic fluctuations, we can find the contribution of a cluster to the susceptibility by
multiplying the perturbative result by $s^2$, where $s$ is the size of the cluster.  
One factor of $s$ arises from the fact that the cluster represents $s$ terms in the original sum (\ref{susc}),
and the other follows from the fact that the effective field coupling to the cluster phase is enhanced $s$ times 
when $s$ phases rotate together.  When harmonic fluctuations are taken into account,
both of these factors of $s$ should be replaced by a renormalized factor which we denote as $b$.  
This $b$-factor accounts for the fact that quantum fluctuations weaken the phase coherence of the cluster.
For a bare site, $b = 1$, and when two sites $j$ and $k$ merge, the renormalized $b$-factor for the cluster $C$ is:
\begin{equation}
\label{bfactorcluster}
b_C = b_j c_{DW,j} + b_k c_{DW,k}
\end{equation}
where $c_{DW,j}$ and $c_{DW,k}$ are the Debye-Waller factors (\ref{dwj}).
Hence, the total contribution of the cluster to the susceptibility, before the normalization by $\frac{1}{L^2}$, is:
\begin{equation}
\label{clustersusc} 
\chi_C = \frac{b^2_C}{U_C}
\end{equation}
where $b_C$ and $U_C$ are the $b$-factor and charging energy of the cluster respectively.
Further details of this calculation can be found in Appendix B.

\begin{figure}[h]
\begin{minipage}[b]{0.4cm}
       {\bf (a)}
       
       \vspace{3.3cm}
\end{minipage}
\begin{minipage}[t]{7.9cm}
       \includegraphics[width=7.8cm]{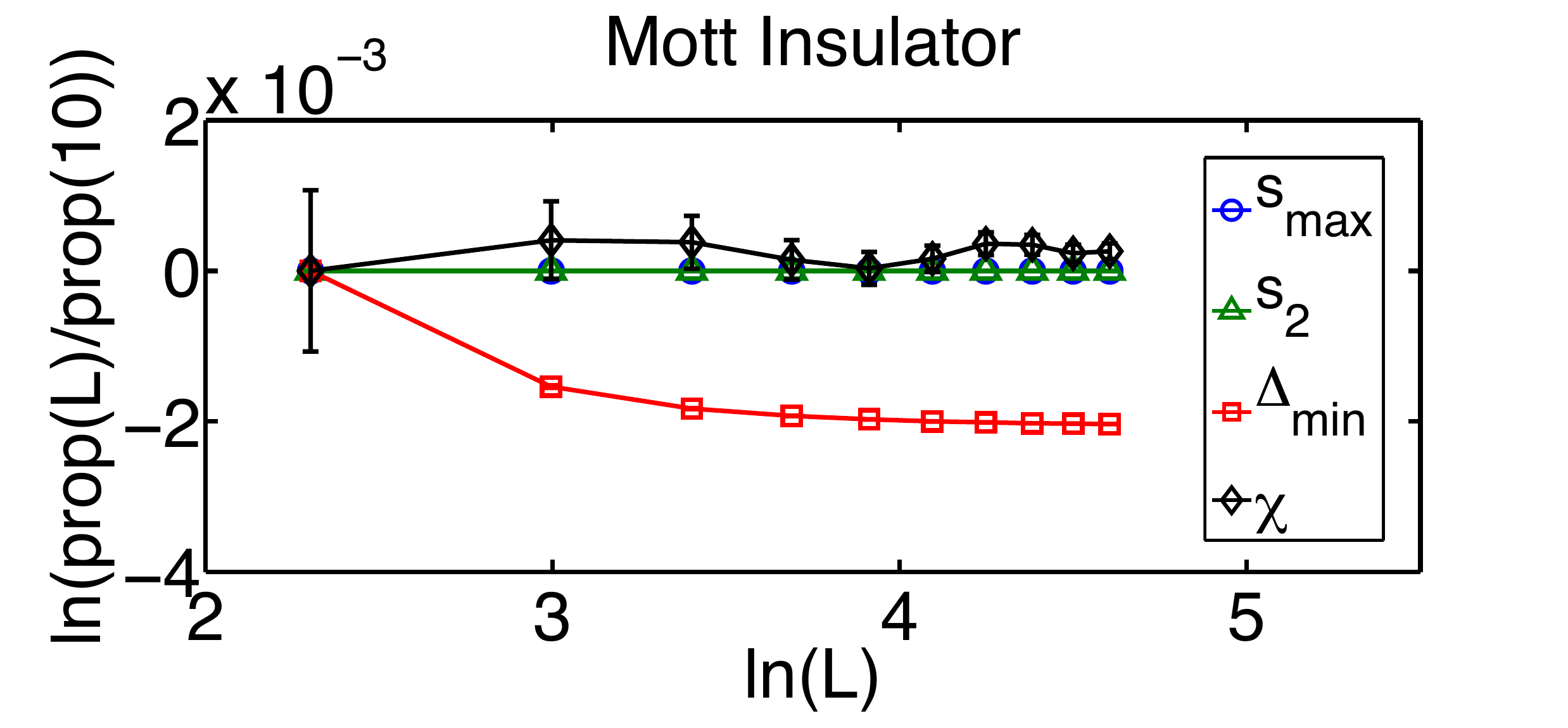}
\end{minipage}\\
\begin{minipage}[b]{0.4cm}
       {\bf (b)}

       \vspace{3.3cm}
\end{minipage}
\begin{minipage}[t]{7.9cm}
       \includegraphics[width=7.8cm]{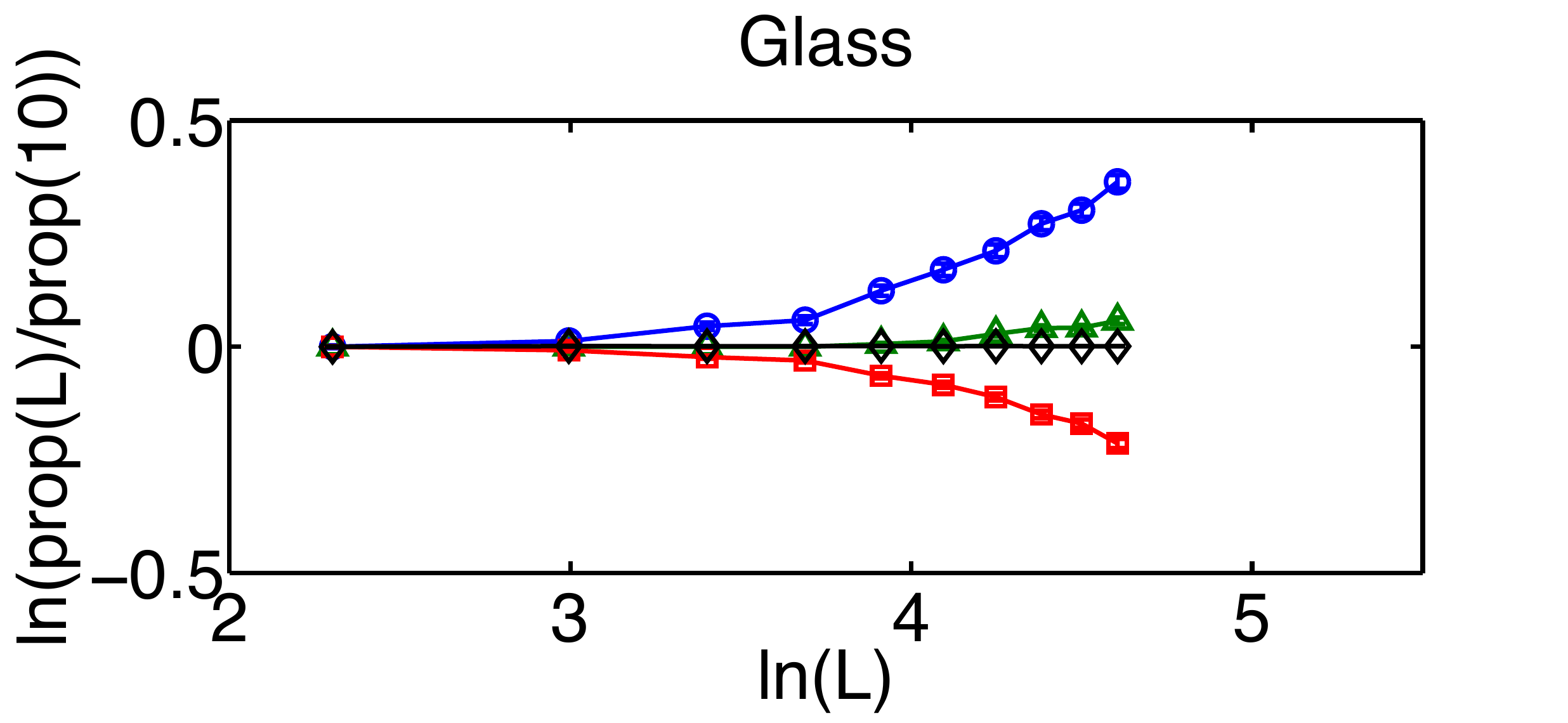}
\end{minipage}\\
\begin{minipage}[b]{0.4cm}
       {\bf (c)}

       \vspace{3.3cm}
\end{minipage}
\begin{minipage}[t]{7.9cm}
       \includegraphics[width=7.8cm]{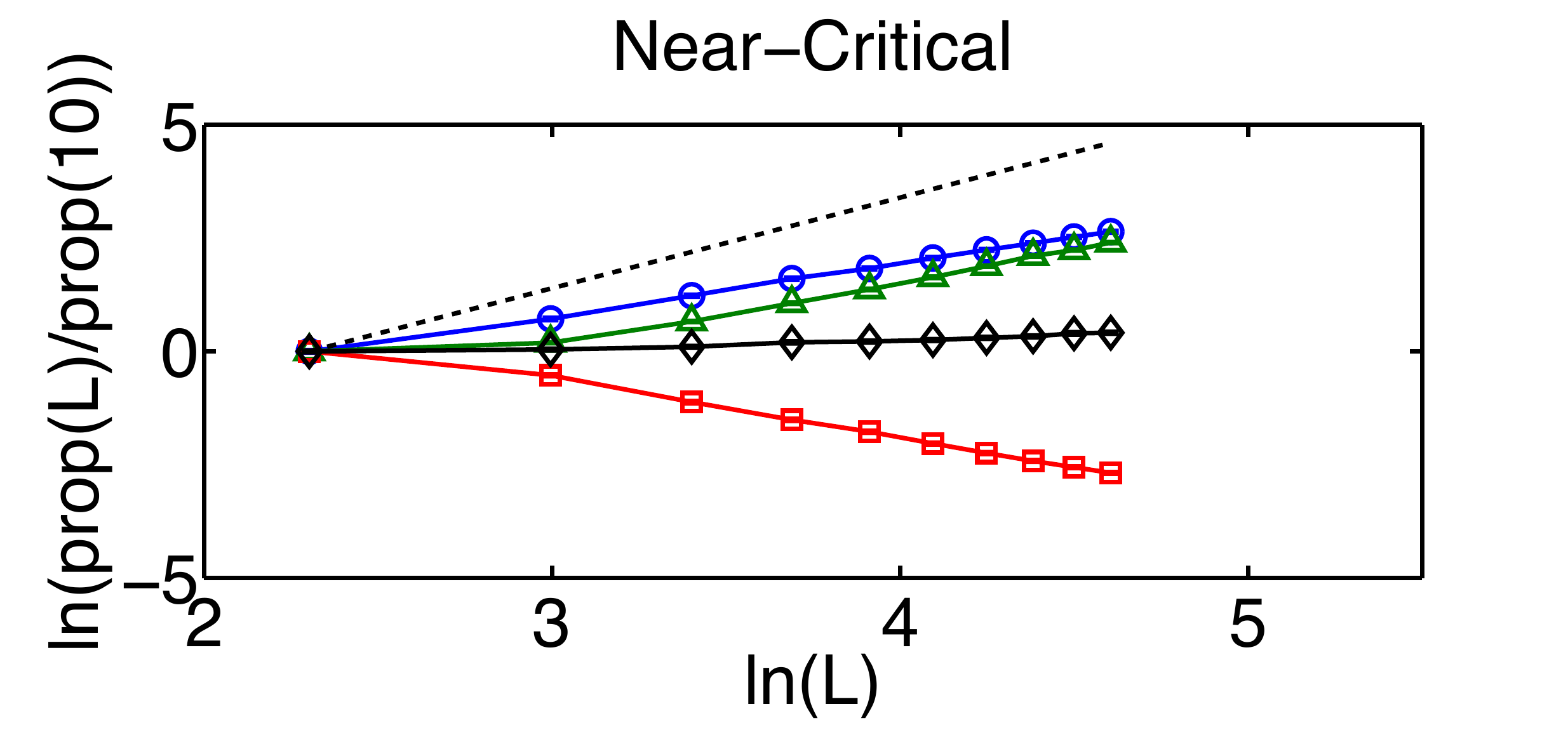}
\end{minipage}\\
\begin{minipage}[b]{0.4cm}
       {\bf (d)}

       \vspace{3.3cm}
\end{minipage}
\begin{minipage}[t]{7.9cm}
       \includegraphics[width=7.8cm]{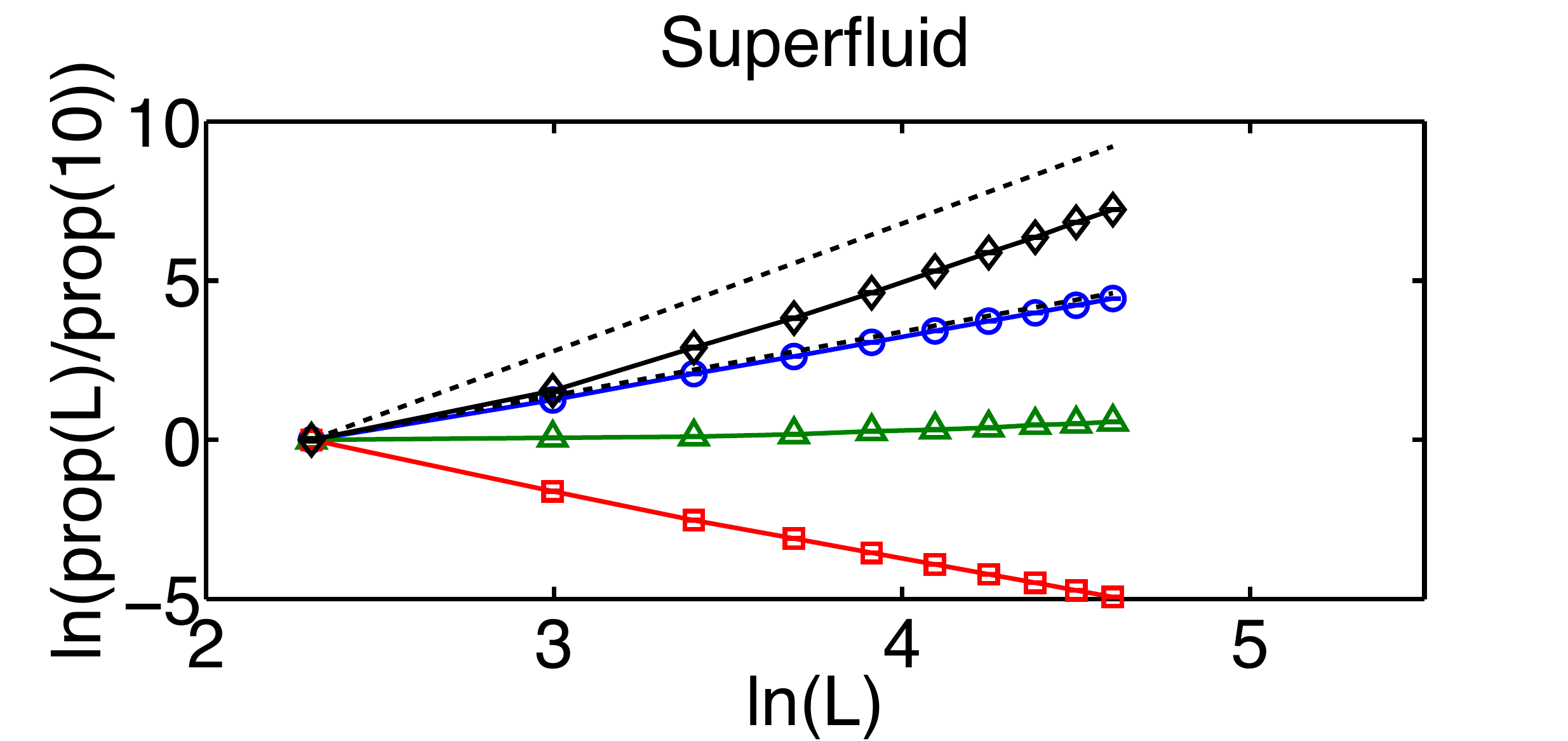}
\end{minipage}\\

\caption{Four characteristic behaviors of $s_{\text{max}}$, $s_2$, $\Delta_{\text{min}}$, and $\chi$ as a function of system size.
Here, $P_i(U)$ and $P_i(J)$ are both bimodal, and $J_h$, the center of the higher peak of the Josephson coupling distribution, is used
as the tuning parameter.  All quantities have been normalized by their value for $L = 10$, and data is shown for $L = 10$ to $L = 100$.  
In panel (a), the quantities reflect the purely local physics of the Mott insulator.  In panel (b), $s_{\text{max}}$ and $s_2$ grow subextensively
with system size, with what appears to be a power law.  The gap also decays with a slow power, and the susceptibility remains constant.
In panel (c), all quantities show power law behavior out to $L = 100$.  The reference line shows $L^2$ growth. 
Panel (d) reflects the macroscopic clusters of the superfluid phase.  The cluster size $s_{\text{max}}$ is parallel to the $L^2$ reference line, and the
susceptibility $\chi$ is parallel to the $L^4$ reference line for large $L$.}
\label{physprop}
\end{figure}

\indent For each lattice size, we obtain an estimate for the four quantities $s_{\text{max}}$, $s_2$, $\Delta_{\text{min}}$, and $\chi$.  
Then, we examine how these estimates vary as we raise $L$.  For certain types of distributions, computational limitations
force us to work on relatively small lattices.  This occurs, for example, when both $P_i(U)$ and $P_i(J)$ are bimodal, and we 
study this case in Figure \ref{physprop}.

\indent In panel (a) of Figure \ref{physprop}, there is no cluster formation whatsoever.  Hence, $s_{\text{max}} = s_2 = 1$.  This results in
a gap $\Delta_{\text{min}}$  that remains asymptotically constant as it cannot be lower than the lower bound of the initial charging energy
distribution.  The susceptibility $\chi$ also remains asymptotically constant.

\indent Next, in panel (b), we find a regime in which link decimations do occur and clusters do form.  Moreover, $s_{\text{max}}$ and $s_2$
grow with system size, with what appears to be a power law for the largest lattice sizes that we explore.  This growth is, however, subquadratic
in $L$, meaning that $s_{\text{max}}$ grows subextensively with lattice size.   Meanwhile, $\Delta_{\text{min}}$
decays with a power slower than $L^{-2}$, and the susceptibility $\chi$ remains constant with $L$.

\indent In panel (c), all quantities show power law behavior out to $L = 100$, including the susceptibility which appears to grow with a very slow
power.  The growth of $s_{\text{max}}$ is still slower than $L^2$, so the largest cluster is still subextensive.

\indent Finally, in panel (d), we find a regime in which $s_{\text{max}}$ grows as $L^2$, reflecting the formation
of macroscopic clusters that scale extensively with the size of the lattice.  The gap $\Delta_{\text{min}}$ decays as $L^{-2}$, and the susceptibility 
shows an approximately $L^4$ growth for large $L$.  Perhaps surprisingly, although $s_2$ grows with system size, it does so more slowly than it does in
panel (c).

\begin{figure}[htb]
\begin{minipage}[b]{0.4cm}
       {\bf (a)}
       
       \vspace{3.3cm}
\end{minipage}
\begin{minipage}[t]{7.9cm}
       \includegraphics[width=7.8cm]{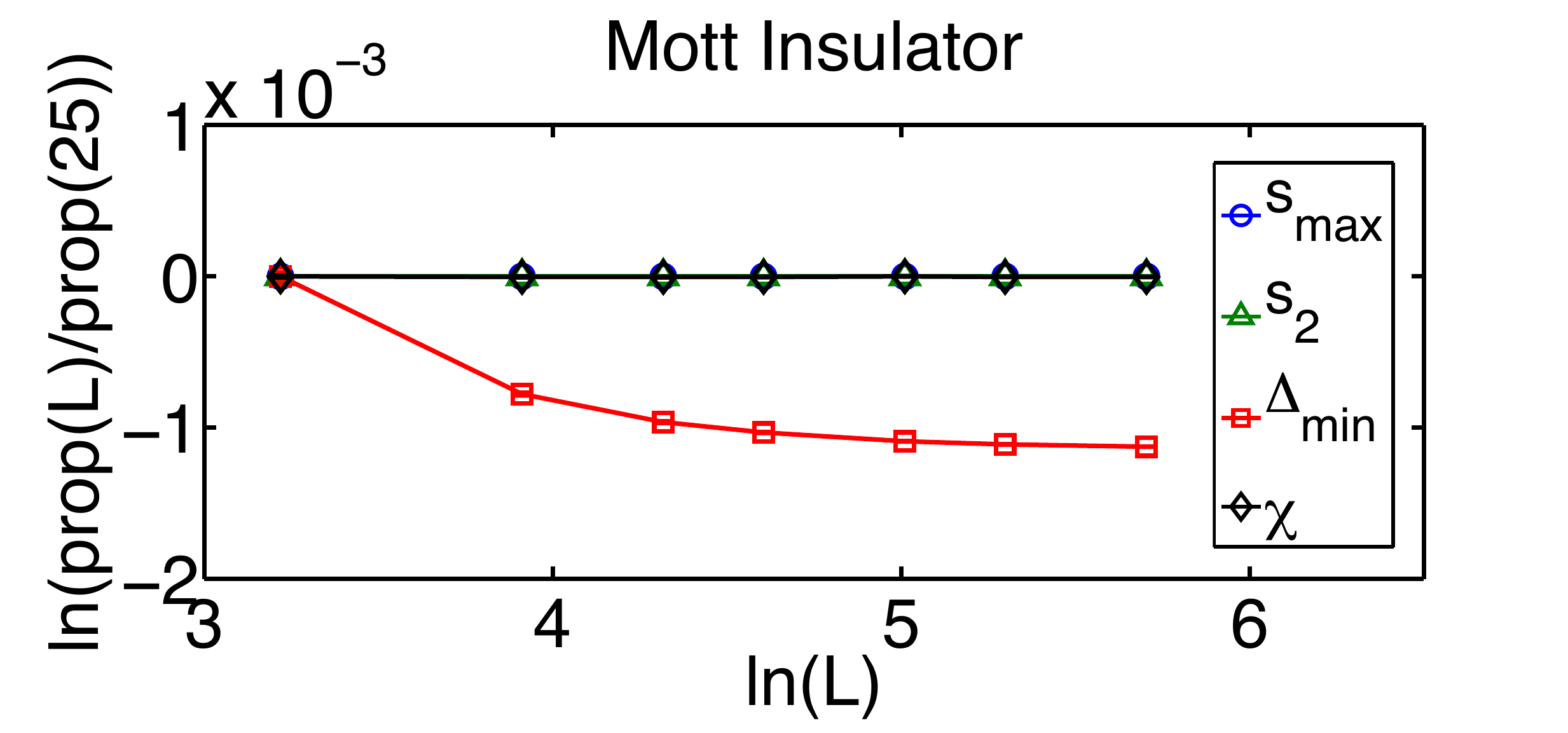}
\end{minipage}\\
\begin{minipage}[b]{0.4cm}
       {\bf (b)}

       \vspace{3.3cm}
\end{minipage}
\begin{minipage}[t]{7.9cm}
       \includegraphics[width=7.8cm]{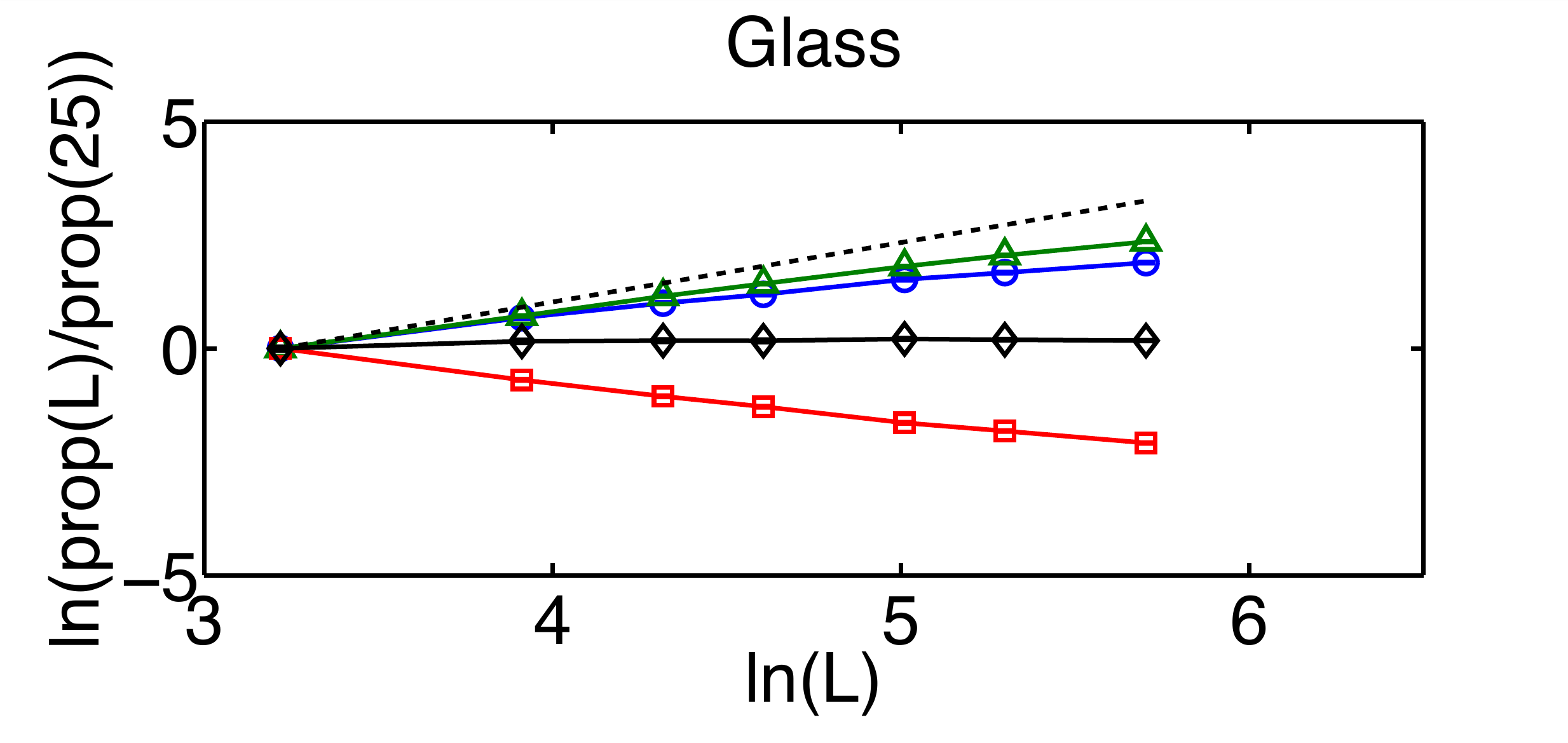}
\end{minipage}\\
\begin{minipage}[b]{0.4cm}
       {\bf (c)}

       \vspace{3.3cm}
\end{minipage}
\begin{minipage}[t]{7.9cm}
       \includegraphics[width=7.8cm]{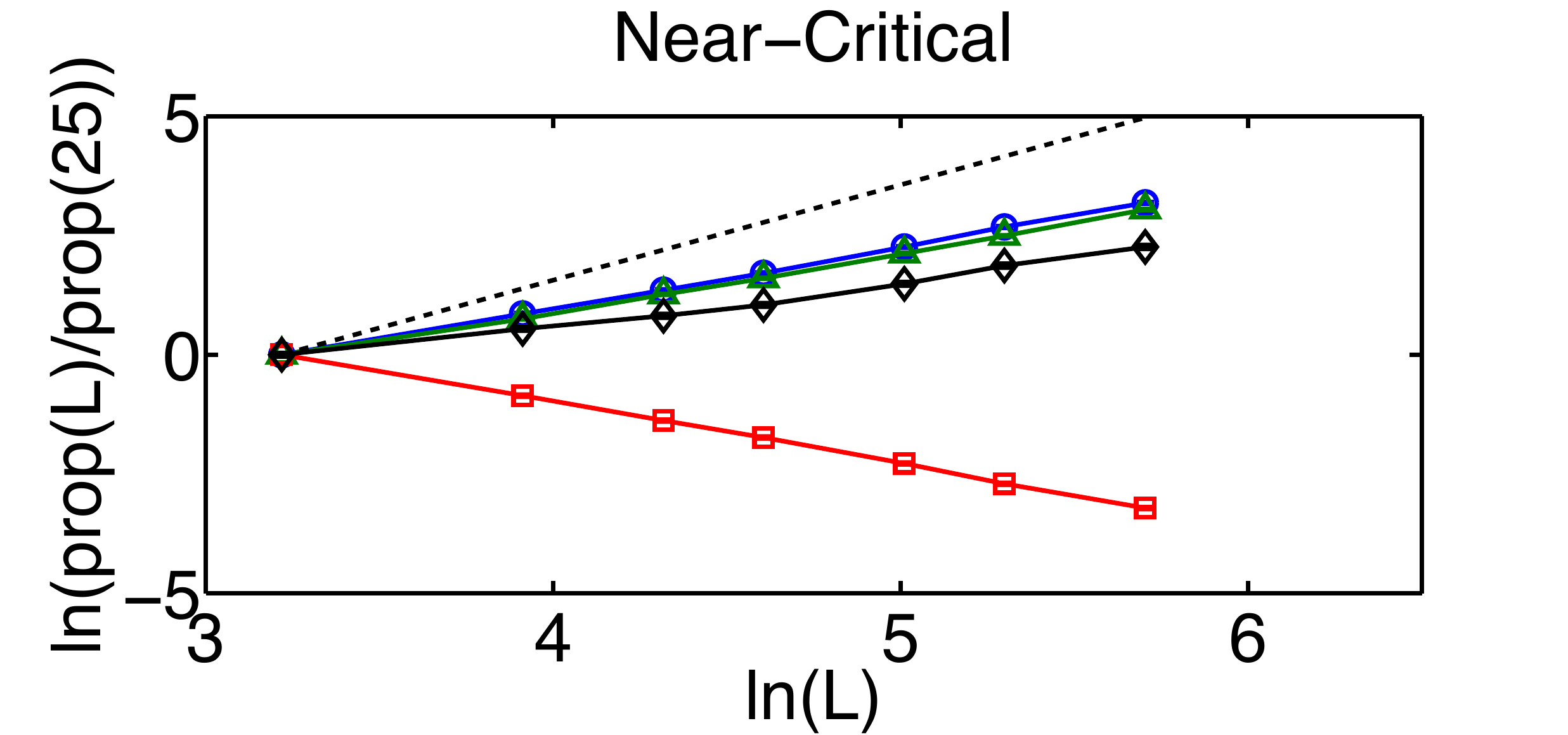}
\end{minipage}\\
\begin{minipage}[b]{0.4cm}
       {\bf (d)}

       \vspace{3.3cm}
\end{minipage}
\begin{minipage}[t]{7.9cm}
       \includegraphics[width=7.8cm]{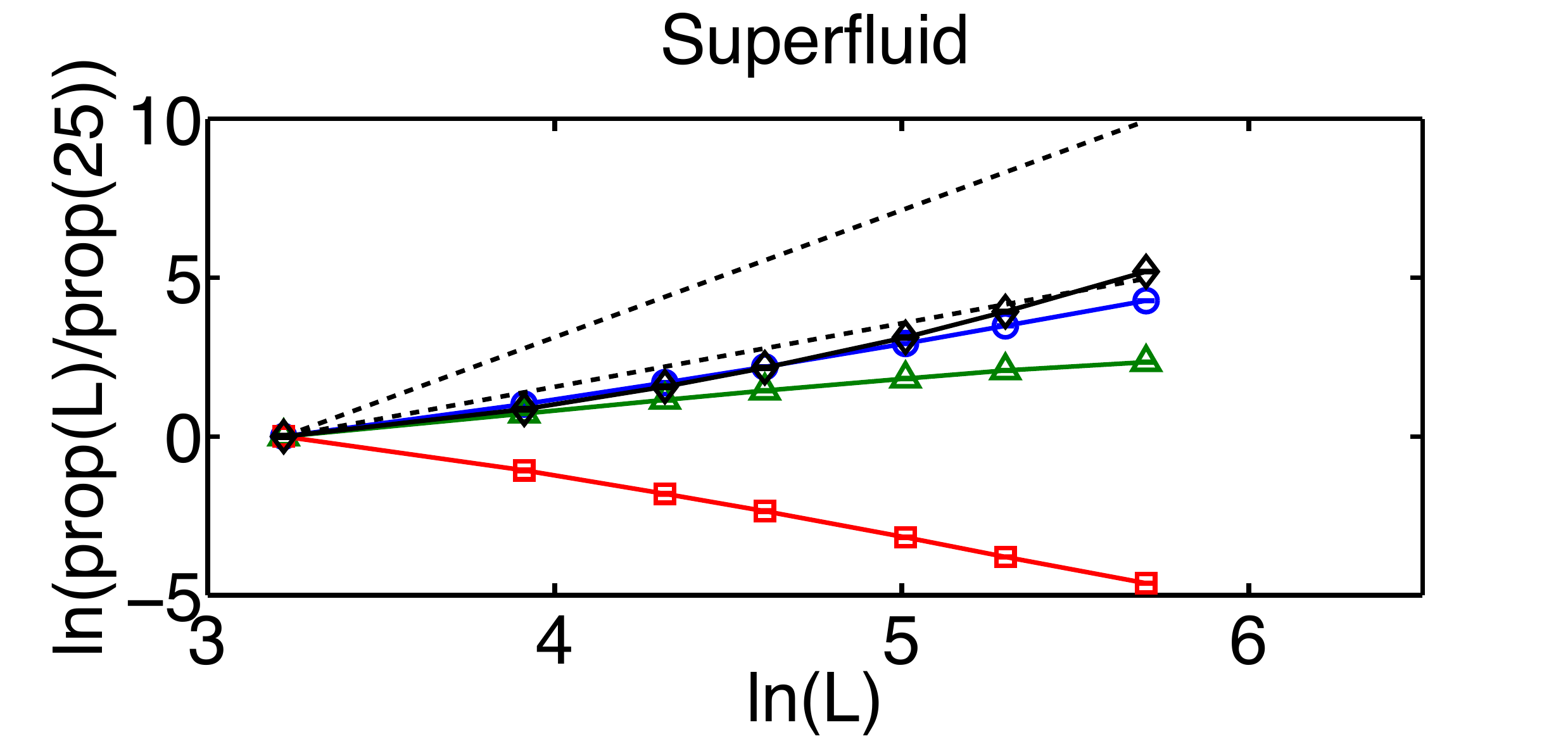}
\end{minipage}\\

\caption{Four characteristic behaviors of $s_{\text{max}}$, $s_2$, $\Delta_{\text{min}}$, and $\chi$ as a function of system size.
The initial distributions are those described in Appendix C. All quantities have been normalized by their value for $L = 25$, and data is shown for $L = 25$ to $L = 300$.  
In the four panels, $U_0 = 400$, $9.2$, $8.97$, and $8.8$ respectively.
Panel (a) reflects the purely local physics of the Mott insulator.  Panel (b) shows a glassy regime characterized
by rare-regions clusters that grow subextensively in system size.  The reference line shows the power law that $s_{\text{max}}$ obeys
near criticality.  This nearly critical regime is shown in in panel (c).  The reference line here shows $L^2$ growth.  
Finally, panel (d) shows the superfluid regime, in which $s_{\text{max}}$ is asymptotically parallel to the $L^2$ reference line. 
The susceptibility $\chi$ is expected to grow as $L^4$ for very large $L$, but it does not quite reach this behavior (indicated with a reference line) for $L \leq 300$.}
\label{physpropaux}
\end{figure}

\indent We now turn to a class of distributions for which we can reach larger lattice sizes.  In particular, we return to the data 
set described in Appendix C, in which $P_i(U)$ is Gaussian and $P_i(J) \propto J^{-1.6}$.   Data for this choice of distributions is
shown in Figure \ref{physpropaux}.  

\indent Panel (a) of this figure qualitatively reproduces the features of panel (a) of Figure \ref{physprop}.
Panel (b) of the new figure, on the other hand, differs from panel (b) of Figure \ref{physprop} in an important way.  For large $L$,
the power law behaviors of $s_{\text{max}}$, $s_2$ and $\Delta_{\text{min}}$ are lost, and all three quantities vary more slowly.  
In Figure \ref{physprop}, this effect may have been hidden by the use of smaller system sizes.

\indent  If the parameters are tuned such that the corresponding flow propagates very close to the unstable fixed point, then 
we \textit{do} find a regime in which all quantities show nearly power law behavior out to $L = 300$.  This regime is depicted in panel (c) of
Figure \ref{physpropaux}.

\indent Tuning past this point, we enter a regime in which macroscopic clusters form.  Panel (d) of Figure \ref{physpropaux} 
shows the behavior of the four quantities in this regime, and we see that most of the essential features of the corresponding 
panel of Figure \ref{physprop} are reproduced.  An important feature of the plot in panel (d) is that we can clearly see that 
the behavior of $s_2$ in this regime is closer to that observed in panel (b) than in the intervening panel (c).

\indent The plots in Figures \ref{physprop} and \ref{physpropaux} are suggestively labeled with their corresponding phase identifications.  We will 
provisionally use these labels for convenience in referring to these regimes, in advance of presenting arguments for these 
classifications in Section \ref{sec:phases}.  In the flow diagrams that we presented earlier, choices of initial distributions that correspond
to the Mott insulating and glassy behaviors shown in panels (a) and (b) of Figures \ref{physprop} and \ref{physpropaux} flow to the stable insulating
region where $\frac{\bar{J}}{\bar{U}} \rightarrow 0$.  The superfluid behavior in panel (d) of the figures corresponds to flows
towards the high $\frac{\bar{J}}{\bar{U}}$ regime of the flow diagrams.  The pure power law behavior of panel (c) emerges for flows
that propagate very close to the proposed unstable fixed point.  This suggests that this fixed point may control the glass to superfluid transition 
of the disordered rotor model. 

\indent For now, we assume that this is the case and investigate more closely the behavior of physical
quantities in the vicinity of this proposed transition.  Having provided evidence of the universality that emerges
near the critical point, here and in the remainder of this paper, we will focus exclusively on the choice of distributions detailed in Appendix C.  
In Figure \ref{collapsesmax}, we show that plots of physical quantities vs.\ tuning parameter, taken for different
$L$, can be collapsed onto universal curves.  We will use this scaling collapse in Section \ref{sec:phases} to determine the critical 
exponents governing the putative transition between glassy and superfluid phases.

\begin{figure}[h]
\begin{minipage}[b]{0.4cm}
       {\bf (a)}
       
       \vspace{3.3cm}
\end{minipage}
\begin{minipage}[t]{7.9cm}
       \includegraphics[width=7.8cm]{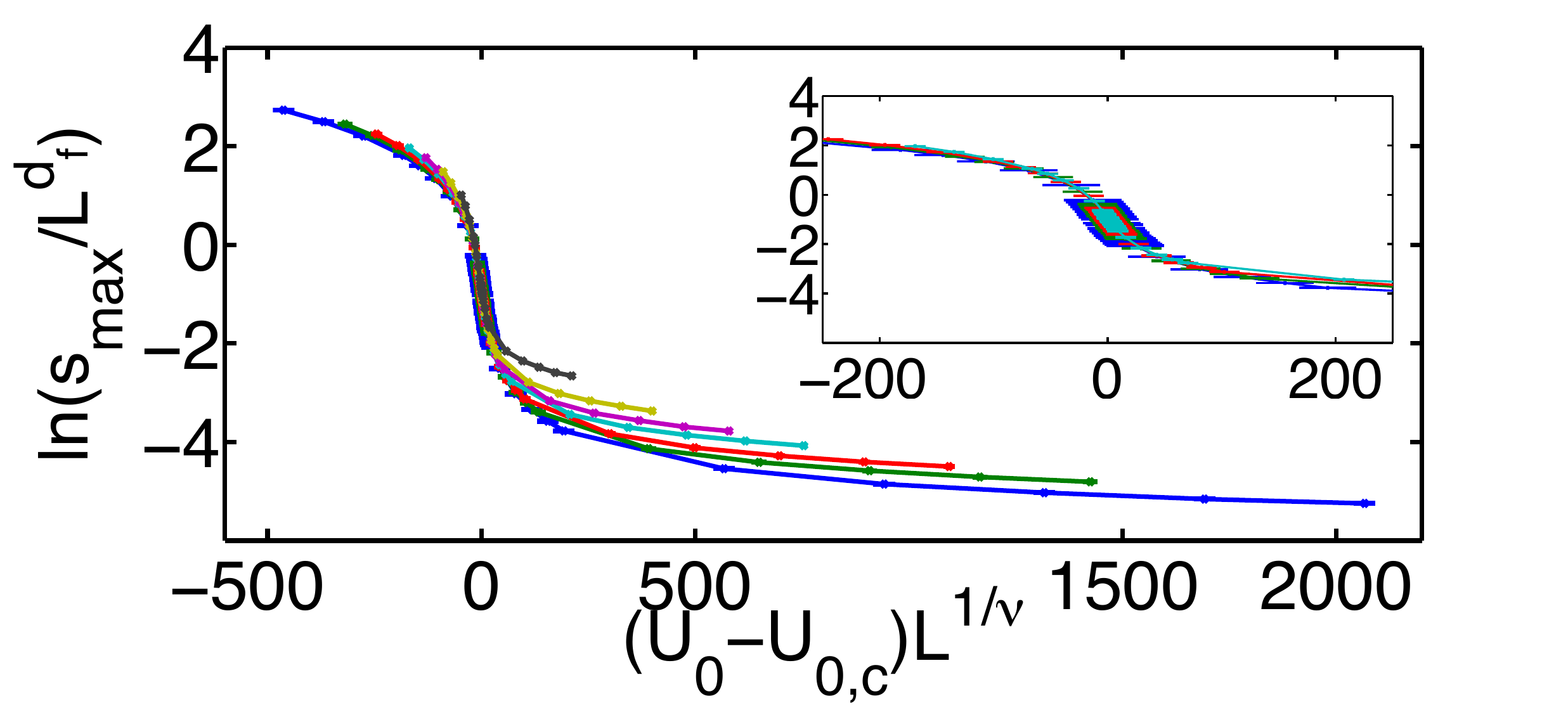}
\end{minipage}\\
\begin{minipage}[b]{0.4cm}
       {\bf (b)}

       \vspace{3.3cm}
\end{minipage}
\begin{minipage}[t]{7.9cm}
       \includegraphics[width=7.8cm]{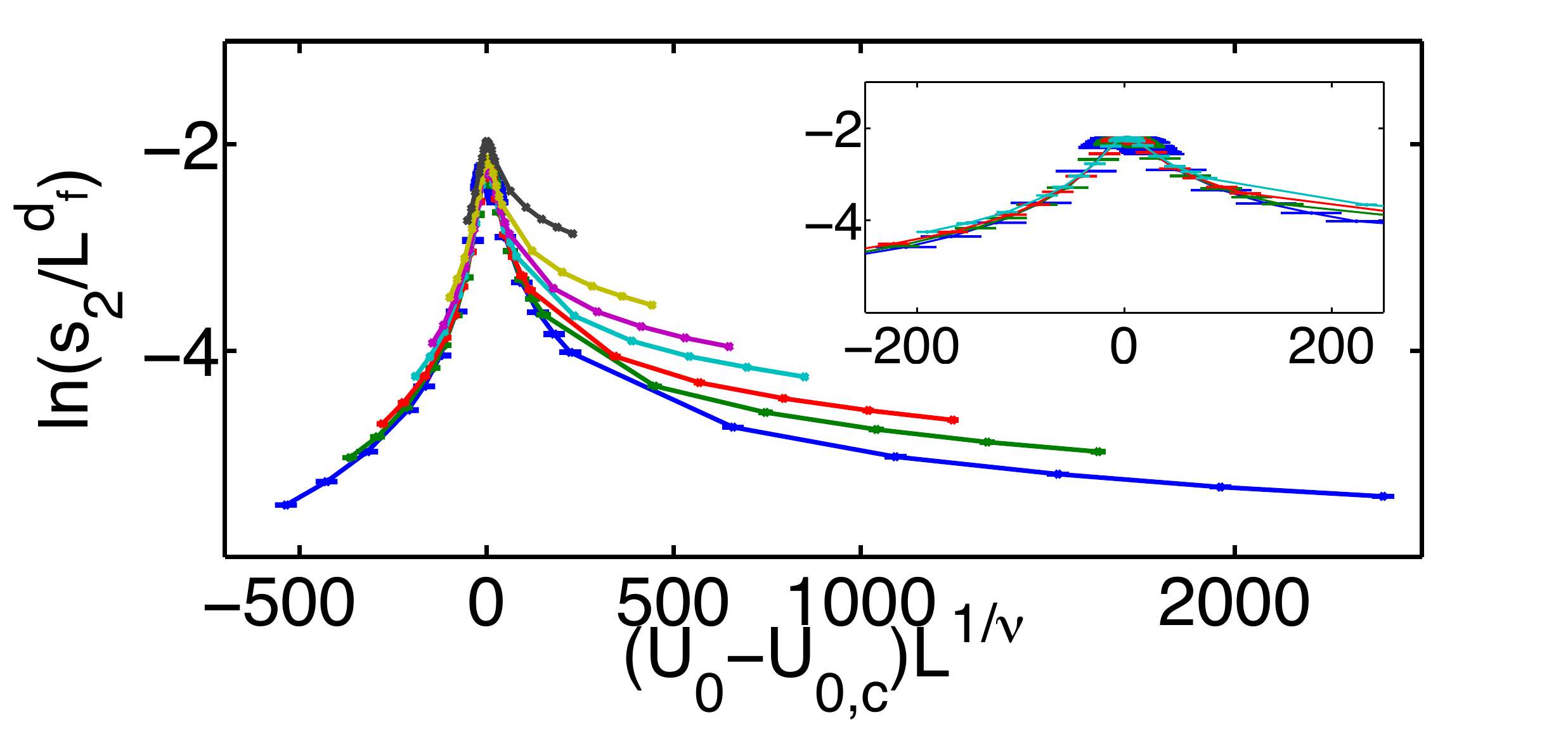}
\end{minipage}\\

\caption{Panel (a) shows the scaling collapse of $s_{\text{max}}$ as a function of tuning parameter, and panel (b) shows a similar collapse of $s_2$.  
Each line corresponds to a different value of the lattice size.  We show data for $L = 25$, $50$, $75$, $100$, $150$,
$200$, and $300$.  The insets show magnified views of the vicinity of the critical point for the four largest lattice sizes.  
The error bars, which are difficult to see in the larger plots, are clearly visible in the insets.  To see the values of the
exponents $\nu$ and $d_f$ used in each panel, consult equations (\ref{nuestimate1}), (\ref{dfestimate1}), (\ref{nuestimate2}),
and (\ref{dfestimate2}).}
\label{collapsesmax}
\end{figure}

\subsection{Cluster Densities and $b$-factors}

\indent The physical property that distinguishes the Mott glass from the Bose glass is the compressibility.  Later,
we will show that, in order to calculate the compressibility, it is insufficient to consider just the minimum charging gap.  
Within each sample, the RG may form several clusters, each of which implies a local gap for adding 
and removing bosons.  We will need to monitor all of these local gaps to find the compressibility.  

\indent More specifically, in this section, we will calculate the density (number per unit area) of clusters of a given size 
and of clusters with gaps in a given range of energy.  We call these densities $\rho(s)$ and $\rho(\Delta)$ respectively.  
The latter quantity gives a ``density-of-states" for addition of single bosons 
or holes to the system.  For a single sample corresponding to a specific choice of initial distributions, we monitor the size and 
local charging gap for all the clusters formed during the renormalization, excluding bare sites that are not clustered by the RG.  
In the case of the local charging gap, we again estimate this quantity as the charging energy of a cluster at the time of decimation.  
In principle, we could also include perturbative corrections to this local gap, but we omit these and do not anticipate that they would 
affect the behavior of the density at low $\Delta$.
We pool data for $10^3$ samples, choose a discretization of energy to determine a histogram bin size, 
compute a histogram of gaps and a histogram of cluster sizes, and finally normalize these histograms by the total simulated surface
area: $L^2$ times the number of samples.

\indent Our study of these densities will bring into focus the crucial difference between two types of clusters formed by the RG:
rare-regions clusters and the macroscopic clusters that characterize the superfluid phase.  We will, therefore, also take the
opportunity to examine how the $b$-factors, which quantify the effect of harmonic fluctuations on the susceptibility, vary as 
a function of $s$ for the two types of clusters.

\subsubsection{The Charging Gap Density $\rho(\Delta)$}

\indent Note that $\rho(s)$ and $\rho(\Delta)$ are not particularly interesting for choices of distributions and parameters that yield
the Mott insulating behavior from Figures \ref{physprop} and \ref{physpropaux}.  The profile of $\rho(\Delta)$ will be identical 
to the profile of the initial charging energy distribution, and because we choose this distribution to be
bounded from below by some positive $U_{\text{min}}$, it can be shown that this always corresponds to an incompressible phase.
Hence, we begin by focusing on the glassy regime.  

\begin{figure}
\centering
\includegraphics[width=8cm]{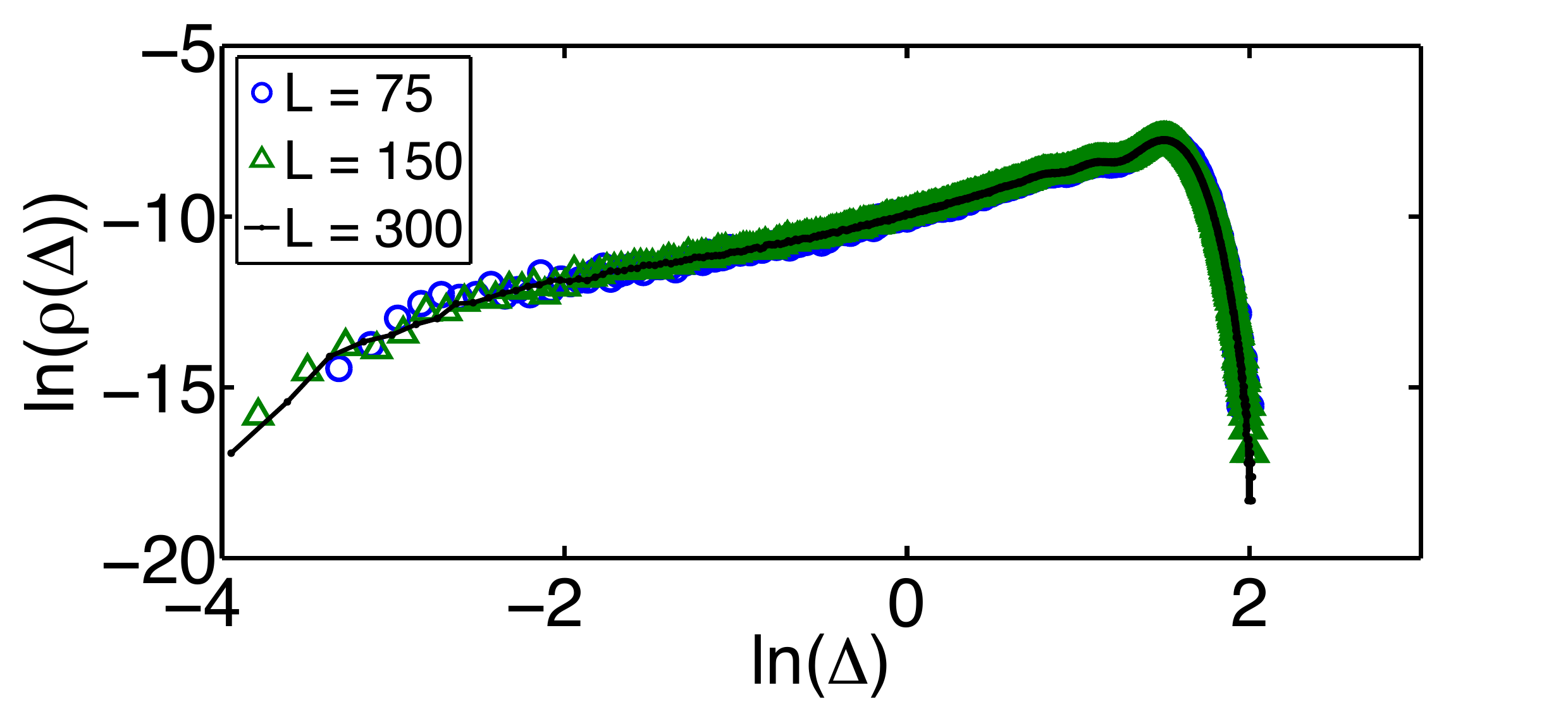}
\caption{The number per unit area of clusters with local gap near $\Delta$ for a choice of parameters in the glassy phase.
The initial distributions are those described in Appendix C, with $P_i(U)$ Gaussian and $P_i(J) \propto J^{-1.6}$.  
The value of the tuning parameter is $U_0 = 8.8$, and this puts the system on the glassy side of the transition
at $U_0 \approx 8.97$. Data is shown for $L = 75$, $150$, and $300$ lattices.  The density decays faster than
a power law at small $\Delta$.}
\label{rhodeltaglass}
\end{figure}

\indent Figure \ref{rhodeltaglass} shows the gap density for a choice of parameters in the glassy phase.  As the size of 
the lattice is raised, the density profile remains essentially invariant at large $\Delta$, but smaller gaps, corresponding to larger clusters,
begin to appear.  However, these smaller gaps simply fill out a decay to $0$ as $\Delta \rightarrow 0$.

\begin{figure}
\begin{minipage}[b]{0.4cm}
       {\bf (a)}
       
       \vspace{3.3cm}
\end{minipage}
\begin{minipage}[t]{7.9cm}
       \includegraphics[width=7.8cm]{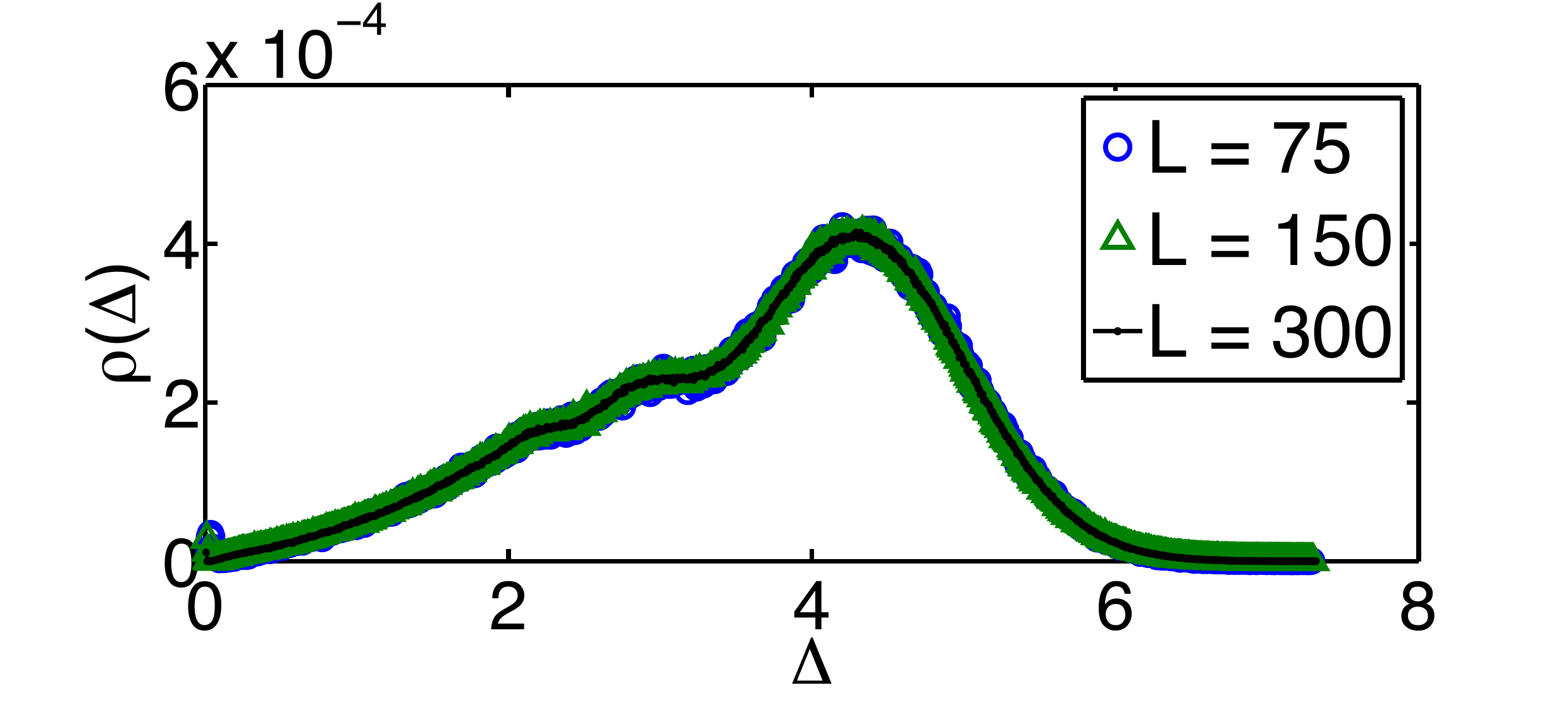}
\end{minipage}\\
\begin{minipage}[b]{0.4cm}
       {\bf (b)}

       \vspace{3.3cm}
\end{minipage}
\begin{minipage}[t]{7.9cm}
       \includegraphics[width=7.8cm]{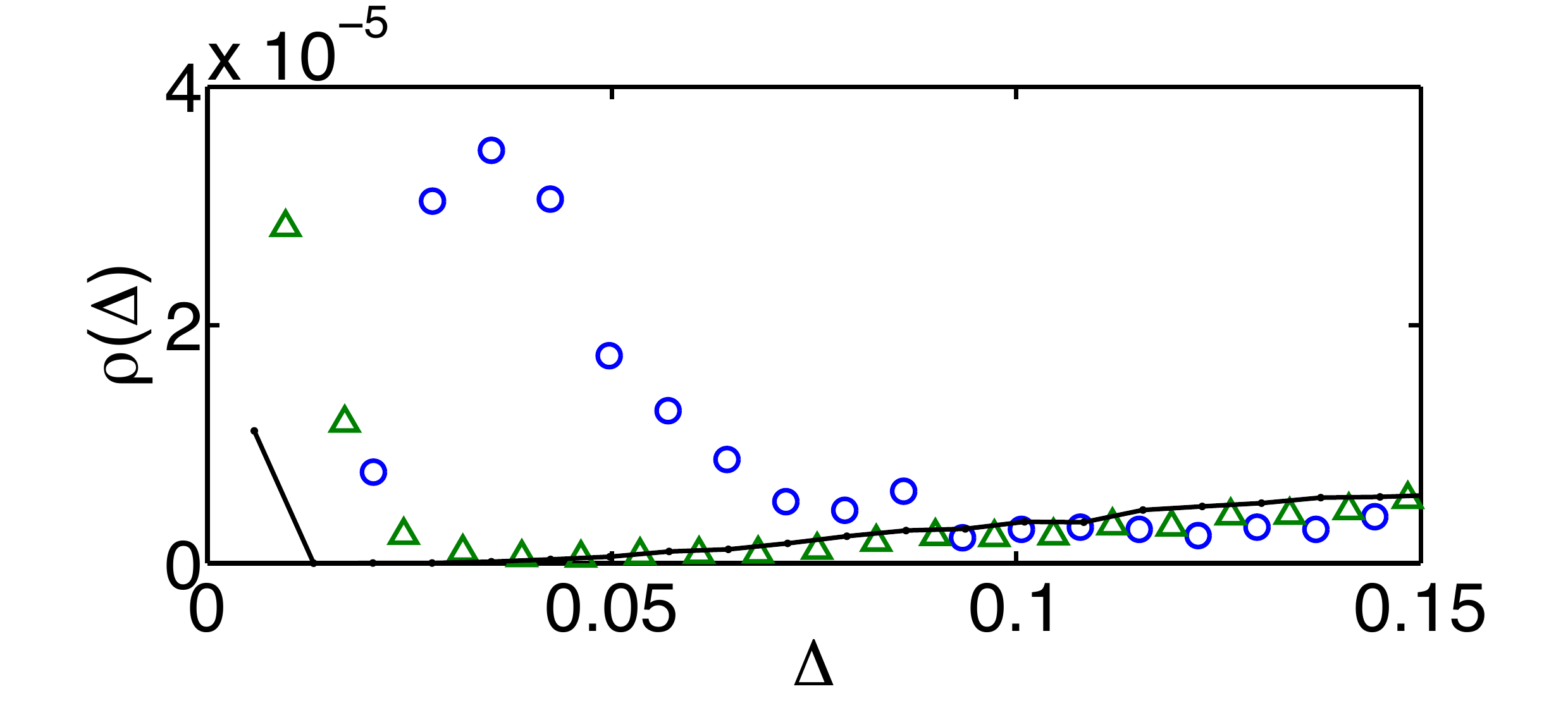}
\end{minipage}\\

\caption{In panel (a), the density (number per unit area) of clusters with a given gap $\Delta$ for a choice of parameters in the
superfluid phase.  The initial distributions are those described in Appendix C, with $P_i(U)$ Gaussian and $P_i(J) \propto J^{-1.6}$.  
The value of the tuning parameter is $U_0 = 9.2$, and this puts the system on the superfluid side of the transition
at $U_0 \approx 8.97$. Data is shown for $L = 75$, $150$, and $300$ lattices.  The density profiles exhibit two peaks.  
The broad peak that is visible in panel (a) remains invariant with $L$.  To expose the second peak, we provide a magnified
view of the low $\Delta$ part of the density in panel (b).  This peak simultaneously shrinks and propagates towards $\Delta = 0$ as the system size is raised.}
\label{rhodeltasf}
\end{figure}

\indent Now, we turn our attention to the putative superfluid phase.  The gap density in this phase is shown in Figure \ref{rhodeltasf}.  
In panel (a), there is an invariant piece to the distribution, but at very low $\Delta$, a second peak appears as well. 
Panel (b) of Figure \ref{rhodeltasf} shows a magnified view of this low $\Delta$ peak.  
This peak propagates towards $\Delta = 0$ as the system size is raised.  Accompanying the propagation is a shrinking: the integrated weight of the 
low $\Delta$ peak shrinks as $L^{-2}$.  The consequences of these two effects need to be taken into account carefully
to calculate the compressibility.

\subsubsection{The Cluster Size Density $\rho(s)$}

\begin{figure}
\begin{minipage}[b]{0.4cm}
       {\bf (a)}
       
       \vspace{3.3cm}
\end{minipage}
\begin{minipage}[t]{7.9cm}
       \includegraphics[width=7.8cm]{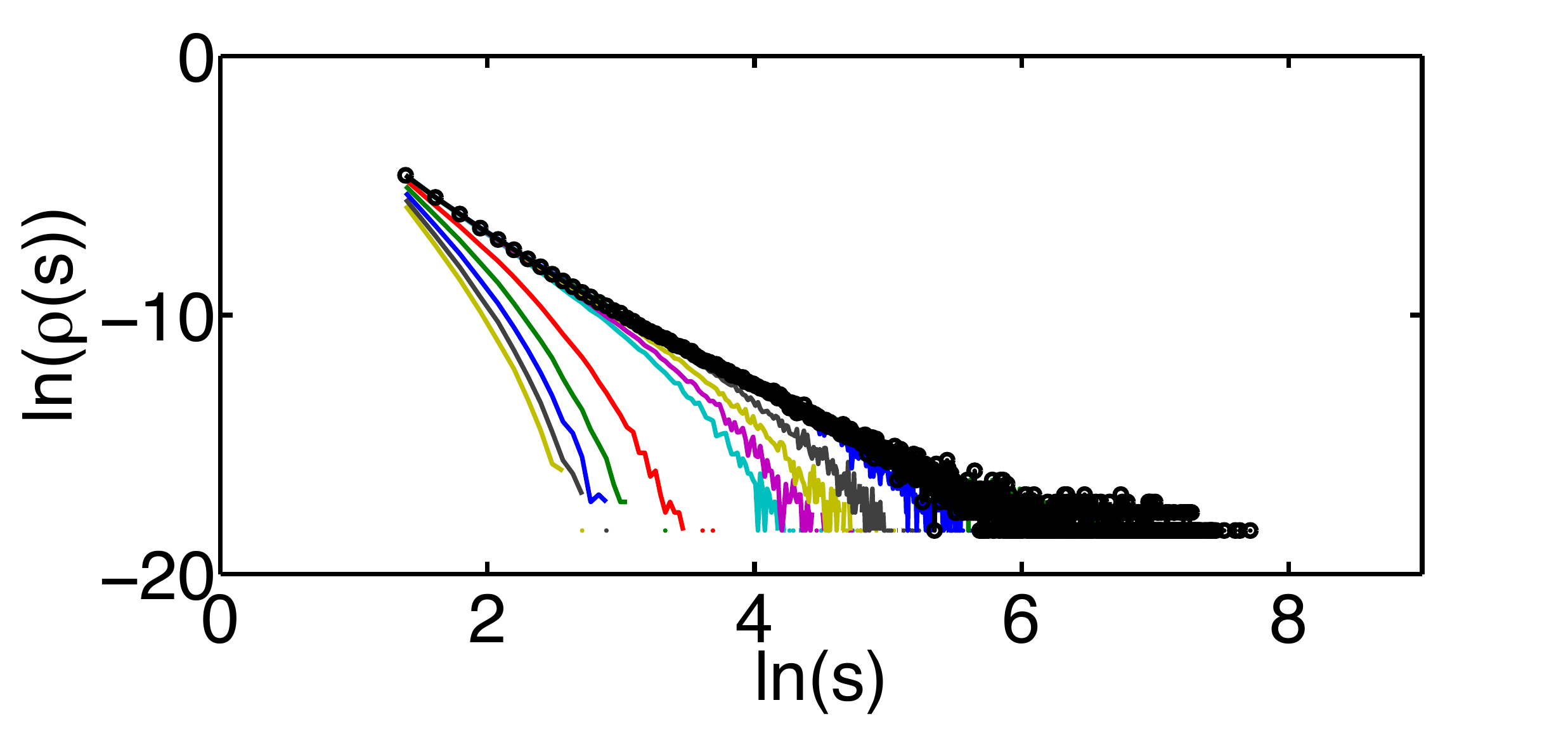}
\end{minipage}\\
\begin{minipage}[b]{0.4cm}
       {\bf (b)}

       \vspace{3.3cm}
\end{minipage}
\begin{minipage}[t]{7.9cm}
       \includegraphics[width=7.8cm]{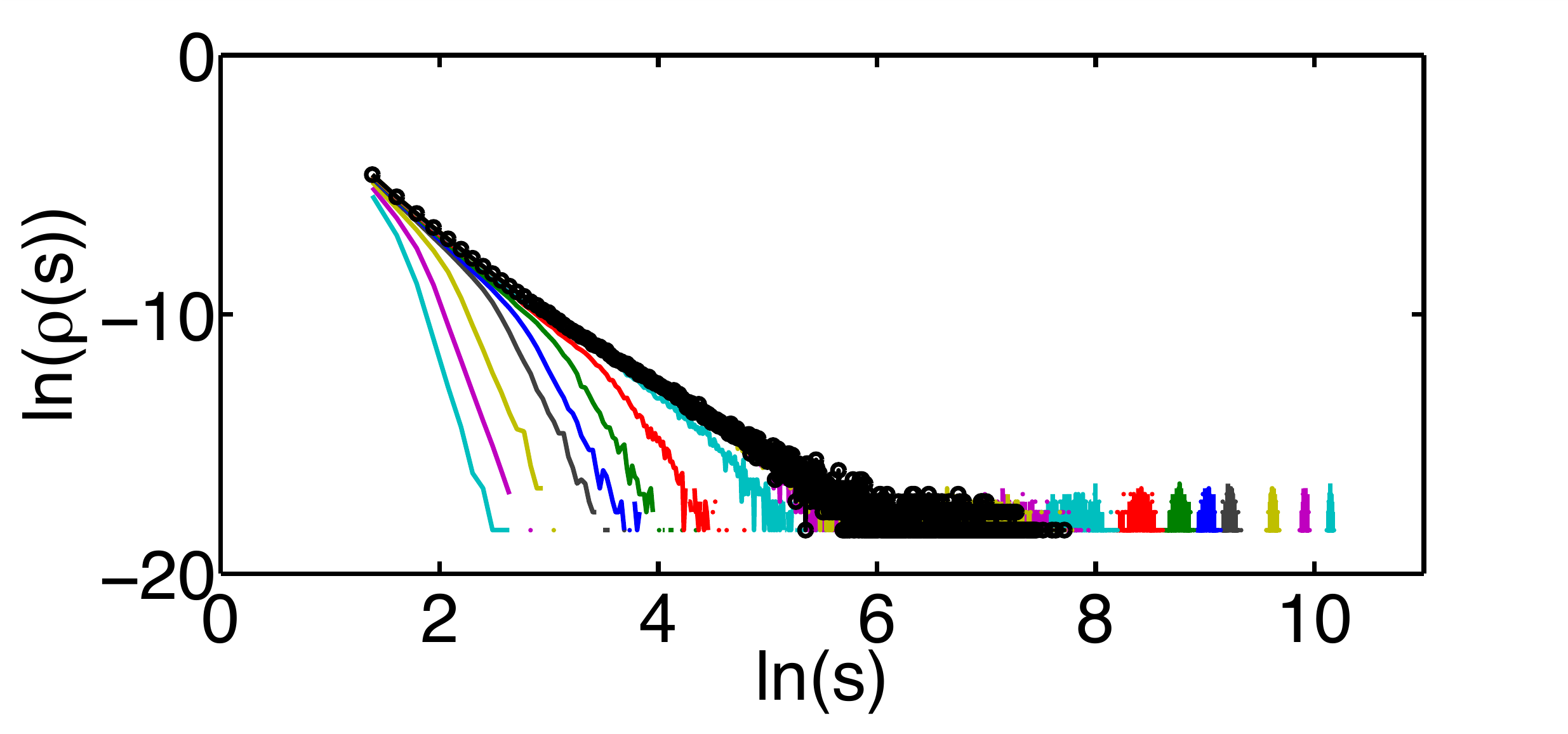}
\end{minipage}\\

\caption{Sweeps of $\rho(s)$ as the system is tuned through the superfluid-insulator transition on $L = 300$ lattices.   
The initial distributions are those described in Appendix C, 
with $P_i(U)$ Gaussian and $P_i(J) \propto J^{-1.6}$.   The tuning parameter is $U_0$, the center of $P_i(U)$.
Panel (a) shows the sweep from deep in the glassy phase ($U_0 = 20$) to the transition ($U_0 = 8.97$).  The closest data set to
the transition is indicated by the black line with large data point markers.  This line is reproduced in panel (b), which shows the sweep from the transition into the 
superfluid phase (up to $U_0 = 6.5$).  In the superfluid phase, the density plot is characterized by a peak at large $s$ and a remnant decay at 
low $s$.}
\label{rhossweep}
\end{figure}

\indent We now consider how $\rho(s)$, the density of clusters of size $s$, 
varies as we sweep through the glassy regime and into the superfluid.  
Panel (a) of Figure \ref{rhossweep} shows the approach to the transition from the glassy side.  
Very close to the transition at $U_0 \approx 8.97$, $\rho(s)$ exhibits a striking power law decay.  Proceeding into
the proposed glassy phase, the power law decay persists at small $s$.  However, this behavior is cut off by some scale $\tilde{s}$,
beyond which $\rho(s)$ decays very rapidly, essentially exponentially.    

\indent In panel (b), we proceed in the other direction from the putative transition, into the superfluid regime.  
A peak, corresponding to the macroscopic clusters, appears at large $s$.
The macroscopic cluster in each sample is dressed by rare-regions clusters,
and these clusters are represented by the remnant decay at small $s$.  While the large $s$ peak is
related to the low $\Delta$ peak in Figure \ref{rhodeltasf}, the remnant decay at small $s$ is 
the analogue of the high $\Delta$ feature that stays invariant with system size.  The low $s$ decay in panel (b) of Figure \ref{rhossweep}
qualitatively resembles the decay well inside the glassy regime.  In summary, $\rho(s)$ exhibits a pure power law decay in the vicinity of 
the proposed glass-superfluid transition; tuning away from criticality in either direction, and excluding the macroscopic clusters of the superfluid phase, 
the power law form of $\rho(s)$ only survives up to a scale $\tilde{s}$.  For $s > \tilde{s}$, clusters become exponentially rare.  

\begin{figure}
\centering
\includegraphics[width=8cm]{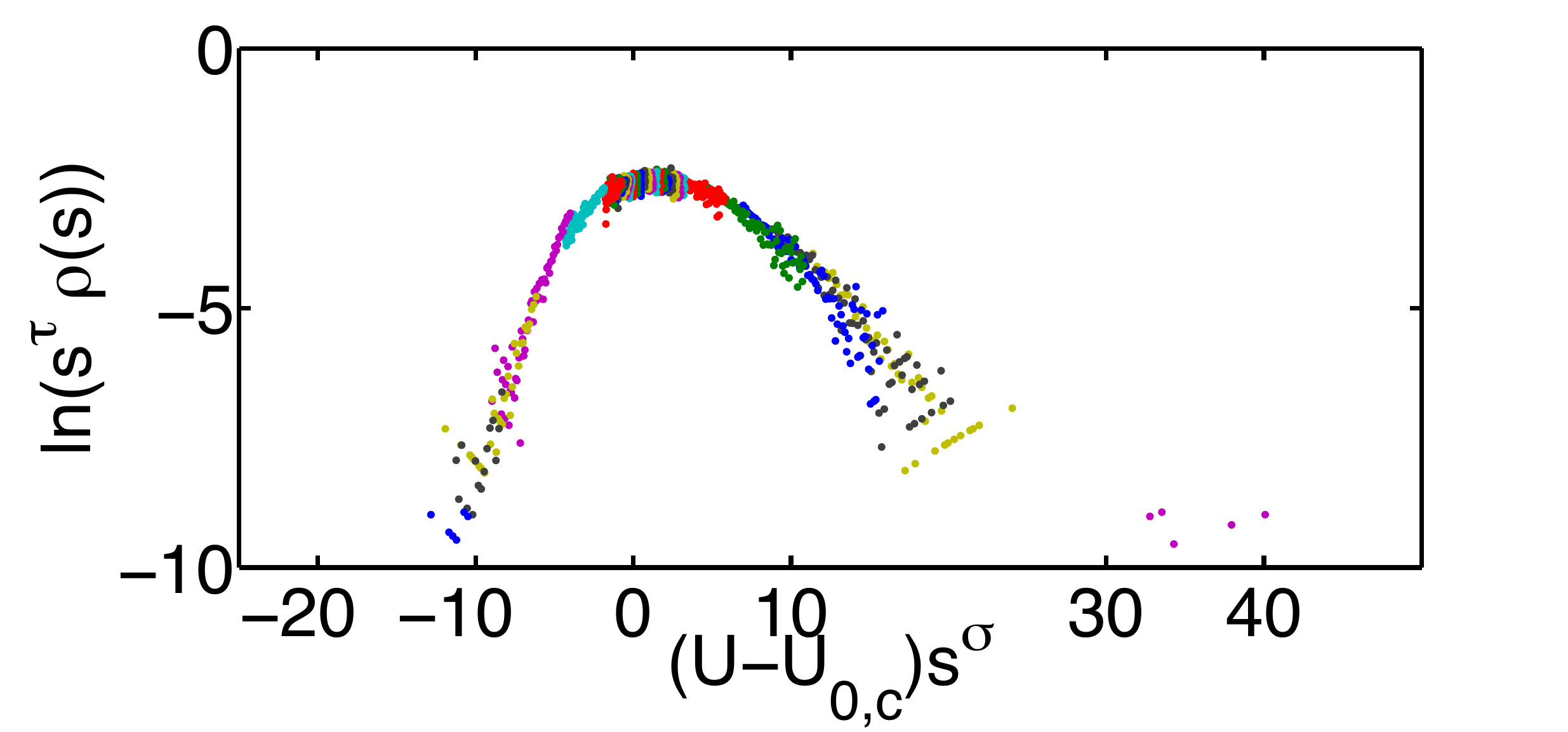}
\caption{Scaling collapse of the $\rho(s)$ curves from Figure \ref{rhossweep}.  Small cluster sizes  ($s < 30$) need to be
discarded, because they are non-universal.  Large cluster sizes ($s > 100$) need to be discarded, because they are noisy.
Then, the remaining $\rho(s)$ curves, taken for different values of the tuning parameter, collapse onto a universal curve. }
\label{rhoscollapse}
\end{figure}

\indent A type of scaling collapse can be performed for the $\rho(s)$ curves from Figure \ref{rhossweep}, and this collapse is shown in Figure \ref{rhoscollapse}.  We will
see below that this collapse gives a complementary set of critical exponents which are related by scaling
to those that can be extracted from Figure \ref{collapsesmax}.

\subsubsection{Susceptibility $b$-factors}

\indent The data presented above for the cluster densities $\rho(s)$ and $\rho(\Delta)$ highlights the difference
between the rare-regions and macroscopic clusters generated by the RG.  A study of how the $b$-factors for the 
clusters vary as a function of $s$ can bring out another difference between the two types of clusters.  Recall that
the $b$-factor was introduced in equation (\ref{clustersusc}) to quantity the effect of harmonic fluctuations on the
susceptibility of a superfluid cluster.  As such, understanding the behavior of the $b$-factors will also be essential
in explaining the behavior of $\chi$ in the various phases of our model.

\begin{figure}[h]
\begin{minipage}[b]{0.4cm}
       {\bf (a)}
       
       \vspace{3.3cm}
\end{minipage}
\begin{minipage}[t]{7.9cm}
       \includegraphics[width=7.8cm]{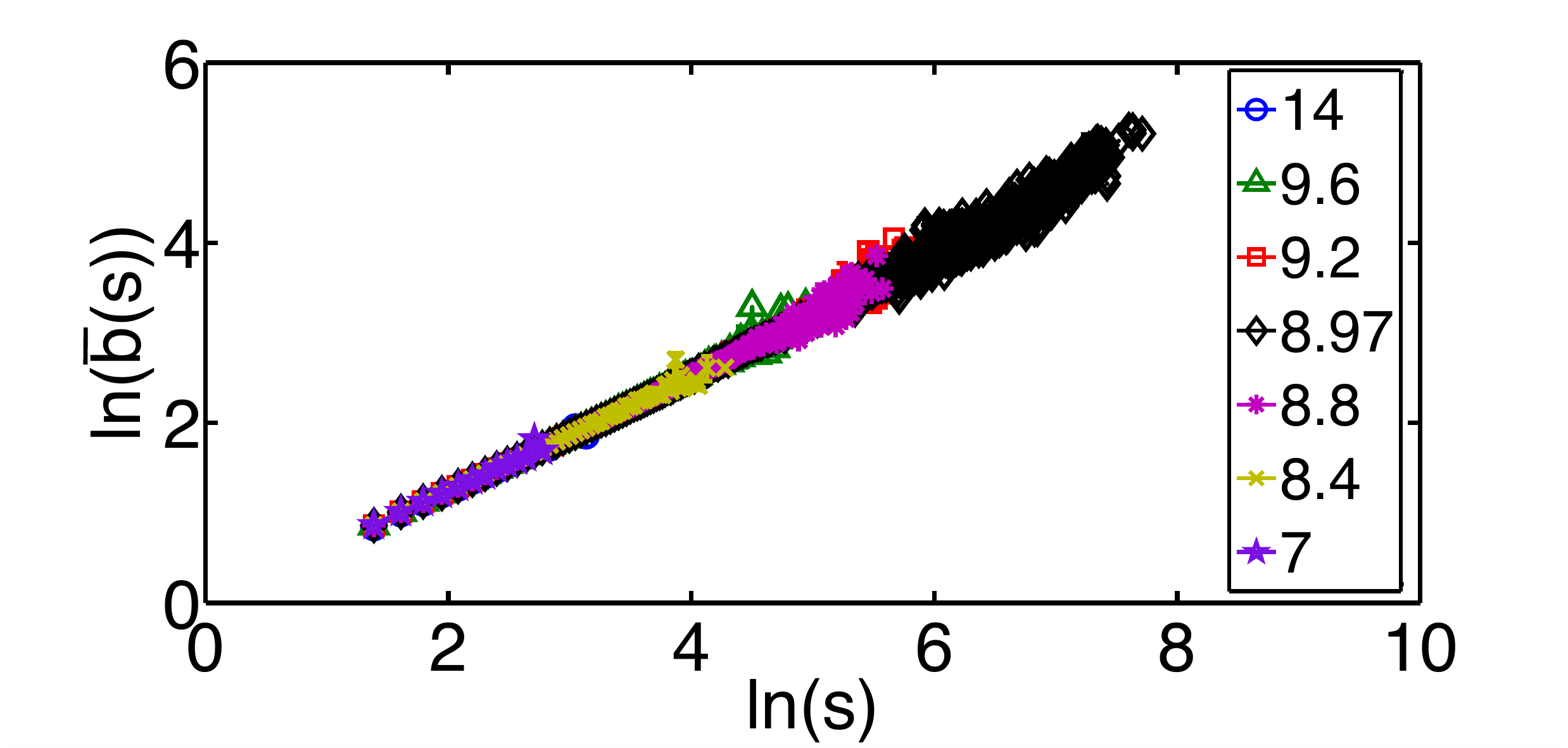}
\end{minipage}\\
\begin{minipage}[b]{0.4cm}
       {\bf (b)}

       \vspace{3.3cm}
\end{minipage}
\begin{minipage}[t]{7.9cm}
       \includegraphics[width=7.8cm]{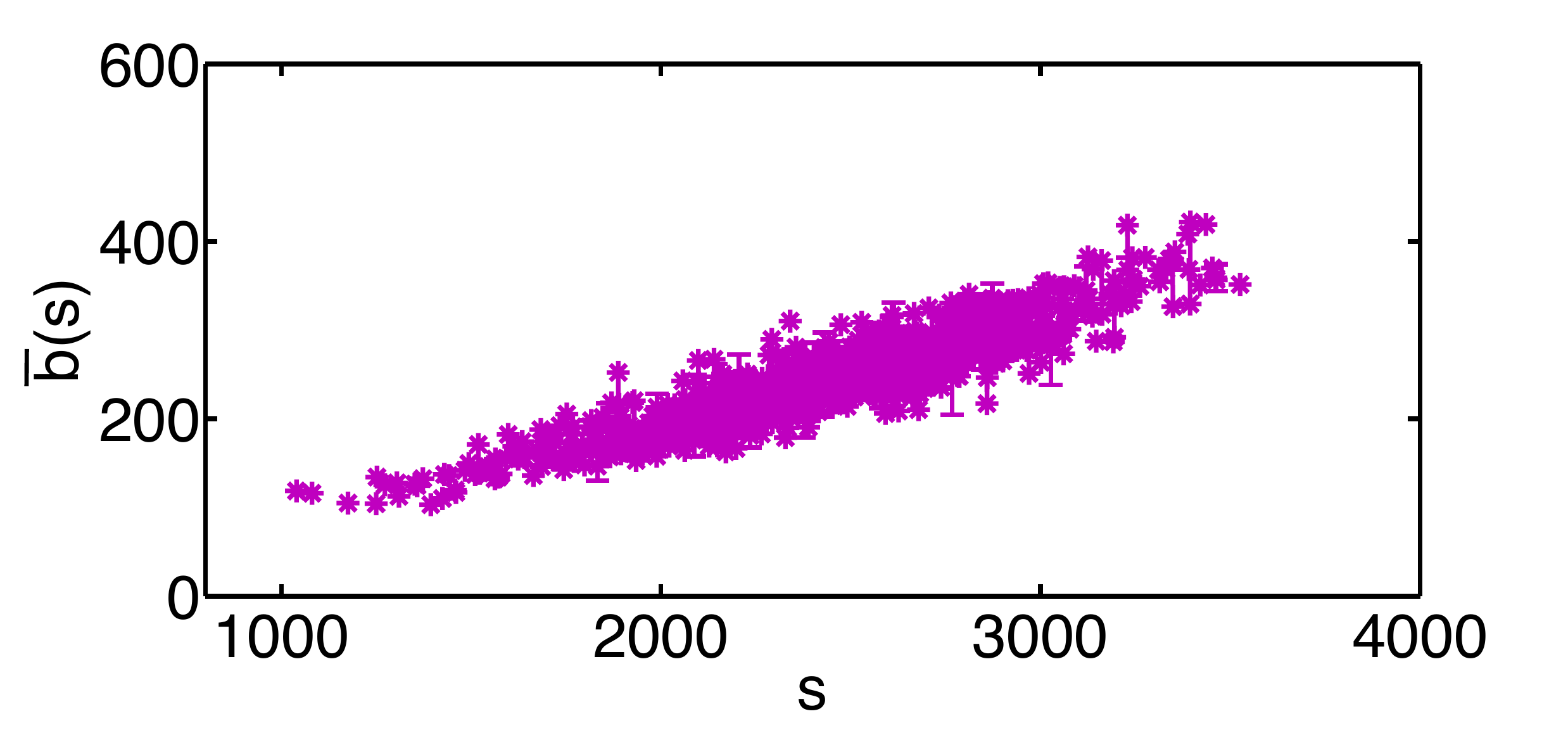}
\end{minipage}\\

\caption{The behavior of the mean $b$-factor for clusters of size $s$ as a function of $s$.  In panel (a),
we show data for the glassy regime and for the non-macroscopic clusters of the superfluid regime.  
The initial distributions are those described in Appendix C, and data is shown for seven values of
the tuning parameter $U_0$ on $L = 300$ lattices.  The log-log plot shows power law behavior of $\bar{b}(s)$ vs.\ $s$.
In panel (b), we show data for the macroscopic clusters when $U_0 = 8.8$.  The plot indicates that $\bar{b}(s) \sim s$
for macroscopic clusters.  In both plots, the error bars indicate the error on the mean $\bar{b}(s)$ over all 
clusters of size $s$.  For some of the largest and smallest values of $s$ in each plot,
the absence of an error bar indicates that only one cluster of that size was generated in all of the samples.}
\label{bfactors}
\end{figure}

\indent In Figure \ref{bfactors}, we plot the average value of $b$ for a cluster of size $s$ and plot it against $s$.  
Again, we work with $L = 300$ lattices and the choice of distributions described in Appendix C.
Panel (a) shows data for the glassy phase and for the non-macroscopic clusters of the superfluid phase.  We see that $\bar{b}$
varies with $s$ as a power law:
\begin{equation}
\label{bspower}
\bar{b}(s) \sim s^\zeta
\end{equation}
and that the power is consistent for seven different choices of the tuning parameter $U_0$.  We will provide an estimate of $\zeta$ 
in Section \ref{sec:phases}.  Panel (b) of Figure \ref{bfactors} shows 
data for the macroscopic clusters when $U_0 = 8.8$.  Now, $\bar{b}(s) \propto s$. 
This behavior can be anticipated from a simple picture of macroscopic cluster growth, which we will discuss in Section \ref{sec:phases}.

\section{The Phases and Quantum Phase Transitions of the Disordered Rotor Model}
\label{sec:phases}

\indent Having collected representative numerical data in the previous section, we now assess what the data
tells us about the phases and phase transitions of our model.  Our first task will be to confirm the association of phases with the
behaviors of physical properties that we observed in Figures \ref{physprop} and \ref{physpropaux}.  To this end, we will have
to preemptively introduce one of the main conclusions of the discussion below: that our data points to a superfluid-insulator
transition driven by a percolation-type process.  The rare-regions clusters of the glassy phase find one another, and their phases
cohere, producing a macroscopic cluster of superfluid ordering and driving the transition to long range order and global 
superfluidity.

\indent Motivated by this intuitive picture of the transition, the logic of the discussion below will be the following:
in the proposed glassy and superfluid regimes of Figure \ref{physpropaux}, the cluster size density $\rho(s)$
exhibits the universal features that we would expect from non-percolating and percolating phases of standard models of
percolation. We can use these correspondences to extrapolate the behaviors seen in Figure \ref{physpropaux} 
to the thermodynamic limit, in the process showing that these phases will indeed have the properties expected of 
glassy and superfluid phases.  Furthermore, we can analyze the critical point and extract the critical exponents that 
govern the superfluid-insulator transition.  After characterizing this transition, we will finally return to the question of the
identity of the glassy phase and determine if a Mott glass is present in the disordered rotor model (\ref{rot}). 

\indent Before proceeding, we should clarify that, although our RG produces clusters with size distributions 
similar to models of percolation, our transition is not the standard percolation transition. Indeed, the 
exponents that we recover are significantly different from the percolation exponents on a 2D square lattice. 
However, the analogy to percolation allows us to easily identify the relations between the various exponents
and the scaling functions for the observables.

\subsection{Phases of the Model}

\subsubsection{Mott Insulator}

\indent We briefly digress to describe the phase of our model in which the percolation picture is irrelevant, simply because
there are no regions of superfluid ordering.  In a Mott insulating phase, the system wants to pin the number fluctuation to zero on 
each site to avoid the energetic costs of charging.  Hence, $s_{\text{max}}$ and $s_2$ simply stay pinned at one as $L$ is raised.  
Meanwhile, $\Delta_{\text{min}}$ should asymptote to a constant, reflecting the fact that the Mott insulator is gapped.  A phase without any cluster formation can be described
by completely local physics.  This means that the susceptibility can be approximated by disorder averaging a single site problem.  
In other words, $\chi$ should also stay constant as the system size is increased. Thus, in a Mott
insulating phase, we expect exactly the behavior seen in panel (a) of Figures \ref{physprop} and \ref{physpropaux}.

\subsubsection{Glass}

\indent In non-percolating phases of standard models of percolation, the cluster size density is expected to go as:
\begin{equation}
\label{percolationrhos}
\rho(s) = c s^{-\tau}f\left(\frac{s}{\tilde{s}}\right)
\end{equation}
where $c$ is a constant.   The function $f(x)$ is expected to be approximately constant for $x < 1$ and to rapidly decay for $x > 1$\cite{dietrich1994introduction}.  
Equation (\ref{percolationrhos}) is consistent with what we have observed in our 
proposed glassy phase in panel (a) of Figure \ref{rhossweep} and is also consistent with the expectation that, in a Griffiths phase,
the frequency of occurrence of large rare-regions decays exponentially in their size\cite{vojta2010quantum}.  In our numerics, $f(x)$ seems to follow a pure exponential 
decay $f(x) \sim e^{-x}$, but the conclusions we present below would be qualitatively unchanged if, for example, $f(x) \sim e^{-x^\lambda}$.

\indent With the form of equation (\ref{percolationrhos}) in hand, we can now formulate a simple argument for the asymptotic behavior of 
$s_{\text{max}}$.  In particular, the order-of-magnitude of $s_{\text{max}}$ is set by the condition\cite{dietrich1994introduction}:
\begin{equation}
\label{smaxglasscondition}
L^2 \sum^{L^2}_{s = s_{\text{max}}} \rho_L(s) \approx 1
\end{equation}
If the left hand side of equation (\ref{smaxglasscondition}) is less than $1$, then it is unlikely that even one cluster of size greater than or 
equal to $s_{\text{max}}$ will be present in a sample of size $L^2$.  In equation (\ref{smaxglasscondition}), $\rho_L(s)$ is the finite size approximant
to the function $\rho(s)$ that appears in equation (\ref{percolationrhos}).  The upper limit of the sum in equation (\ref{smaxglasscondition})
is taken as $L^2$ because larger clusters simply cannot appear in the finite size sample.  With enough sampling of the random distributions,
it should, in principle, be able to accurately represent the approximant $\rho_L(s)$ out to $s = L^2$.  The data indicates that $\rho_L(s)$ will
simply reproduce $\rho(s)$ out to this value of $s$, so in the calculations below, we can replace the approximant $\rho_L(s)$ by $\rho(s)$.
This will \textit{not} be possible in the superfluid phase.

\indent For large $L$, where we also expect $s_{\text{max}} > \tilde{s}$, we can use equation (\ref{smaxglasscondition}) to find $s_{\text{max}}$ by computing
\begin{eqnarray}
L^2 \sum^{L^2}_{s = s_{\text{max}}} \rho(s)  & \approx & L^2 \int_{s_{\text{max}}}^{L^2} ds \rho(s) \nonumber \\
                                                                                        & = & cL^2 \int_{s_{\text{max}}}^{L^2} ds \exp{\left[-\frac{s}{\tilde{s}}-\tau\ln(s)\right]} \nonumber \\
                                                                                        & \approx & cL^2 \int_{s_{\text{max}}}^{L^2} ds \exp{\left[-\frac{s}{\tilde{s}}\right]} \nonumber \\
                                                                                        & = & c\tilde{s}L^2 \left(e^{-\frac{s_{\text{max}}}{\tilde{s}}} - e^{-\frac{L^2}{\tilde{s}}}\right)
\label{smaxglassconditioncalc}
\end{eqnarray}
Setting this expression equal to $1$ and inverting for $s_{\text{max}}$, we find that, asymptotically in $L$:
\begin{equation}
\label{smaxglassresult}
s_{\text{max}} \sim \ln{L}
\end{equation}
For large clusters, the link decimation rule for addition of charging energies (\ref{linkdec}) implies that $U \sim s^{-1}$, and therefore:
\begin{equation}
\label{deltaminglassresult}
\Delta_{\text{min}} \sim \frac{1}{\ln{L}}
\end{equation}

\indent An entirely analogous condition to equation (\ref{smaxglasscondition}) can be formulated for $s_2$.  We simply replace the right 
hand side of equation (\ref{smaxglasscondition}) with two, indicating that we want to find the value of $s$ such that
there are likely to be two clusters of that size or greater in a typical sample.  However, the remainder of the calculation is 
qualitatively unaffected by this change.  Therefore, $s_2$ should also grow logarithmically in this regime:
\begin{equation}
\label{s2glassresult}
s_{2} \sim \ln{L}
\end{equation}

\indent Finally, we turn to the susceptibility.  This can be computed as follows:
\begin{eqnarray}
\chi & = & \frac{1}{L^2} \sum_{C} \frac{b^2_C}{U_C} \nonumber \\
       & \sim & \frac{1}{L^2} \sum_{C} s_C b^2_C \nonumber \\
       & \sim & \frac{1}{L^2} \sum^{s_{\text{max}}}_{s=1} \rho(s)L^2 (\bar{b}(s))^2 s \nonumber \\
       & = & \sum^{s_{\text{max}}}_{s=1} \rho(s) (\bar{b}(s))^2 s
\label{chiglasscalculation}
\end{eqnarray}
In this calculation, the sum over $C$ is a sum over clusters, with $s_C$ being the size of the cluster.
Then, the sum over $s$ is, as before, a sum over cluster sizes, and $\bar{b}(s)$ is the average value of the
$b$-factor for a cluster of size $s$.  Figure \ref{bfactors} shows that, in the glassy regime, $\bar{b}$ varies as a 
power of $s$ and that this power remains the same all the way up to criticality.  While we do not have a complete
understanding of this behavior, we can still understand the asymptotic behavior of $\chi$ by reasoning that 
$\bar{b}$ can, at most, grow linearly in $s$.  This follows from an interpretation of the $b$-factor as the effective
number of rotors that order with the field in a cluster of size $s$.
Since $\rho(s)$ decays exponentially for large $s$ while $\bar{b}(s)$ grows at most as a power, 
the sum (\ref{chiglasscalculation}) converges, and $\chi$ should be asymptotically constant:
\begin{equation}
\label{chiglassthermlimit}
\lim_{L \rightarrow \infty} \chi  =  \chi_\infty 
\end{equation}

\indent All of these behaviors are consistent with what has been observed numerically in panel (b) of Figure \ref{physpropaux}.
Moreover, since logarithmic behavior can be difficult to discriminate from a slow power law at low $L$, 
they are also consistent with panel (b) of Figure \ref{physprop}.  Thus, under the numerically justified assumption that
this regime corresponds to a non-percolating phase, we can deduce that, as $L$ increases, 
arbitrarily large rare-regions of superfluid ordering will appear, driving the gap to zero.  However, the typical size of these regions
grows extremely slowly (i.e.\ logarithmically) with system size.  The fraction of sites occupied by the largest
cluster in a typical sample vanishes as $L \rightarrow \infty$, so there is no long range order.  The behavior of this phase 
for large $L$ corresponds to what we should expect for a glassy phase.

\subsubsection{Critical Region}

\indent At the critical point of a percolation transition, the cluster size scale in equation (\ref{percolationrhos}) 
is expected to diverge as:
\begin{equation}
\label{definesigma}
\tilde{s} \sim |g-g_c|^{-\frac{1}{\sigma}}
\end{equation}
where $g$ is the tuning parameter for the transition.  This divergence is related to the divergence of
a correlation length which indicates the typical linear extent of the largest clusters:
\begin{equation}
\label{percolationnu}
\xi \sim |g-g_c|^{-\nu}
\end{equation}
By equation (\ref{definesigma}), $\rho(s)$ is a power law at criticality:
\begin{equation}
\label{rhoscritical}
\rho(s) = c s^{-\tau}
\end{equation}
This means that the critical point is characterized by a scale invariant, fractal structure of clusters\cite{dietrich1994introduction}.
In turn, this implies that $\xi$ and $\tilde{s}$ are related by a fractal dimension:
\begin{equation}
\label{tildesxirelation}
\tilde{s} = \xi^{d_f}
\end{equation}
Equations (\ref{definesigma}), (\ref{percolationnu}), and (\ref{tildesxirelation}) together imply:
\begin{equation}
\label{scaling2}
\sigma = \frac{1}{\nu d_f}
\end{equation}
We will use this scaling law in our analysis of the transition below\cite{dietrich1994introduction}.

\indent For the present purposes, note that equation (\ref{rhoscritical}) is once again consistent with 
what we have observed numerically in Figure \ref{rhossweep}.  Now, the calculation for $s_{\text{max}}$ becomes:
\begin{equation}
L^2 \sum^{L^2}_{s = s_{\text{max}}} \rho(s)  \approx cL^2 \int_{s_{\text{max}}}^{L^2} ds s^{-\tau} = 1
\label{smaxcritconditioncalc}
\end{equation}
which, when inverted, yields:
\begin{eqnarray}
\label{lnsmaxcrit}
\ln{s_{\text{max}}} & = & \frac{2}{\tau-1}\ln{L} - \frac{1}{\tau-1}\ln{\left(\frac{\tau-1}{c}\right)} \nonumber \\
                          & \quad & + \ln{\left(1+\frac{c}{(\tau-1)L^{2(\tau-2)}}\right)}
\end{eqnarray}
Asymptotically, as long as $\tau > 2$, this simply corresponds to a power law growth of $s_{\text{max}}$:
\begin{equation}
\label{smaxcrit}
s_{\text{max}} \sim L^\frac{2}{\tau-1} 
\end{equation}
On the other hand, since $d_f$ is the exponent that connects length and cluster size scales at the transition, equation (\ref{smaxcrit}) 
yields another scaling relation:
\begin{equation}
\label{scaling1}
d_f = \frac{2}{\tau-1}
\end{equation}
Equations (\ref{scaling2}) and (\ref{scaling1}) are the usual scaling laws connecting exponents at a percolation transiiton\cite{dietrich1994introduction}.

\indent In accordance with equation \ref{smaxcrit}, the gap should close as:
\begin{equation}
\label{deltamincrit}
\Delta_{\text{min}} \sim L^{-\frac{2}{\tau-1}} = L^{-d_f}
\end{equation}
Furthermore, as in the glassy regime, the qualitative behavior of the second largest cluster size $s_2$ should be identical to that
of $s_{\text{max}}$:
\begin{equation}
\label{s2crit}
s_{2} \sim L^\frac{2}{\tau-1} 
\end{equation}

\indent Turning to the susceptibility, we find that it no longer converges to a finite value.  At criticality, power law behavior
of $\bar{b}(s)$ follows naturally from scale invariance.  When $\rho(s) \sim s^{-\tau}$ and $\bar{b}(s) \sim s^\zeta$:
\begin{eqnarray}
\label{chicrit}
\chi & = & \sum_{s=1}^{s_{\text{max}}} \rho(s)(\bar{b}(s))^2s \nonumber \\
                                                    & \sim & \int_{1}^{s_{\text{max}}} ds s^{1+2\zeta-\tau} \nonumber \\
                                                    & \sim & s^{2+2\zeta-\tau}_{\text{max}} \nonumber \\
                                                    & \sim & L^{\frac{4+4\zeta-2\tau}{\tau-1}} \nonumber \\
                                                    & = & L^{d_f(1+2\zeta)-2}
\end{eqnarray}
From Figure \ref{bfactors}, we can estimate the exponent $\zeta$:
\begin{equation}
\label{zetaestimate}
\zeta \approx 0.68 \pm 0.01
\end{equation}
With another choice of initial distributions ($P_i(U)$ bimodal and $P_i(J) \propto J^{-1.6}$), we have found a similar value
for $\zeta$.  If $d_f(1+2\zeta) > 2$, then $\chi$ asymptotically diverges as a power law, as seen in Figure \ref{physpropaux}.  
We will provide an estimate of $d_f$ shortly in equation (\ref{dfestimate1}).  For now, we note that the power observed for 
$\chi$ vs.\ $L$ in Figure \ref{physpropaux} is consistent with this estimate of $d_f$ and the estimate for $\zeta$ that is given
above.  In Figure \ref{physprop}, the relatively small system sizes likely put us out of the scaling regime for $\chi$,
and this is probably responsible for the extremely slow growth of $\chi$ with $L$.

\subsubsection{Superfluid}

\indent The percolating phase is characterized by the presence of a macroscopic cluster that
grows with the size of the lattice, so trivially:
\begin{equation}
\label{smaxsf}
s_{\text{max}} \propto L^2
\end{equation}
and therefore:
\begin{equation}
\label{deltamin}
\Delta_{\text{min}} \propto L^{-2}
\end{equation}
It is possible to construct an argument along the lines of condition (\ref{smaxglasscondition}) for the
behavior of $s_{\text{max}}$ in equation (\ref{smaxsf}), but in this case, it is important \textit{not} to substitute the 
infinite lattice density $\rho(s)$ for the finite size approximant $\rho_L(s)$.  The subtlety lies in the 
high $s$ peak observed in the density plots in panel (b) of Figure \ref{rhossweep}.  Consistent with their 
low $\Delta$ counterparts in the plots of the gap density, the location of these peaks propagates as $L^2$ 
towards high $s$ as $L$ is raised.  Simultaneously, the integrated weight of the peaks shrinks
as $L^{-2}$, reflecting the fact that there is only one macroscopic cluster in each sample.  
Balancing the shrinking and propagation, it is possible to see that, in order to achieve the condition (\ref{smaxglasscondition}),
$s_{\text{max}}$ must scale as shown above in equation (\ref{smaxsf}).

\indent The reasoning above has important implications for the behavior of $s_2$.  Since the weight of
the high $s$ peak of $\rho_L(s)$ shrinks as $L^{-2}$, the second largest cluster must be drawn from the
remnant low $s$ decay.  The physical picture behind this low $s$ decay is the following: suppose we 
remove the sites belonging to the macroscopic cluster from the original lattice.  In doing so, we remove 
some of the lowest charging energies and highest Josephson couplings from the bare lattice.  
Consequently, the remnant lattice is globally insulating. Nevertheless, rare-regions of superfluid ordering can form 
exactly as in the glassy phase.  It follows that $s_2$ will grow with $L$ as in the glassy phase:
\begin{equation}
\label{s2sf}
s_{2} \sim \ln{L}
\end{equation}
This is responsible for the peak in $s_2$ at criticality.

\indent To calculate the susceptibility, we first take into account the contribution of the macroscopic cluster.
The behavior of $\bar{b}(s)$ for a macroscopic cluster can be inferred from a simplified picture of a large cluster
merging with single neighbors.  As the macroscopic cluster grows, its charging energy becomes smaller, driving the
Debye-Waller factor for the cluster to one.  At the same time, the cluster's Josephson couplings to other sites
grow through link addition processes.  This means that the Debye-Waller factor for another site that is merging
with the macroscopic cluster also approaches unity.  Therefore, the $b$-factor addition rule (\ref{bfactorcluster}) 
approximately becomes $\tilde{b}_C = b_C+1$, and $b_C \sim s_C$ follows.  Then:
\begin{eqnarray}
\chi_{mac} & = & \frac{1}{L^2} \times \frac{(\bar{b}(s_{\text{max}}))^2}{\Delta_{\text{min}}} \nonumber \\
                    & \sim & \frac{1}{L^2} \times \frac{s_{\text{max}}^2}{\Delta_{\text{min}}} \nonumber \\
                    & \sim & L^4
\label{chimac}
\end{eqnarray}
The rare-regions clusters dressing the macroscopic cluster add a subleading contribution to 
the susceptibility, which we call $\chi_{rr}$.  The reasoning of equation (\ref{chiglassthermlimit}) indicates that 
this contribution should be asymptotically constant.  Thus, the quartic behavior of equation (\ref{chimac}) is the 
correct asymptotic behavior.  Finite size corrections from $\chi_{rr}$ will modify this behavior, however, and this
is probably why we do not quite see $\chi$ reach the $L^4$ behavior in panel (d) of Figure \ref{physpropaux}. 

\indent We have now provided arguments for the behaviors observed in each panel of that figure, and we summarize 
this information in Table \ref{physproptable}.

\begin{table}
\centering
\begin{tabular}{|r|r|r|r|r|}
\hline
 & Mott Insulator & Glass & Critical & Superfluid\\
\hline
$s_{\text{max}}$ & const. & $\ln(L)$ & $L^{d_f}$ & $L^2$  \\
\hline
$s_2$ & const. & $\ln(L)$ & $L^{d_f}$ & $\ln(L)$  \\
\hline
$\Delta_{\text{min}}$ & const. & $\frac{1}{\ln(L)}$ & $L^{-d_f}$ & $L^{-2}$  \\
\hline
$\chi$ & const. & const. & $L^{d_f(1+2\zeta)-2}$ & $L^4$  \\
\hline
\end{tabular}
\caption{Large lattice size (L) behavior of physical properties in the three phases and at the critical point of the glass-superfluid transition.}
\label{physproptable}
\end{table}

\subsection{Quantum Phase Transitons}

\subsubsection{Glass-Superfluid Transition}

\indent Earlier, we remarked in passing that systems exhibiting the behaviors that we have
now identified as Mott insulating and glassy eventually propagate towards the putative insulating region to the top left 
of the numerical flow diagrams.  Correspondingly, the systems exhibiting superfluid behavior propagate towards the putative superfluid
region to the bottom right.  We can now verify our tentative identifications of these stable regions of the diagram.  We can also determine
that the unstable fixed point controls the transition between the glass and the superfluid, the superfluid-insulator
transition of our disordered rotor model.  This allows us to draw the schematic picture of the flow diagram that we
presented in Figure \ref{flows}.

\indent We will now focus on the critical region and extract critical exponents governing the glass-superfluid transition.  
Estimating these exponents requires formulating scaling hypotheses for the behavior of physical quantities in the critical region.  
In the case of $s_{\text{max}}$\cite{dietrich1994introduction}:
\begin{equation}
\label{scalinghypothesissmax}
s_{\text{max}} = L^{d_f} h_1\left(\frac{L}{\xi}\right) = L^{d_f}\tilde{h}_1(L^{\frac{1}{\nu}}(g-g_c))
\end{equation} 
Exactly at criticality, $s_{\text{max}} \sim L^{d_f}$ asymptotically, so plotting $\frac{s_{\text{max}}}{L^{d_f}}$ vs.\  $(g-g_c)$ generates a crossing of
the curves for different lattice sizes. Slightly away from criticality, the power law behavior should persist if $L < \xi$.  For $L > \xi$, the system recognizes that it is not critical and we
should see a crossover to logarithmic growth on the glassy side and $L^2$ growth on the superfluid side.  Hence, $\frac{L}{\xi}$ is the important ratio near criticality, and this motivates
scaling form (\ref{scalinghypothesissmax}).  The scaling form, in turn, implies that we can produce scaling collapse by plotting 
$\frac{s_{\text{max}}}{L^{d_f}}$ vs.\  $(g-g_c)L^{\frac{1}{\nu}}$.  This is what we have done in panel (a) of Figure \ref{collapsesmax}.
This scaling collapse leads to the estimates:
\begin{equation}
\label{nuestimate1}
\nu \approx 1.09 \pm 0.04
\end{equation}
\begin{equation}
\label{dfestimate1}
d_f \approx 1.3 \pm 0.2
\end{equation}

\indent These estimates and error bars are obtained through the following procedure:  first, to find an estimate of the critical value of the tuning parameter $g$,
we examine the behavior of the sample average of $s_2$ vs.\ g.  Since $s_2$ varies as a power law in $L$ exactly at criticality and grows
more slowly within the phases, $s_2$ should exhibit a maximum at $g_c$.  We can find the error $\Delta g_c$ on our estimate of $g_c$ by taking the window of values of 
$g$ around $g_c$ for which the sample average of $s_2$ is within one error bar of the maximum.   To proceed to obtain estimates for $\nu$ and $d_f$, we now 
partition our data into bins of size $N_b$ samples.  For example, the first bin may consist of the first $N_b = 50$ samples for each value of the tuning parameter 
and each lattice size $L$.  Our immediate goal is to find the best values of $\nu$ and $d_f$ for this subset of our data.  
We first focus on the data for $g = g_c$ and do linear regression to find the best exponent that describes the power law growth of the average value of $s_{\text{max}}$ with $L$.
This gives an estimate of $d_f$ for the bin.  Next, for two lattice sizes (typically, $L = 150$ and $L = 300$), we compute an average of $s_{\text{max}}$
over the samples in the bin for several values of the tuning parameter around $g_c$.  Then, using the previous estimate of $d_f$ for the bin and the scaling
hypothesis (\ref{scalinghypothesissmax}), we test various values of $\nu$ until we achieve the best collapse.  Thus, we also obtain an estimate of $\nu$ for the bin.  
From the distribution of estimates for the different bins, we can find mean values and error bars for $d_f$ and $\nu$.  However, these error bars do not
take into account the error on our estimate of the critical point.  To propagate this error, we need to repeat the critical point estimation procedure using 
$g_c + \Delta g_c$ and $g_c - \Delta g_c$ as our estimates of the critical point.  

\indent We have repeated the scaling collapse of $s_{\text{max}}$ for a different choice of initial distributions: bimodal $P_i(U)$ and $P_i(J) \propto J^{-1.6}$.
Ultimately, we have been able to recover estimates for $\nu$ and $d_f$ which are 
consistent with (\ref{nuestimate1}) and (\ref{dfestimate1}):
\begin{equation}
\label{nuestimate3}
\nu \approx 1.1 \pm 0.1
\end{equation}
\begin{equation}
\label{dfestimate3}
d_f \approx 1.2 \pm 0.2
\end{equation}

\indent A completely analogous scaling hypothesis can be made for $s_2$:
\begin{equation}
\label{scalinghypothesis2}
s_{2} = L^{d_f} h_2\left(\frac{L}{\xi}\right) = L^{d_f}\tilde{h}_2(L^{\frac{1}{\nu}}(g-g_c))
\end{equation} 
Then, the exponents needed to produce the collapse in panel (b) of Figure \ref{collapsesmax} are:
\begin{equation}
\label{nuestimate2}
\nu \approx 1.06 \pm 0.09
\end{equation}
\begin{equation}
\label{dfestimate2}
d_f \approx 1.31 \pm 0.07
\end{equation}
Since all the estimates (\ref{nuestimate1})-(\ref{dfestimate2}) are consistent, we will proceed using our tightest
estimates on these exponents: (\ref{nuestimate1}) and (\ref{dfestimate2}).

\indent We now note that a scaling hypothesis can also be formulated for the density $\rho(s)$.  From equation (\ref{percolationrhos}), we see that,
for fixed lattice size $L$, $s^\tau\rho(s)$ should depend only on the combination $\frac{s}{\tilde{s}} \sim s(g-g_c)^\frac{1}{\sigma}$.
Hence, by plotting $s^\tau\rho(s)$ vs.\ $s(g-g_c)^\frac{1}{\sigma}$ and tuning until the curves for different choices of $g$ collapse,
we ought to be able to extract estimates for $\tau$ and $\sigma$.  On the other hand, $\tau$ and $\sigma$ are related to $\nu$ and
$d_f$ through the scaling relations (\ref{scaling1}) and (\ref{scaling2}), so from the estimates (\ref{nuestimate1}) and
(\ref{dfestimate2}), we can infer:
\begin{equation}
\label{tauestimate1}
\tau \approx 2.53 \pm 0.08
\end{equation}
\begin{equation}
\label{sigmaestimate1}
\sigma \approx 0.70 \pm 0.04
\end{equation}
In Figure \ref{rhoscollapse}, we attempt to produce collapse of $\rho(s)$ for $L = 300$ lattices using these estimates of $\tau$ and
$\sigma$.  To produce this plot, we need to discard data points for small cluster sizes ($s < 30$), because they are non-universal, and for large
cluster sizes ($s > 100$), because they are noisy.  Once we do this, the collapse works fairly well, indicating that we have found a consistent
set of critical exponents obeying the necessary scaling relations.

\subsubsection{Comments on the Insulator-Insulator Transition}

\indent Before proceeding further, we should note that our numerics do not accurately capture the boundary between the Mott insulator
and the Mott glass.  Several authors have argued that we should expect glassy behavior to occur whenever the ratio of the largest 
bare Josephson coupling to the lowest bare charging energy $\frac{J_{\text{max}}}{U_{\text{min}}}$ exceeds the value of the
clean transition, because this condition allows for the presence of regions in which the system locally looks like it is in the superfluid phase~\cite{weichman2008particle,pollet2009absence}.  
However, in the strong disorder RG treatment, some distributions that satisfy this criterion still produce Mott insulating 
behavior out to the largest lattice sizes that we investigate.  Since the glassy phase occurs due to rare-regions or Griffiths effects, 
a finite size system will only look glassy if it is large enough for the rare-regions to appear.  This suggests that some choices of parameters
which yield Mott insulating behavior on finite size lattices may actually yield glassy behavior in the thermodynamic limit.  
Of course, this difficulty necessarily afflicts all numerical methods (except those that rely on mean-field type approximations\cite{bissbort2009stochastic}), 
since they are confined to operate on finite size systems.  

\indent We will not comment on this transition further, but we take this opportunity to refer the reader to papers by Kr\"{u}ger et al.\
and Pollet et al., which present two viewpoints on the Mott insulator to Bose glass transition~\cite{kruger2010two,pollet2009absence}.

\subsection{Identifying the Glass}

\indent Finally, we return to the question of the identity of the glassy phase.  Is the phase a Bose glass or a Mott glass? 
A definitive diagnosis requires a measurement of the compressibility:  
\begin{equation}
\label{compress}
\kappa = \frac{1}{L^2} \sum_j \left. \frac{\partial \langle \hat{n}_j \rangle}{\partial \mu}\right|_{\mu = 0}
\end{equation}
The compressibility is more subtle to measure than the quantities that we have already discussed.
Any finite size system is gapped and therefore incompressible.  On the other hand, in the thermodynamic limit, the gap can
vanish and the compressibility need not be zero.

\indent How can we measure the compressibility of the glassy phase in the RG?  In Figures \ref{rhodeltaglass} and \ref{rhodeltasf},
we presented data for the density (number per unit area) of clusters with a given gap $\Delta$.  
With this density profile in hand, we can calculate the density of particles introduced to the system by a small chemical potential
shift $\mu$:
\begin{eqnarray}
\rho_{\text{ex}} & = & \int_0^{\mu} d\Delta \rho(\Delta)n(\Delta) \nonumber \\
                 & = & \int_0^{\mu} d\Delta \rho(\Delta)\lfloor \frac{\mu}{2\Delta}-\frac{1}{2} \rfloor \nonumber \\
                 & \approx & \int_0^{\mu} d\Delta \rho(\Delta)\left( \frac{\mu}{2\Delta}-\frac{1}{2} \right)
\label{particledensityglass}
\end{eqnarray}
Here, $n(\Delta)$ is the number of particles added to a cluster with gap $\Delta$ if the chemical potential is $\mu$.
If $\rho(\Delta)$ stays finite as $\Delta \rightarrow 0$, the integral is divergent, and the system is infinitely compressible.
Suppose alternatively that $\rho(\Delta)$ vanishes as $\Delta^\beta$ for small $\Delta$.  Then:
\begin{equation}
\label{compglasscalculation}
\int_0^{\mu} d\Delta \Delta^\beta \left( \frac{\mu}{2\Delta}-\frac{1}{2} \right)  = \frac{\mu^{\beta+1}}{2\beta(\beta+1)}
\end{equation}
The derivative of the integral vanishes at $\mu = 0$ for $\beta > 0$.  Thus, the system is incompressible.  Comparing
to the data shown in Figures \ref{rhodeltaglass}, we see that there is no numerical evidence for a finite glass compressibility;
the gap density appears to vanish even faster than a power law as $\Delta \rightarrow 0$.  This is consistent with the behavior of 
$\rho(s)$ in equation (\ref{percolationrhos}), since $\Delta$ is expected to scale as $s^{-1}$.  Hence, the numerics imply that the Mott glass 
intervenes between the Mott insulator and the superfluid in this model.

\indent At first glance, the preceding argument may be troubling.  Due to the shrinking of the low $\Delta$ peak
in panel (b) of Figure \ref{rhodeltasf}, the gap density $\rho(\Delta)$ also appears to vanish as $\Delta \rightarrow 0$ in the superfluid phase.  
The caveat is that it is necessary to more carefully evaluate the competing effects of the shrinking and
the propagation.  The low $\Delta$ peak in Figure \ref{rhodeltasf} represents the macroscopic superfluid clusters
that form in the superfluid phase.  These clusters do not appear in proportion to the surface area of the sample, as is the
case for rare-regions clusters; instead, one such macroscopic cluster appears in each sample.  
Therefore, the density of macroscopic clusters will go as $\frac{1}{L^2}$, and this is responsible
for the shrinking of the low $\Delta$ peak.  The propagation of the peak, meanwhile, reflects the fact that the gap closes as $L^{-2}$.
For a fixed choice of $\mu$, the number of bosons that will be added to these macroscopic clusters scales as: 
\begin{equation}
\label{compsfcalculation}
\frac{\mu}{2\Delta}-\frac{1}{2} \propto \mu L^2
\end{equation}
for large $L$.  Then, the \textit{density} of particles introduced to the system is:
\begin{equation}
\label{compsf}
\rho_{\text{ex}} \propto \frac{1}{L^2} \times  \mu L^2 = \mu
\end{equation}
This directly implies that the compressibility (\ref{compress}) is a constant in the thermodynamic limit, so we recover
the expected result that the superfluid phase is compressible.

\section{Conclusion}
\label{sec:conclusions}

\indent While the interplay of disorder and interactions in bosonic systems has attracted considerable interest for nearly 
three decades, the random boson problem remains a fertile source of intriguing physics.  In this paper,
we have investigated a particular model of disordered bosons, the two-dimensional rotor (or Josephson junction) model.
Our strong disorder RG analysis suggests the presence of an unstable finite disorder fixed point of the RG flow,
near which the coupling distributions flow to universal forms.  Furthermore, the strong disorder renormalization procedure
indicates the presence of three phases of the model: the Mott insulating and superfluid phases of the clean model are
separated in the phase diagram by an intervening glassy phase.  The unstable fixed point governs the transition between
the superfluid and this glassy phase, and the transition is driven by a kind of percolation.  
The RG procedure also provides evidence that this glassy phase is, in fact, the incompressible Mott glass.

\indent Our work is a numerical extension into two dimensions of the one-dimensional study by Altman, Kafri, Polkovnikov,
and Refael\cite{altman2004pha}.   The 2D fixed point, however, differs from the 1D fixed point in an important way.  The 1D fixed point occurs at \textit{vanishing} interaction
strength (all charging energies $U_j = 0$).  Thus, it corresponds to a completely classical model and reveals that the superfluid-insulator
transition can be tuned by varying the width of the Josephson coupling distribution at arbitrarily small interaction strength.  
The 2D fixed point is, in contrast, fully quantum.  Indeed, in the critical distributions generated by the strong disorder RG, the charging energy 
distribution is peaked near the RG scale while the Josephson coupling distribution is peaked well below.  

\indent On the other hand, the fixed point that we have identified in this paper is similar to its 1D counterpart in that it does not exhibit
infinite randomness.  Finite disorder fixed points are not optimal settings for strong disorder renormalization analyses, because the procedure
does not become asymptotically exact near criticality and is, in this sense, an uncontrolled approximation.  We have proceeded with
such an analysis anyway.  In doing so, we have found a robust fixed point controlling the superfluid-insulator transition and phases
exhibiting reasonable physical properties.  While this may be surprising given the perils of the method, we have attempted to argue for the
appropriateness of the method, as an \textit{approximation}, through an analysis of the RG steps in light of the forms of the fixed point
distributions $P_{\text{univ}}(U)$ and $P_{\text{univ}}(J)$.  We certainly acknowledge that there are other subtleties due to the lack of infinite randomness; 
for example, the notion of a superfluid cluster is not completely sharp, and consequently, percolation of superfluid clusters can
only be an approximate picture of the transition\cite{huse2011pc}.  Nevertheless, the structure of the fixed point Josephson distribution
(\ref{univdistJ}) suggests that the picture may be a good approximation, and we take this opportunity to remind the reader that
we extensively discuss the reliability of the RG, in light of the properties of the fixed point, in Appendix D.  Moreover, the self-consistency of our numerical results, 
especially the striking universality and robustness of the unstable fixed point, gives us confidence that our strong disorder RG analysis provides useful information
about the system.  With the potential caveats in mind, we therefore turn to exploring connections with other theoretical, numerical, and experimental work.

\indent The Mott glass phase of the two-dimensional model is the straightforward analogue of the phase found in one dimension
by Altman et al.  The charging gap, the energy needed to add or remove a boson from the system, vanishes due to the presence 
of arbitrarily large rare-regions of superfluid ordering.  However, there is no true long range order because these rare-regions grow
subextensively with system size.  If a small chemical potential shift is turned on, it becomes energetically preferable to add 
bosons somewhere in the system, specifically in the largest of the rare-region superfluid clusters.  Nevertheless, these clusters do 
not occur with sufficient number to produce a finite density of bosons, and the glass remains incompressible.   In a Monte Carlo study of 
a model similar to ours, Prokof'ev and Svistunov previously found evidence for a glassy phase in which the compressibility vanishes for this reason\cite{prokof2004superfluid}.  
Moreover, the Mott glass that was identified by Roscilde and Haas in a related spin-one model also relies on the same mechanism\cite{roscilde2007mott}.  
The original proposal of Giamarchi, Le Doussal, and Orignac is, however, fundamentally distinct\cite{giamarchi2001competition}.  
In their Mott glass, the charging gap remains finite, guaranteeing a vanishing of the compressibility; however, gaplessness is achieved 
through the closing of a mobility gap for transport of particles between nearby insulating and superfluid patches.
Sengupta and Haas have argued that particle-hole symmetry, a crucial ingredient in the formation of our Mott glass, is not necessary
for the realization of the phase through this alternative mechanism\cite{sengupta2007quantum}.

\indent In the superfluid phase, true long range order emerges because the largest cluster scales with the size of the lattice.  In
this sense, this cluster is macroscopic.  Despite this, near the transition, the macroscopic cluster may only occupy a small fraction of the total number of 
lattice sites.  Because the clustering procedure can merge sites that are not nearest neighbors in the bare lattice, the fraction of insulating 
sites may actually exceed the standard 2D square lattice percolation threshold.  Even with such a large fraction of insulating sites, a superfluid
phase can still exist because virtual tunneling processes can carry supercurrent through the insulating sites, allowing for macroscopic superfluidity
on the ``depleted" lattice that forms when the insulating sites are removed from the lattice by site decimation.

\indent Nevertheless, the Mott glass to superfluid transition of our model should be contrasted with transitions that arise
when disorder is introduced to a 2D square lattice by bond or site depletion~\cite{roscilde2007mott,fernandes2011complex,bray2006ordering,
wang2006low,wang2010low,sandvik2006quantum}.  Models of the latter type only have the opportunity to form long range ordered phases
when the underlying lattice is percolating.  This percolation is purely classical and exhibits all the critical properties expected of standard
site or bond percolation on a square lattice\cite{dietrich1994introduction}.  The superfluid-insulator transition is, in general, distinct from this transition;
once the underlying lattice percolates, the bosonic model defined on that lattice may still exhibit Mott insulating, glassy, and superfluid phases\cite{roscilde2007mott}.  
In contrast, the only percolation-type process in our model is the one that actually drives the superfluid-insulator transition.  
The critical properties of this transition differ dramatically from those of classical 2D square lattice percolation, because
the transition is not a purely geometric phenomenon.  Instead, there are quantum tunneling processes
overlaid on top of a geometric structure\cite{moore2011pc}.

\indent We have remarked that several experimental groups are currently working on studying disordered bosonic
systems in ultracold atoms~\cite{fallani2007ultracold,schulte2005routes,billy2008direct,white2009strongly}.  Dirty and granular superconductors
provide another experimental context that may be relevant to the physics of this paper.  The question of the applicability
of bosonic pictures to the 2D superconductor-insulator quantum phase transition is long-standing.  Recently, this
problem has been addressed numerically by Bouadim et al., who used quantum Monte Carlo simulations to study a 
fermionic model of the transition and found that bosonic physics emerges at criticality\cite{bouadim2010single}.  This issue 
has also been addressed experimentally by Crane et al.  These authors studied indium oxide thin films and measured a superfluid stiffness
in the insulating state, indicating that Cooper pairs may survive into the insulating region\cite{crane2007survival}.  This leaves open the possibility
that the transition is driven by percolation of superconducting regions, a possibility that has also been explored in experiments on granular superconductors by Frydman et 
al.\ and Sherman et al.\ ~\cite{frydman2002universal,sherman2011superconducting}.  During the preparation of this manuscript, we learned of a recent experiment by Allain et al.\ on tin-decorated graphene.
Intriguingly, this experiment may point to a superconductor-insulator transition that is bosonic in nature, driven by percolation, and characterized by critical exponents
similar to those identified by our work\cite{allain2011gate}.

\indent Motivated by the link between the Mott glass and particle-hole symmetry, Roscilde and Haas have proposed a different class of experimental 
systems in which Mott glass physics may be present: nickel based spin-one antiferromagnets.  
The advantage in these spin systems is that it may be easier to realize particle-hole symmetry in the guise of $Z_2$ symmetry\cite{roscilde2007mott}.
Very recently, Yu et al.\ have followed up on this proposal with an experimental investigation of Bose and Mott glass phases in bromine-doped
dichloro-tetrakis-thiourea-nickel\cite{yu2011bose}.  This system is three-dimensional, so the character of the transition would naturally differ from 
what we have calculated in this paper.

\indent One immediate extension of our study would be to consider adding random filling offsets to 
the disordered rotor model, as Altman et al.\ did in a follow-up to their work on the 1D model\cite{altman2010superfluid}.
The intuition from 1D suggests that such a model would contain a Bose glass phase.  On the other hand, 
very recent Monte Carlo work by Wang et al.\ indicates that the Mott glass survives the substitution of exact
particle-hole symmetry with \textit{statistical} particle-hole symmetry\cite{wang2011incompressible}.  
In one dimension, Altman et al.\ found that the universality class of the transition (but not the identity of the glassy phase)
is independent of the symmetry properties of the random filling offsets, but the situation may differ in $d > 1$; this
remains to be understood.

\indent Another interesting extension may be to study the rotor model defined on random networks.  Suppose we do not begin
with a square lattice but rather with a generalized network of mean connectivity $z = 4$.  At its critical point, would such a 
model flow to the same fixed point as the model defined on a square lattice?  The fact that the strong disorder RG modifies 
the initial lattice structure into a more general network suggests that this may be the case for at least some types
of random networks.  Next, suppose we vary the mean connectivity from $z = 4$.  Is there a range of
connectivities for which random network models access our fixed point?  

\indent Perhaps it would be better to precede such investigations with a better characterization of the fixed point itself.
In one dimension, Altman et al.\ were able to write down master equations for the RG flow, solve them to find fixed point distributions, and then verify
numerically that these distributions are stable\cite{altman2004pha}.  In two dimensions, a direct analytical approach is more difficult, 
and it remains to be seen whether such an approach is tractable.  
Our work provides suggestive numerical evidence regarding the forms of the universal distributions
that characterize the critical point of the disordered rotor model (\ref{rot}).   A Lyapunov analysis of these distributions, in which the RG
is used as a tool to identify irrelevant directions in the space of possible distributions, could be a useful step in clarifying the critical 
forms still further.  Then, analytically \textit{verifying} these forms as attractor solutions of the RG flow may be an easier task than analytically 
identifying them would have been in the absence of any numerical guidance.

\indent Recent work by Vosk and Altman suggests yet another direction for connecting the results of the RG to experiment.  
These authors have derived the 1D version of the rotor model as an effective description of continuum bosons.  In doing so, they have
established a connection between the strong disorder RG treatment of Altman, Kafri, Polkovnikov, and Refael 
and cold atom experiments on rubidium-87.  Remarkably, the distributions that Vosk and Altman derive are of the
same form as the fixed point distributions found by the strong disorder RG~\cite{altman2004pha,vosk2011superfluid}.  
If such a treatment can be extended to the 2D case, that would be very valuable, both as a clarification of the critical 
behavior and as an indication of the relevance of the results to current experiments.

\acknowledgments

It is our pleasure to acknowledge useful discussions with Anton Akhmerov, Ehud Altman, Ravin Bhatt, Fran\c{c}ois Cr\'{e}pin, Thierry Giamarchi,
Walter Hofstetter, David Huse, Joel Moore, Olexei Motrunich, Anatoli Polkovnikov, Nikolai Prokof'ev, Srinivas Raghu, Boris Svistunov, Nandini Trivedi, and Ari Turner.  
We particularly thank Bryan Clark for pointing out how to estimate errors on the critical exponents that we obtain from scaling collapse
and an anonymous referee for helpful comments on sources of error to incorporate into our estimates.
SI would like to acknowledge Robert Dondero for helpful suggestions for resolving problems with the RG code.
SI would also like to express gratitude to the organizers of the 2010 Boulder School for Condensed Matter and Materials Physics, 
the 2011 Carg\`{e}se School on Disordered Systems, and the 2011 Princeton Summer School for Condensed
Matter Physics.  SI and DP both thank the 2010 International Centre for Theoretical Sciences School in Mysore.
This material is based upon work supported, in part, by the National Science Foundation under Grant No. 1066293 
and the hospitality of the Aspen Center for Physics.  DP acknowledges financial support by the Lee A. DuBridge Fellowship, 
and GR acknowledges support from the Packard Foundation.

\appendix

\section{Sum Rule vs.\ Maximum Rule}

\indent In this appendix, we present a short argument for why it may be preferable to use the \textit{sum rule} (see equations (\ref{specsitedec}) and
(\ref{speclinkdec})) instead of the \textit{maximum rule} (see equation (\ref{maxrule})).  In dimensions greater than one, it should be easier
to form ordered (e.\ g.\ superfluid) phases.  Indeed, the transition for the clean rotor model occurs when $\frac{J}{U}$ is 
substantially smaller than one\cite{teichmann2009bose}.
Suppose we begin with the clean model at its critical point and then disorder it by increasing some Josephson couplings
and decreasing some charging energies.  Suppose further that we do this such that that the greatest Josephson coupling $J_{\text{max}}$
is still less than the weakest charging energy $U_{\text{min}}$.  Then, using the maximum rule, the strong disorder renormalization
procedure will predict no cluster formation at all.  In other words, it will predict the ground state to be a Mott insulator, and 
this result seems inconsistent with the location of the clean transition.  As mentioned previously, several authors have argued that,
if $\frac{J_{\text{max}}}{U_{\text{min}}}$ exceeds the ratio $\frac{J}{U}$ at the clean transition, then we should expect glassy behavior, because
there can be rare-regions in which the system looks locally superfluid\cite{pollet2009absence}.  
With the sum rule, the procedure has a mechanism to circumvent this inconsistency.  The Josephson coupling scale can 
actually grow through decimation and compete with the charging energy scale.  Thus, there can be cluster formation, and the 
procedure can find ground states that are glassy or superfluid, even when all Josephson couplings of the bare model are less 
than the minimum bare charging energy.  This indeed occurs in the numerics, as we have noted while presenting the numerical
data above.

\indent The notion of the Josephson coupling scale increasing through the RG may be a source of concern to some readers.
The increase actually corresponds to the generation of multiple effective couplings between two sites through different
paths in the lattice.  This still happens when the maximum rule is used, but it is hidden through the discarding of certain paths.
If the coupling through each path is treated as an independent Josephson coupling, then the Josephson coupling scale
does decrease as the renormalization proceeds.  However, when it is time to determine the next decimation step, 
it is necessary to consider all of the couplings between any two sites.  For this reason, it makes sense to sum all the couplings
into one effective coupling between the sites.

\section{Measuring Physical Properties in the RG}

\indent Here, we work out two examples of how estimates of physical properties can be extracted from the RG procedure.

\subsection{Particle Number Variance}

\indent First, consider the quantity:
\begin{equation}
\label{variance}
\sum_j \langle \hat{n}^2_j \rangle
\end{equation}  
This particle number variance gives the mean squared number fluctuation away from the large filling, summed
over all sites in the lattice.  When normalized by $\frac{1}{L^2}$, we find numerically that this quantity stays
constant as the system size is increased for all choices of distributions and parameters.  As such, this
quantity is completely uninteresting for discriminating between phases of the model, but we do calculate it for
comparison to exact diagonalization in Appendix D. 

\indent The calculation of the particle number variance (\ref{variance}) is most straightforward 
when clusters do not form, so let us first consider the case where some site $X$ is not clustered 
with any other site during the renormalization.
At some stage in the procedure, the site is decimated away.   The number fluctuation
on site $X$ is locked to zero at leading order with corrections incorporated in second 
order perturbation theory.  An approximation to the ground state
value of $ \langle \hat{n}^2_X \rangle$ can be obtained from the perturbative expansion of the state.  In particular:
\begin{equation}
\label{Nsqapprox}
\langle \hat{n}^2_X \rangle \approx \frac{1}{2}\sum_k \frac{J^2_{Xk}}{(U_X+U_k)^2}
\end{equation}

\indent When clusters do form, the calculation is trickier, but it can be performed by carefully keeping
track of how the operator that we are targeting is written in terms of the cluster and relative number
operators introduced in link decimation.  To illustrate this, suppose we are trying to measure the
operator:
\begin{equation}
\label{Nsqcalculation1}
\langle a_j \hat{n}^2_j + a_k \hat{n}^2_k \rangle = a_j \langle \hat{n}^2_j \rangle + a_k \langle \hat{n}^2_k \rangle
\end{equation}
The factors $a_j$ and $a_k$ are just numbers.  In (\ref{variance}), all these factors are one, but
we will motivate the inclusion of more general $a$ factors here shortly.  If sites $j$ and $k$ are merged into a cluster, then
we switch to the operators $n_C$ and $n_R$ to find:
\begin{eqnarray}
a_j \hat{n}^2_j + a_k \hat{n}^2_k & = & \frac{a_j U^2_k + a_k U^2_j}{(U_j+U_k)^2} \hat{n}^2_C + \frac{a_j U_k - a_k U_j}{U_j+U_k}\hat{n}_R\hat{n}_C \nonumber \\
                                                           & \quad &   + (a_j+a_k)\hat{n}^2_R
\label{Nsqcalculation2}                                                           
\end{eqnarray}
During link decimation, the relative coordinate is specified, so the expectation value of the final term 
can be found immediately from the harmonic approximation:
\begin{equation}
\label{Nsqcalculation3}
(a_j+a_k)\langle \hat{n}^2_R \rangle \approx (a_j+a_k)\frac{1}{2\gamma^{\frac{1}{2}}}
\end{equation}
Furthermore, the harmonic theory also predicts that the expectation value of the term linear in $\hat{n}_R$ 
will vanish.  The calculation of the term proportional to $\hat{n}^2_C$ must be deferred to later
in the renormalization procedure.  Thus, we keep the operator $\hat{n}^2_C$ in the portion of 
the sum (\ref{variance}) that remains to be evaluated, where it appears just like the $\hat{n}^2$
for a bare site, but multiplied by a renormalized $a_C$ coefficient.  This was the motivation for 
including the $a$ factors; though the bare values of these factors are all equal,  different values can
be generated through cluster formation.  If the cluster is merged with another cluster
in a future link decimation, we repeat the procedure above.  When the cluster is finally decimated
in a site decimation, the cluster's contributions to the sum (\ref{variance}) are calculated through 
equation (\ref{Nsqapprox}) and then multiplied by the appropriate $a$ factor.

\indent The procedure outlined above can run into a difficulty that we can anticipate by
thinking about the two-site problem.  Suppose two sites, labelled $1$ and $2$, are connected
by a Josephson coupling $J_{12}$.  Furthermore, suppose $U_1 > U_2$, but both charging energies
are greater than $J_{12}$.  Then, we would decimate site $1$ first and obtain an estimate for the site's
contribution to the particle number variance (\ref{variance}) through equation (\ref{Nsqapprox}).
Next, we would decimate site $2$ and find that it does not contribute at all to (\ref{variance}) because
there are no remaining links.  However, the contribution of site $2$ should, in fact, be equal to that of
site $1$, so our estimate is off by a factor of two.  We can verify this by adopting cluster 
and relative coordinates (\ref{ldclustreloperators}) and then calculating (\ref{variance}) by 
doing perturbation theory for the relative coordinate.  To partially resolve this difficulty, we can keep
track of which sites are unclustered by the RG process.  At the end, we can return to the original lattice
and calculate the contributions of these unclustered sites to (\ref{variance}) using the bare couplings.
For sites that \textit{are} clustered by the RG, we reason that the main contribution to (\ref{variance})
comes from internal fluctuations of the cluster, so this correction may not be so important.

\indent It is possible to raise another objection to our calculation of (\ref{variance}).
The perturbation theory that leads to the result in (\ref{Nsqapprox}) incorporates the charging energies
on sites neighboring site $X$.  However, the perturbation theory leading to the RG rule (\ref{sitedec}) 
does not.  How do we resolve this contradiction?  When we calculate effective Josephson couplings in 
the RG, what we want to calculate is the effective rate of tunneling, once a boson has left one neighboring site
and before it arrives at another, through the link-site-link system.  For this purpose, it is appropriate to treat the sites neighboring site $X$
as fictitious charging energy-free islands.  On the other hand, when calculating observable quantities like (\ref{variance}),
it is important to account for the fact that the ability to move a particle or hole from site $X$ to a neighboring site 
also depends on how hard it is to charge the neighboring site.

\subsection{Susceptibility}

\indent In our studies of physical properties of the phases of the disordered rotor model, we considered
the susceptibility $\chi$.  We defined this quantity in equation (\ref{susc}) as the linear response of 
the system to a uniform ordering field (\ref{Hchipert}) coupling to $\cos(\hat{\phi_j})$.  

\indent To calculate $\chi$, we consider how to calculate the expectation value:
\begin{equation}
\label{chiaux}
\sum_j b_j \langle \cos(\phi_j)\rangle
\end{equation}
in the presence of an infinitesimal ordering field $h$.
As with the $a$-factors in the calculation of the particle number variance, all the bare $b_j = 1$.
When clusters form, the effective $b$-factor for the cluster can differ from unity.

\indent If a bare site is decimated, then perturbation theory in $h$ gives:
\begin{equation}
\label{baresitechi}
\langle \cos(\phi_X) \rangle = \frac{h}{U_X}
\end{equation}
Since $b_X = 1$ for a bare site, (\ref{baresitechi}) is the contribution of site $X$ to the quantity (\ref{chiaux}).

\indent When a link decimation joins two sites into a cluster, the corresponding terms in the sum (\ref{chiaux}) merge as:
\begin{eqnarray}
\label{chimerger}
b_j \cos(\hat{\phi}_j) +  b_k \cos(\hat{\phi}_k) & = & b_j \cos(\hat{\phi}_C+\mu_j \hat{\phi}_R) \nonumber \\
                  & \quad & +b_k \cos(\hat{\phi}_C-\mu_k \hat{\phi}_R) \nonumber \\
                  & \approx & b_j \cos(\hat{\phi}_C)\langle\cos(\mu_j \hat{\phi}_R)\rangle \\
                  & \quad & -b_j \sin(\hat{\phi}_C)\langle\sin(\mu_j \hat{\phi}_R)\rangle \nonumber \\
                  & \quad & + b_k\cos(\hat{\phi}_C)\langle\cos(\mu_k \hat{\phi}_R)\rangle \nonumber \\
                  & \quad & + b_k \sin(\hat{\phi}_C)\langle\sin(\mu_k \hat{\phi}_R)\rangle  \nonumber \\
                  & = & (b_j c_{DW,j} + b_k c_{DW,k}) \cos(\hat{\phi}_C) \nonumber
\end{eqnarray}
In this calculation, $\mu_j$ and $\mu_k$ are the ratios introduced in equation (\ref{lddwmu}), and $c_{DW,j}$ and $c_{DW,k}$
are precisely the Debye-Waller factors given in equation (\ref{dwj}).  We can read off the renormalized $b$-factor for the cluster
from the calculation above, and the resulting expression is given in equation (\ref{bfactorcluster}).

\indent Next, it is important to note that the ordering field terms in the Hamiltonian (\ref{Hchipert}) transform in the same way.
In other words, the term coupling to $\cos(\hat{\phi}_C)$ in the Hamiltonian should appear multiplied by $b_C$ after a merger.
Physically, this corresponds to the fact that, when the cluster phase rotates, $s_C$ bare phases rotate semi-coherently.  Complete
coherence is lost due to quantum fluctuations, which are accounted for by Debye-Waller factors.  The factor $b_C$ can be thought of as the
effective number of rotors that are coherently ordering with the field.  The energetic cost of $\phi_C$ straying from the direction
of the ordering field is therefore amplified by this factor.  When the cluster is finally decimated, the perturbation is amplified by this amount.
Furthermore, $\cos(\hat{\phi}_C)$ appears in the sum (\ref{chiaux}) multiplied by $b_C$, so the total contribution of the cluster to the sum is:
\begin{equation}
\label{clusterchi2}
b_C \langle \cos(\phi_C) \rangle = \frac{h b^2_C}{U_C}
\end{equation}
From equations (\ref{baresitechi}) and (\ref{clusterchi2}), the calculation of the linear response to an infinitesimal $h$ (i.e.\ the susceptibility
$\chi$) follows immediately. 

\section{The $P_i(U)$ Gaussian, $P_i(J)$ Power Law Data Set}

\indent In the work reported above, we have explored the superfluid-insulator transition of the
disordered rotor model using several different species of disorder (see equations (\ref{Pgaussian})-(\ref{Pbimodal})). 
In doing so, we have exposed universal features of the critical behavior.  After providing numerical evidence of this universality
however, we focus on data for one particular choice of initial distributions.  In this
appendix, we describe this choice of distributions in greater detail.

\indent The choice of distributions in question first appears in Figure \ref{flowJplUg}.  The
initial distribution of charging energies $P_i(U)$ is taken to be a Gaussian with center at $U_0$
and width $\sigma_U = 2$.  Hence, the form of the distribution is:
\begin{equation}
\label{appendixUdist}
P_i(U) \propto \exp{\left[-\frac{(U-U_0)^2}{8}\right]}
\end{equation}
Recall that the Gaussian distribution is truncated at $3\sigma_U = 6$, so the distribution only has
weight in the interval $U \in (U_0-6,U_0+6)$.  We leave $U_0$ unspecified for the moment, because
it is the parameter that we use to tune through the transition.

\indent The initial Josephson coupling distribution $P_i(J)$ is a power law of the form $J^{-1.6}$.
We choose the cutoffs so that the distribution has weight for $J \in (0.5,100)$.  This is a very wide
power law distribution, so the corresponding flows begin well above the unstable finite disorder
fixed point in the numerical flow diagrams.  Explicitly:
\begin{equation}
\label{appendixUdist}
P_i(J) \approx 0.413 J^{-1.6}
\end{equation}

\begin{figure}
\centering
\includegraphics[width=8cm]{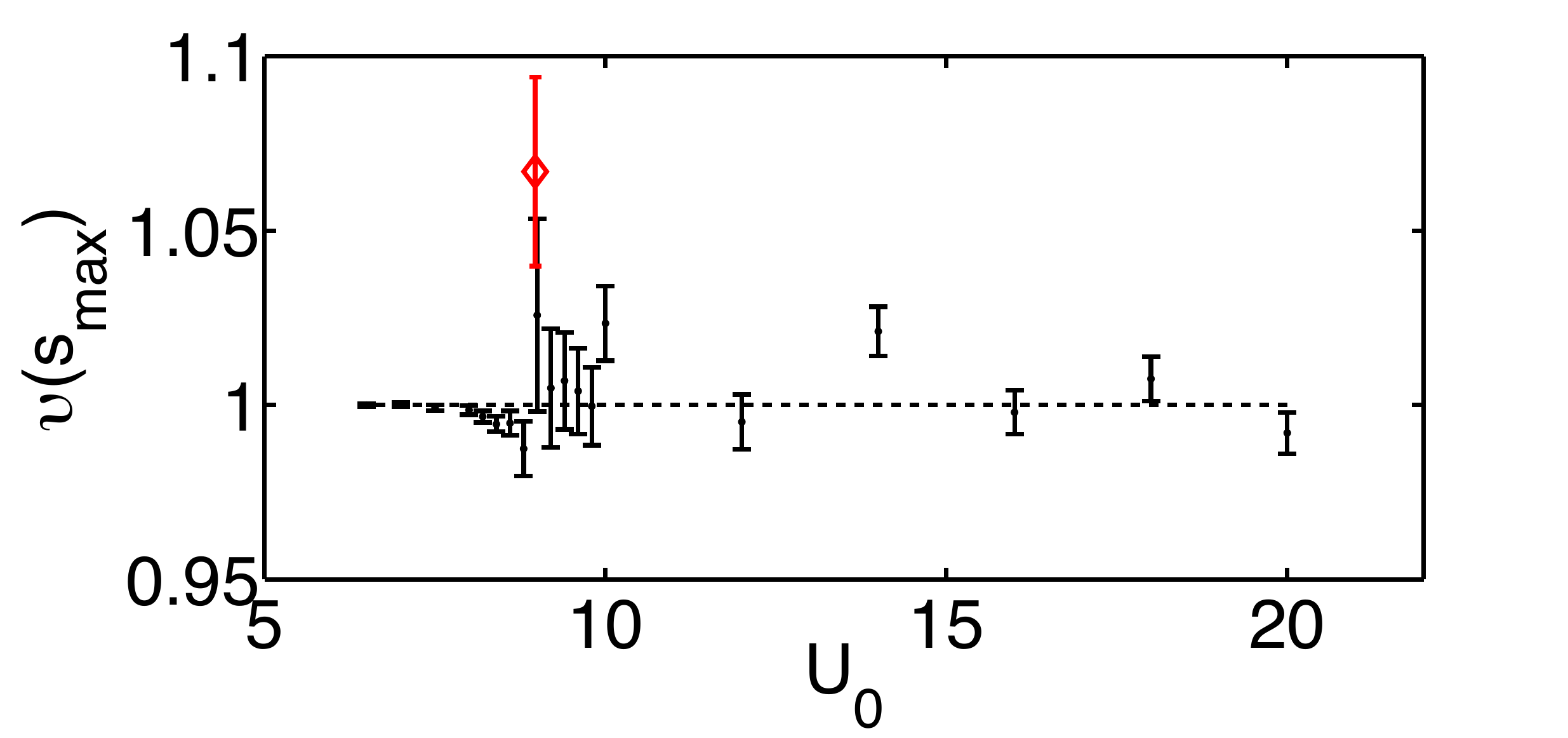}
\caption{A test of the convergence of $s_{\text{max}}$ in the thresholding parameter $\alpha$.  The variable $\upsilon$ is the ratio of the
estimate of $s_{\text{max}}$ for $\alpha_h$, the less conservative value of the thresholding parameter, to $\alpha_\ell$, the more conservative
value.  We plot $\upsilon$ against the tuning parameter $U_0$.  The closest data point to criticality ($U_0 = 8.97$) is indicated with a diamond.  
Note that $\upsilon > 1$ typically indicates convergence since less conservative thresholding (higher $\alpha$) corresponds to throwing away more bonds and,
therefore, biases the system away from cluster formation.  Smaller values of $\alpha_h$ and $\alpha_\ell$ are used in the vicinity of the transition.  
See the text of Appendix C for details.}
\label{convergencesmax}
\end{figure}

\indent For this choice of distributions, we have acquired data for $U_0 = 400$, $20$, $18$, $16$, $14$, $12$, 
$10$, $9.8$, $9.6$, $9.4$, $9.2$, $8.8$, $8.6$, $8.4$, $8.2$, $8$, $7.5$, $7$, and $6.5$.  To more finely target
the transition, we have probed the interval $9.1 \geq U_0 \geq 8.9$ in increments of $0.01$.
We have always acquired data for $L = 25$, $50$, $75$, $100$, $150$, $200$, and $300$.  
In all cases, we have pooled data for $10^3$ disorder samples.  
The peak in the data for $s_2$ vs.\ $U_0$ gives the following estimate of the critical point:
\begin{equation}
\label{appendixUdist}
U_{0,c} \approx 8.97 \pm 0.02
\end{equation}
Close to criticality, it is better to use a lower value of the thresholding parameter $\alpha$.
For several values of $U_0$ such that $10 \geq U_0 \geq 8$, we have run the RG with $\alpha_h = 10^{-5}$ and $\alpha_\ell = 5 \times 10^{-6}$ to test
for convergence.  Further away from criticality, we have instead used $\alpha_h = 5 \times 10^{-5}$ and $\alpha_\ell = 2.5 \times 10^{-5}$.
Figure \ref{convergencesmax} shows a test of the convergence of the maximum cluster size $s_{\text{max}}$ in the thresholding parameter $\alpha$.
We plot:
\begin{equation}
\label{convergenceratio}
\upsilon = \frac{s_{\text{max}}(\alpha_h)}{s_{\text{max}}(\alpha_\ell)}
\end{equation}
vs.\ the tuning parameter $U_0$.  The ratio (\ref{convergenceratio}) is essentially always within two error bars of unity.  We take
this as an indication that physical properties have converged.

\section{Further Comments on the Use of the Strong Disorder RG in a Finite Disorder Context}

\indent This appendix is devoted to exploring, in further detail, the validity of the RG procedure.
We first expand upon the argument, introduced earlier in the paper, for the reliability of the RG near
criticality.  Then, we move away from criticality and assess the performance of the RG in the various phases of 
the disordered rotor model (\ref{rot}).  Next, we focus on each of the RG steps, consider circumstances
in which they may fail to capture important physics, and formulate tests to ensure that the RG is trustworthy
in these situations.  Finally, we present a comparison of the RG to exact diagonalization of small systems.

\subsection{Review of the Argument for the RG at Criticality}

\indent Our confidence in the RG procedure near criticality rests on the form of the critical Josephson
coupling distribution, reported in equations (\ref{univdistJ}) and (\ref{varphiestimate}). Infinite randomness
develops when $P(J) \propto J^{-1}$, and our numerical evidence suggests that the critical distribution of 
the disordered rotor model decays even more strongly\cite{fisher1999phase}.  Nevertheless, the critical 
distribution does not exhibit infinite randomness, because as seen in Figure \ref{universalJscaled}, it is cut
off from below.  Recall that the lower cutoff of the ``Josephson coupling distribution" is set by our 
choice to retain, \textit{in statistics}, only the dominant $2\tilde{N}$ Josephson couplings, where $\tilde{N}$ is
the number of sites remaining in the effective lattice.  Then, the appropriate way to interpret the distributions
in Figure \ref{universalJ} is the following: penetrating any given site in the effective lattice, there
are likely to be on the order of four Josephson couplings drawn from the depicted distribution.  There may be other links
penetrating the site, but these will be even weaker.  We can estimate the typical strength of the four dominant links by 
comparing the median of the critical Josephson coupling distribution to the maximum.  For the closest approach to
criticality with the initial distributions described in Appendix C, we find that the ratio $\frac{J_{\text{typ}}}{\Omega}$
is approximately $0.11 \pm 0.01$ near the fixed point.  Hence, the typical link is quantitatively weak compared to $\frac{J}{U} \approx 0.345$ at the
clean transition\cite{teichmann2009bose}.

\indent The considerations above form a strong argument for the validity of the site decimation RG step.  Here,
we seek out the dominant effective charging energy in the effective lattice, and treat the links penetrating the
site in perturbation theory.  This perturbation theory is likely a very good approximation, because
the Josephson couplings penetrating the site in question are usually extremely weak.

\indent Now, we turn to the link decimation step, in which we seek out the largest Josephson coupling in the lattice and merge the corresponding sites into a cluster. 
Although, the Josephson coupling being decimated is the largest energy scale in the system, there is a high probability that all the other links penetrating the 
two sites being joined will be very small. However, the critical distribution of charging energies is \textit{not} peaked at low $U$.  
The structure of the distributions plotted in Figure \ref{universalU} suggests that it is quite likely that one or both sites being merged will have a 
charging energy of the same order as the RG scale, thus violating the strong disorder hypothesis.   Our treatment of the quantum fluctuations of the 
relative phase of the cluster is based on the harmonic approximation (\ref{lhogs}).  Is this approximation appropriate when the charging energies 
of the two-site problem are comparable in magnitude to the Josephson coupling?  Alternatively, do the quantum fluctuations grow so large that the 
clustering becomes meaningless?  We address this question as follows: using the fact that the remaining links are weak, we isolate the two site 
problem and solve it exactly, treating the remaining links via second order perturbation theory. Comparing the results of the RG with the exact solution, 
we find that, even in this worst case scenario, the RG produces reasonably accurate effective couplings. The evidence for this claim is given in section 3 below.

\subsection{Reliability of the RG in the Phases}

\begin{figure}[h]
\begin{minipage}[b]{0.4cm}
       {\bf (a)}
       
       \vspace{3.3cm}
\end{minipage}
\begin{minipage}[t]{7.9cm}
       \includegraphics[width=7.8cm]{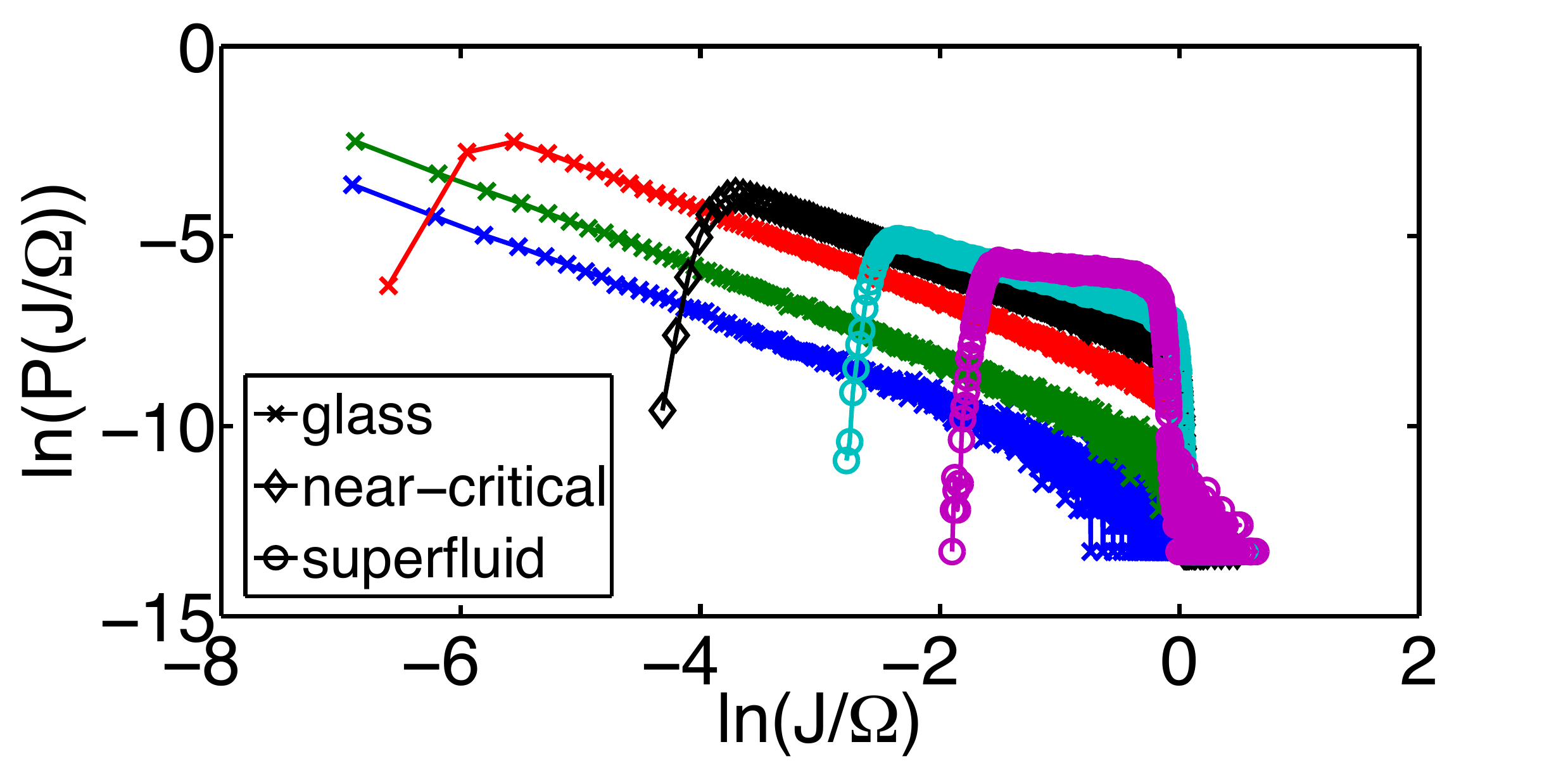}
\end{minipage}\\
\begin{minipage}[b]{0.4cm}
       {\bf (b)}

       \vspace{3.3cm}
\end{minipage}
\begin{minipage}[t]{7.9cm}
       \includegraphics[width=7.8cm]{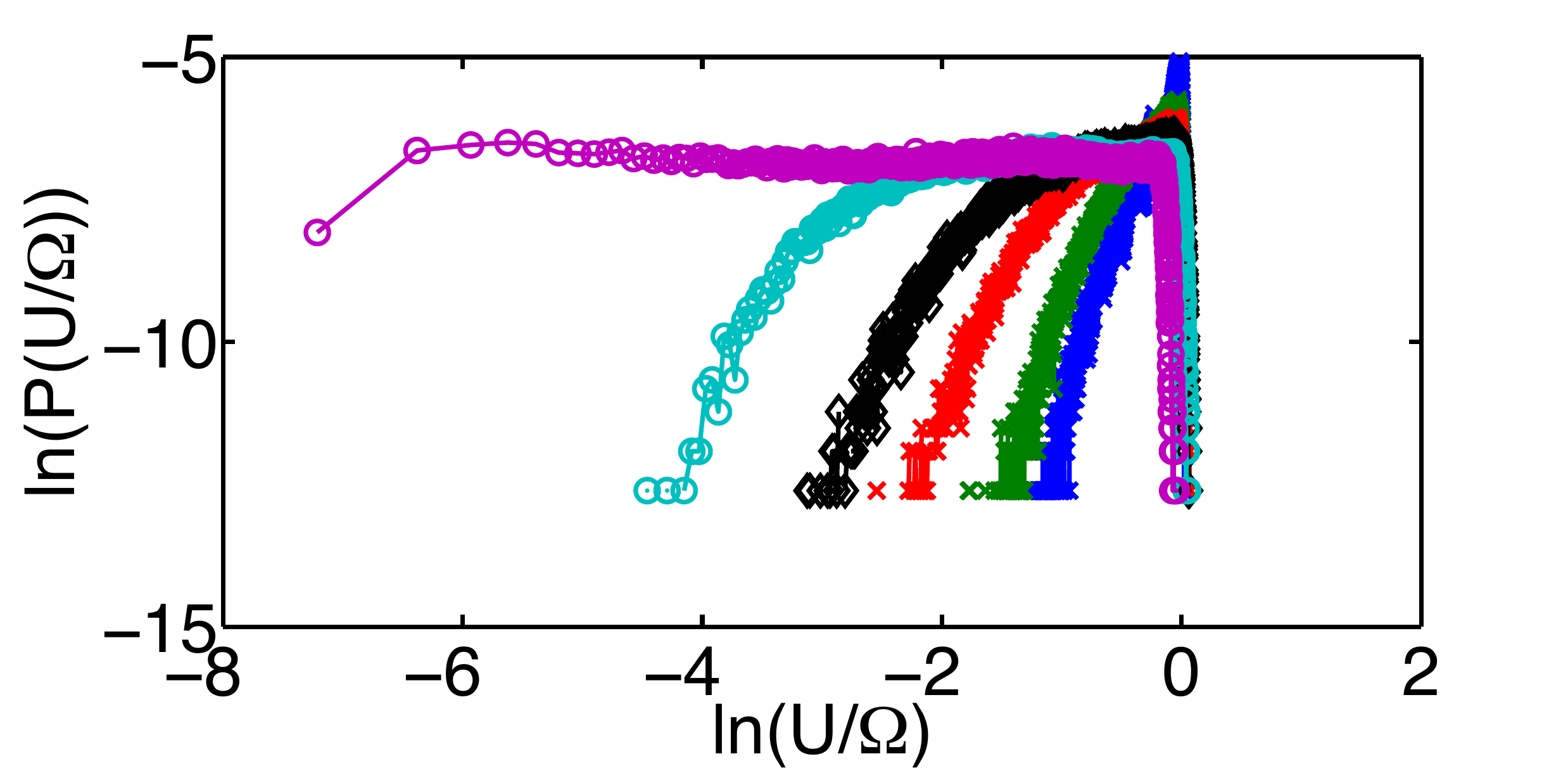}
\end{minipage}\\

\caption{In panel (a), a sweep of renormalized $J$ distribution through the glass and into the superfluid.   The initial distributions are those 
described in Appendix C: $P_i(U)$ is Gaussian and $P_i(J) \propto J^{-1.6}$.  All data is taken for $L = 300$ lattices, and the 
renormalized distribution is computed when $300$ sites remain in the effective lattice.  The values of $U_0$ shown are $18$, $12$,
$9.6$, $8.97$ (near-critical), $8.4$, and $7.5$.  Panel (b) shows the corresponding sweep of the renormalized $U$ distribution at the same
stage of the RG.}
\label{distsweepphases}
\end{figure}

\indent How reliable is the RG when we move away from criticality and into the phases of the disordered rotor model?
To gain some insight into this question, we can consult Figure \ref{distsweepphases}, which expands upon Figures \ref{universalJscaled}
and \ref{universalUscaled} by plotting renormalized $J$ and $U$ distributions away from criticality.  

\indent Proceeding into the glassy regime, the arguments presented above for the validity of the RG near criticality generally become
stronger.  The primary reason for this is that the renormalized $J$ distributions become progressively broader than at criticality.  In the flow
diagrams (Figures \ref{flowg}-\ref{flowJplUb}), this is reflected in the apparent divergence of $\frac{\Delta J}{\bar{J}}$.  Consequently, the
assumption of isolating local degrees of freedom becomes better as we get deeper into the glass.  One complication is that the renormalized $U$
distribution becomes more strongly peaked near the RG scale.  This may pose trouble for the link decimation step and makes it especially important
to consider the reliability of this step when a strong $J_{jk}$ couples two sites with charging energies $U_j \approx U_k \approx J_{jk}$.  As mentioned previously,
we will study this worst-case scenario in part 3 of this appendix and find that the RG still works reasonably well.

\indent Now, we turn to the superfluid phase.  As we proceed away from criticality, the renormalized charging energy distribution becomes flatter and broader.
The broadening of this distribution implies that the likelihood of encountering a strong Josephson coupling that connects sites with comparably strong charging
energies decreases as we get deeper into the superfluid phase.  However, the $J$ distribution \textit{also} seems to become flat deep in the
superfluid, and this is problematic.  For example, during link decimation, it may not be a reasonable approximation to isolate the two-site problem
centered on the strongest Josephson coupling.  As we proceed into the superfluid, it is necessary to be more dubious of the RG results;
nearer to criticality however, the arguments used at the critical point are probably approximately valid.

\indent The plots in Figure \ref{distsweepphases} attempt to elucidate systematics in the behavior of the renormalized distributions
in the insulating and superfluid regimes, but we should mention that this figure should be interpreted with some care.  For the choice
of distributions in Appendix C, flows terminating in the insulating or superfluid regions of the flow diagram nevertheless propagate
in the vicinity of the unstable fixed point along the way.  This can be seen, for instance, in Figure \ref{flowJplUg}.  For certain choices of
initial distributions, there can be flows towards the insulating or superfluid regimes which never propagate anywhere near the unstable
fixed point.  Consequently, the RG never has an opportunity to wash away the details of the initial distributions and allow the universal
properties of the fixed point to emerge.  Hence, the renormalized distributions generated along such flow trajectories are unlikely to
exhibit the clean systematic properties seen in Figure \ref{distsweepphases}.

\subsection{Analysis of Potential Problems with the RG}

\indent Next, we address some potential difficulties with the arguments presented above for the reliability
of the renormalization procedure in the critical region.  These difficulties are rooted in the lack of strong randomness
in the critical charging energy distribution. 

\begin{figure}
\centering
\includegraphics[scale = 0.60]{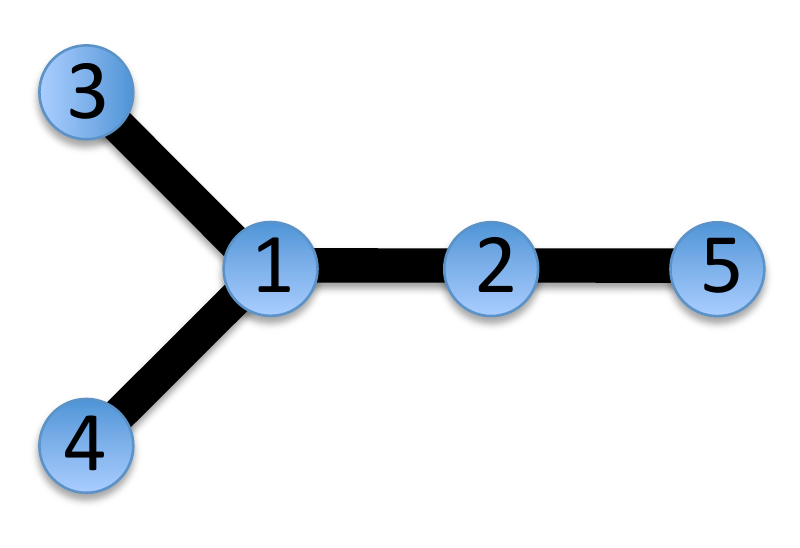}
\caption{The graph structure for the tests reported in Figures \ref{HarmonicBreakdown} and \ref{Resonance}.  The links $J_{13}$, $J_{14}$, and $J_{25}$ are 
assumed to be perturbatively weak.  The charging energies $U_1$ and $U_2$ and the Josephson coupling $J_{12}$ are varied to explore potentially 
troublesome scenarios.}
\label{testgraphstructure}
\end{figure}

\begin{figure}
\centering
\includegraphics[width=8cm]{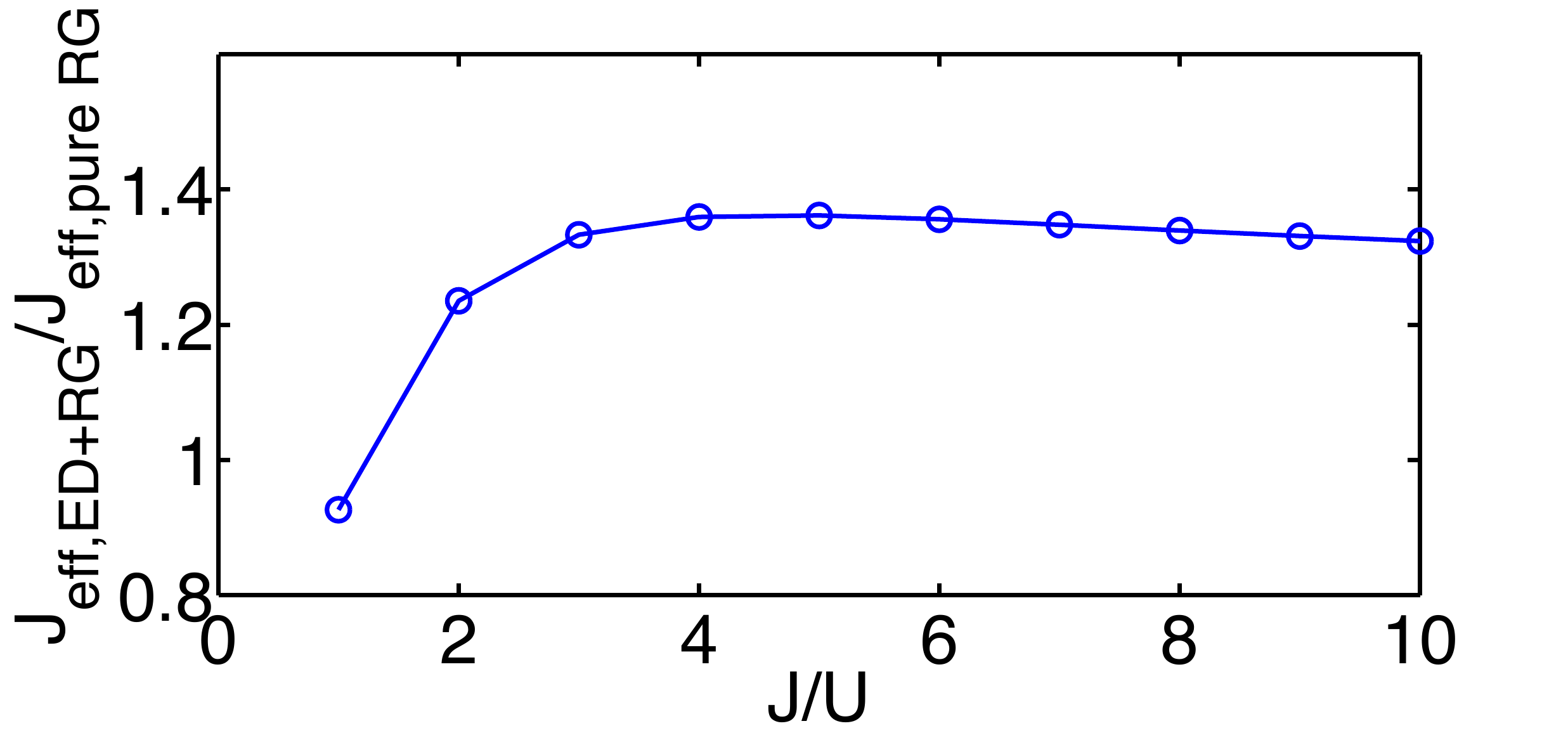}
\caption{In this test, $J = J_{12}$ is assumed to be the strongest coupling in the system, but $U = U_1 = U_2$ may be of the same order.  An effective coupling between
sites $3$ and $5$ is calculated using two methods.  One is the usual RG scheme used in this paper.  Another is a hybrid exact diagonalization and RG scheme:
The two-site problem of sites $1$ and $2$ is exactly diagonalized.  Then, the resulting cluster is decimated away, and perturbation theory is used to calculate
an effective coupling between sites $3$ and $5$.  The two candidate values for the effective coupling $J_{35}$ are compared in the plot, as a function of $\frac{J}{U}$.}
\label{HarmonicBreakdown}
\end{figure}

\indent Consider the lattice geometry shown in Figure \ref{testgraphstructure}.  Suppose that all links except the
Josephson coupling between sites $1$ and $2$ are perturbatively weak.  Now, suppose further that $J_{12}$,
$U_1$, and $U_2$ are comparable in magnitude, but $J_{12}$ is the largest of the three.  We would like to formulate
a test of whether the RG appropriately handles this situation.  Within the RG, a link
decimation would merge sites $1$ and $2$ into a cluster.  All links penetrating sites $1$ and $2$ would be modified
by their corresponding Debye-Waller factors (\ref{dwj}) $c_{\text{DW},1}$ and $c_{\text{DW},2}$.  Then, because all the remaining
links are assumed to be very weak, the cluster of sites $1$ and $2$ will be decimated, producing an effective coupling between
sites $3$ and $5$:
\begin{equation}
\label{test1Jeff35}
\tilde{J}_{35,RG} = c_{\text{DW},1}c_{\text{DW},2} \frac{J_{13}J_{25} (U_1+U_2)}{U_1U_2}
\end{equation}
Another approach to calculating this effective coupling would be the 
following: take the two-site problem of sites $1$ and $2$ and exactly diagonalize it.  Then, to leading order, sites $1$ and $2$ should
be locked into the two-site ground state, with perturbative corrections coming from the Josephson couplings $J_{13}$, $J_{14}$, and $J_{25}$.
This perturbation theory leads to an effective coupling through the sites $1$ and $2$.  This alternative procedure is perhaps
more appropriate, because it does not presuppose the harmonic approximation.  In Figure \ref{HarmonicBreakdown}, we assess how
much of an error we make by adopting the harmonic approximation.  Holding $J$ fixed, we sweep $U = U_1 = U_2$ through $J$, comparing
the RG with the alternative method outlined above.  We see that the usual RG performs reasonably well, 
implying that the harmonic approximation is safe.

\begin{figure}
\centering
\includegraphics[width=8cm]{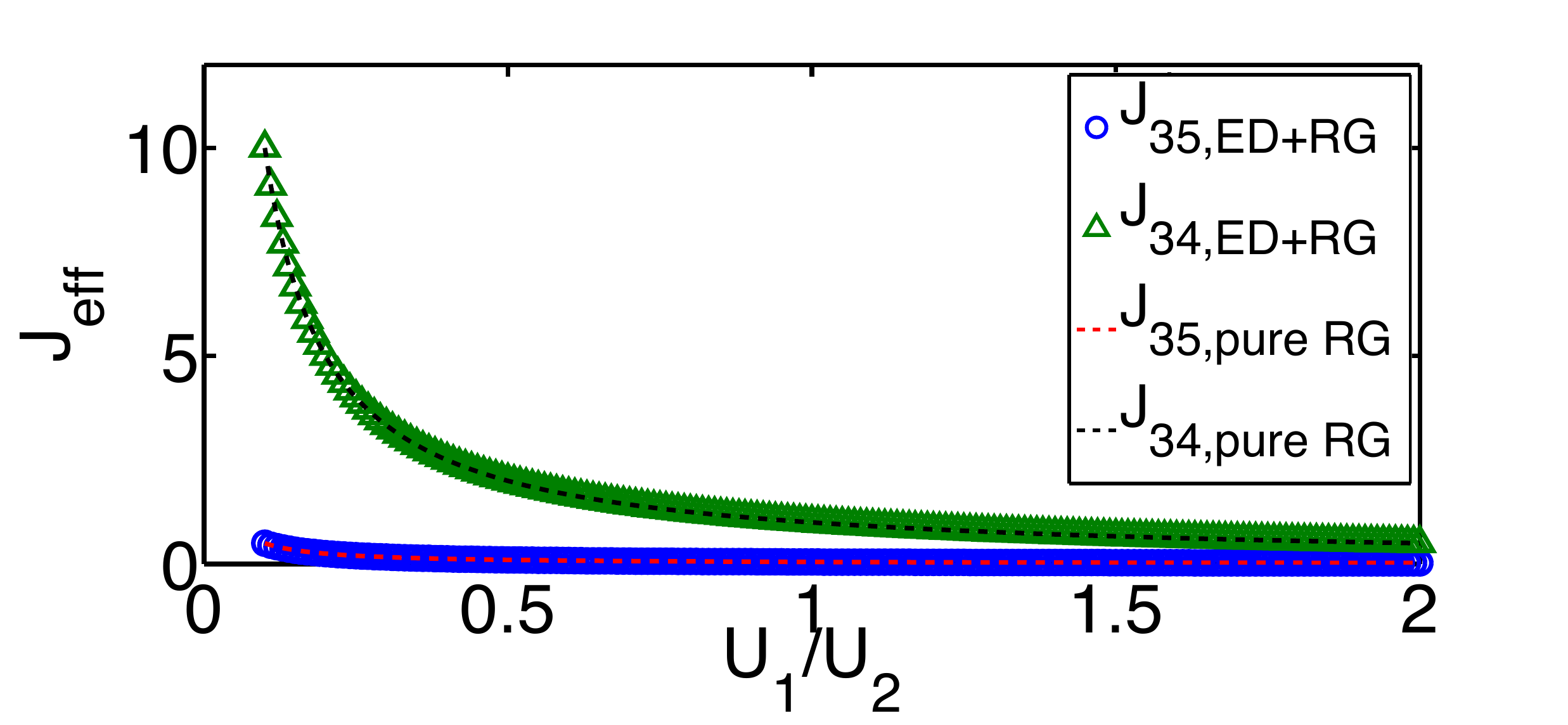}
\caption{In this test, $J_{12} = 0.05$ and $U_2 = 1$.  Hence, $J_{12}$ is relatively quite weak.  However, we vary $U_1$ such that it passes through a regime where $|U_1-U_2| < J_{12}$,
where there may be a danger of resonance effects.
We calculate the effective couplings $J_{34}$ and $J_{35}$ using two schemes. One is the usual RG scheme used in this paper.  Another is a hybrid exact diagonalization and RG scheme:
the two-site problem of sites $1$ and $2$ is exactly diagonalized.  Then, the resulting cluster is decimated away, and perturbation theory is used to calculate
an effective coupling between sites $3$ and $4$ and between sites $3$ and $5$.  The effective couplings predicted by the two methods are compared in the plot, as a function
of $\frac{U_1}{U_2}$.  No resonance effects are observed.}
\label{Resonance}
\end{figure}

\indent Finally, we consider another potentially dangerous scenario.  We return to the lattice shown in Figure \ref{testgraphstructure}.
Now, we assume that $J_{12}$ is greater than all other Josephson couplings, but it too is much weaker than the charging energies $U_1$
and $U_2$.  In particular, $\frac{J_{12}}{U_2} = 0.05$.  Then, we sweep $U_1$ such that it passes through a regime where $|U_1-U_2| < J_{12}$.  
The danger here is that the RG may ignore resonance effects associated with this region.  
Within the usual RG, sites $1$ and $2$ will be decimated in turn to give:
\begin{equation}
\label{test2Jeff35}
\tilde{J}_{35,\text{RG}} \approx \frac{J_{13}J_{12}J_{25} }{U_1U_2}
\end{equation}
\begin{equation}
\label{test2Jeff35}
\tilde{J}_{34,\text{RG}} \approx \frac{J_{13}J_{14}}{U_1}
\end{equation}
where we ignore subleading corrections coming from potential applications of the sum rule, depending on the order of decimation of sites $1$ and $2$.
To consider potential resonance effects, we can also implement the same hybrid exact diagonalization and RG procedure that we used above.
In Figure \ref{Resonance}, we compare the two methods and find excellent agreement. 

\subsection{Strong Disorder RG vs. Exact Diagonalization}

\indent  As a final test of the RG procedure, we now compare the RG to exact diagonalization of small systems.   
We truncate the possible number fluctuation on each site to $n_j = -1, 0, 1$,
interpret these three values as possible z-axis spin projections of a spin-one object, and in doing so, 
arrive at a ``spin-one" model:
\begin{equation}
\label{spinoneXY}
\hat{H} = - \sum_{\langle jk \rangle} \frac{J_{jk}}{2} (\hat{S}_j^+ \hat{S}_k^-+ \hat{S}_j^-\hat{S}_k^+)                           
                           + \sum_{j} U_j \hat{S}^2_{zj}
\end{equation}
The Hilbert space of this spin-one model grows with the size of the lattice as $3^{L^2}$. The particle number
conservation of the rotor model manifests here as total spin conservation along the z-axis.  This means that
we can partition the Hilbert space into different total spin sectors and diagonalize the sectors separately.  For
most ground state expectation values, we just need to diagonalize the total spin zero sector, and to calculate a charging gap, 
the only additional diagonalization needed is for the total spin one sector.  Despite these simplifications, computational limitations restrict us
to studying $3 \times 3$ lattices using CLAPACK.  Testing the RG against exact diagonalization
cannot directly tell us about the reliability of the RG at criticality, because exact diagonalization
is limited to very small system sizes.  However, a comparison with exact diagonalization \textit{can} tell us
how well the RG captures information about small patches of a larger lattice, and this information
is potentially quite valuable for building confidence in the RG.

\begin{figure}
\centering
\includegraphics[width=8cm]{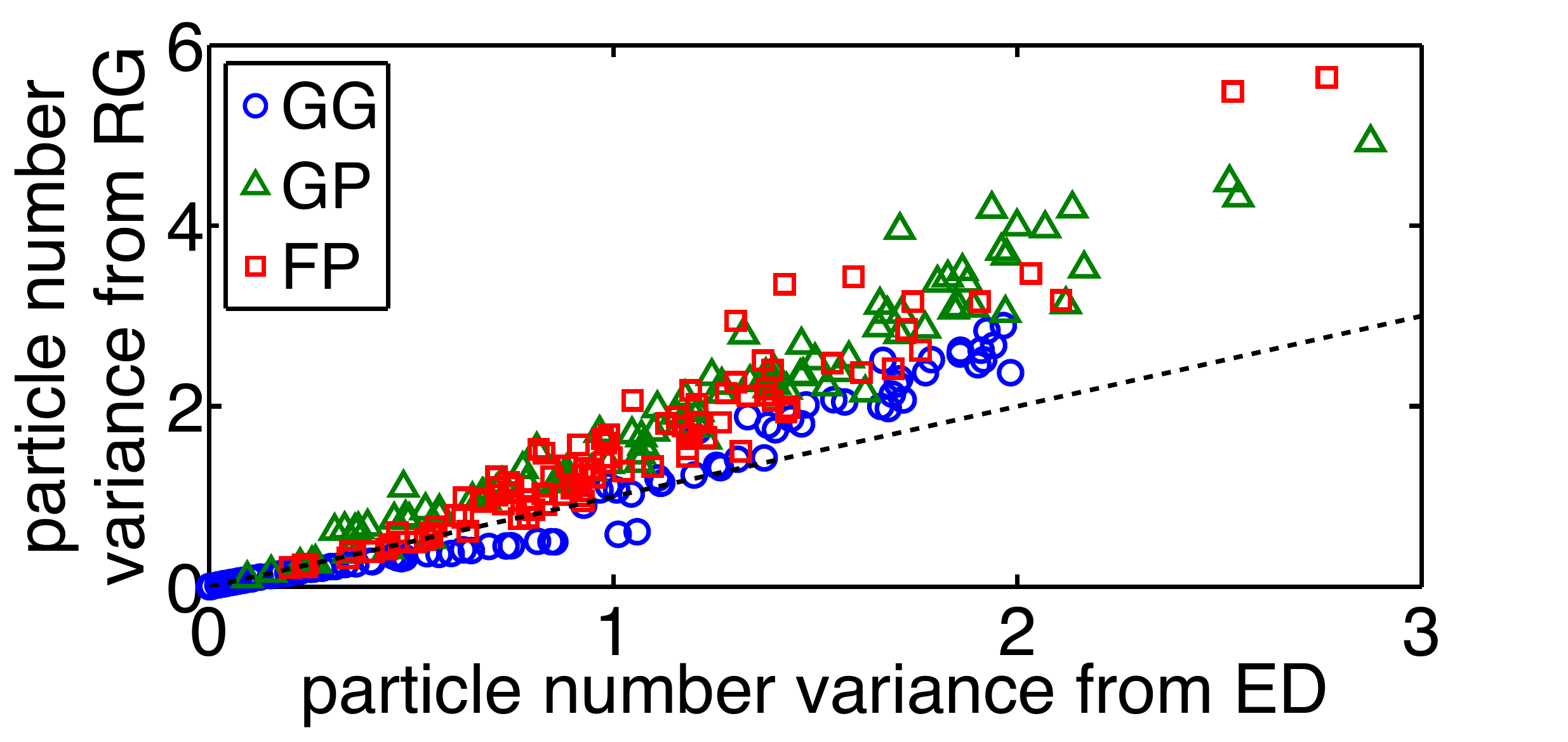}
\caption{A sample-by-sample comparison of the particle number variance predictions from exact diagonalization and from the renormalization
procedure.  The disordered couplings are drawn from three different choices of the distributions, with 100 samples per distribution type.  
See the text of Appendix D for details on the distribution choices GG, GP, and FP.  Also pictured is the coincidence line $y=x$, along which the points would ideally fall for full quantitative agreement.}
\label{scatterNSq}
\end{figure}

\begin{figure}
\centering
\includegraphics[width=8cm]{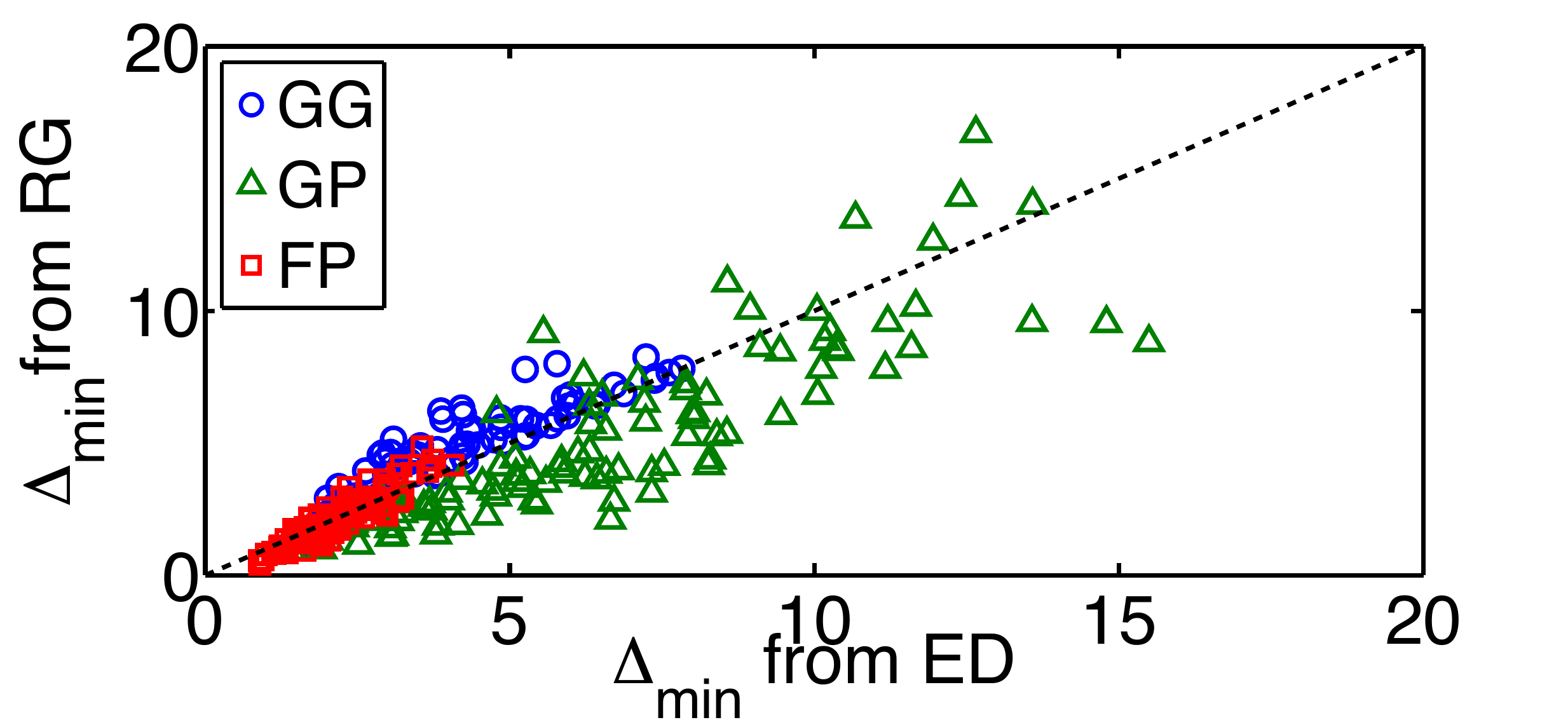}
\caption{Same as Figure \ref{scatterNSq}, but the quantity being compared is the charging gap $\Delta_{\text{min}}$.}
\label{scatterGap}
\end{figure}

\indent Another complication arises precisely due to the Hilbert space truncation: the spin-one model may 
not always approximate the full rotor model well.  This is especially true when there is cluster formation, because 
then the rotor model has more of an opportunity to access particle number fluctuations that exceed $1$ in magnitude.  
In other words, the strong disorder renormalization group and the exact diagonalization of the spin-one model constitute 
\textit{different} approximations to the behavior of our random boson model.  We cannot expect the two
approximations to show full quantitative agreement, but we proceed with the comparison, despite its
limitations, with the hope of at least seeing qualitative correspondence between the two methods.

\indent Our general approach to comparing the RG with exact diagonalization will be to measure physical quantities, on 
individual $3 \times 3$ samples, using both methods and assess if there is a correlation between the predictions.
The first quantity that we compare is the particle number variance (\ref{variance}).  The interested reader may  
consult Appendix B to see how this quantity is calculated during the renormalization procedure.  We also compare the charging gap
$\Delta_{\text{min}}$, the minimum energy needed to add a particle or hole to the system.  
This quantity is typically estimated during site decimation.  The logic behind site decimation is that the charging energy for 
some site $X$ is greater than all other scales in the problem, so the site can be disconnected from the rest of the lattice to leading order.  
Then, the charging energy gives an estimate for the local charging gap on that site.  During the RG, we find many such charging gaps
from the various site decimations.  The minimum among all of these gives an estimate for the charging gap for the whole system.  
This minimum is always given by the charging energy of the final remaining site, so 
an estimate of $\Delta_{\text{min}}$ can be simply obtained by renormalizing down to a single site problem and measuring the charging
energy of the remaining site. For the purposes of comparison to exact diagonalization however, we find that we obtain better quantitative agreement between
the two methods if we renormalize down to an effective two-site problem and then perform exact diagonalization on that system.
Exactly diagonalizing the $n_{\text{tot}} = 0$ and $n_{\text{tot}} = 1$ sectors of this two-site problem then yields a charging gap for the system.
In this exact diagonalization, we need not truncate the on-site number fluctuation to $n_X = -1, 0, 1$.  Instead, in the numerics, we 
typically truncate to $n_X = -100 \ldots 100$. 

\indent In Figures \ref{scatterNSq} and \ref{scatterGap}, we present comparisons for three different data sets.
In the first data set, we use take $P_i(U)$ and $P_i(J)$ to be Gaussian.  We fix $U_0 = 10$ and $\sigma_U = 3$.
Then, we randomly sample $J_0$ in the interval $(0,5)$ and $\sigma_J$ in the interval $(0,\frac{J_0}{3})$.  The
aim is to approximate some of the environments that the RG encounters in runs such as those reported in Figure \ref{flowg}.
The second data set uses the distributions described in Appendix C: $P_i(U)$ is Gaussian and $P_i(J) \propto J^{-1.6}$.  
We randomly sample $U_0 \in (6.5,20)$.  Here, the motivation is to look at the types of environments that the RG encounters 
when it approaches the unstable fixed point from above.   In the final data set, we try to mimic $3 \times 3$ patches that 
the RG might encounter near criticality.  To this end, the initial $J$ distribution is fixed to a power law $P(J) \propto J^{-1.16}$ 
(see equation ($\ref{varphiestimate}$)) and the cutoffs are chosen so that the ratio of $J_{\text{min}}$ to
$J_{\text{max}}$ is approximately that observed in panel (b) of Figure \ref{universalUscaled}.  The distribution $P_i(U)$
is flat with $U_{\text{max}} = J_{\text{max}}$ and with $U_{\text{min}}$ randomly sampled such that the ratio of $U_{\text{min}}$ to $U_{\text{max}}$ lies
in $(e^{-2},e^{-1})$.  Figures \ref{scatterNSq} and \ref{scatterGap} identify these three data sets with the labels GG, GP, and FP respectively.

\indent Both figures show that the predictions of the RG are clearly correlated with the predictions from exact diagonalization,
although the level of quantitative correspondence varies.  For the particle number variance, quantitative agreement is lost at 
higher values of the variance, essentially corresponding to cases in which there is clustering.  One potential source of error 
could be the Hilbert space truncation of the exact diagonalization, although the structure of the harmonic ground state (\ref{lhogs}) makes
it unlikely that this could account for the entire discrepancy.  Nevertheless, these comparisons suggest that the RG is
retaining useful information about the system.

\bibliography{draftrefs}

\begin{thebibliography}{47}
\expandafter\ifx\csname natexlab\endcsname\relax\def\natexlab#1{#1}\fi
\expandafter\ifx\csname bibnamefont\endcsname\relax
  \def\bibnamefont#1{#1}\fi
\expandafter\ifx\csname bibfnamefont\endcsname\relax
  \def\bibfnamefont#1{#1}\fi
\expandafter\ifx\csname citenamefont\endcsname\relax
  \def\citenamefont#1{#1}\fi
\expandafter\ifx\csname url\endcsname\relax
  \def\url#1{\texttt{#1}}\fi
\expandafter\ifx\csname urlprefix\endcsname\relax\def\urlprefix{URL }\fi
\providecommand{\bibinfo}[2]{#2}
\providecommand{\eprint}[2][]{\url{#2}}

\bibitem[{\citenamefont{Crooker et~al.}(1983)\citenamefont{Crooker, Hebral,
  Smith, Takano, and Reppy}}]{crooker1983superfluidity}
\bibinfo{author}{\bibfnamefont{B.}~\bibnamefont{Crooker}},
  \bibinfo{author}{\bibfnamefont{B.}~\bibnamefont{Hebral}},
  \bibinfo{author}{\bibfnamefont{E.}~\bibnamefont{Smith}},
  \bibinfo{author}{\bibfnamefont{Y.}~\bibnamefont{Takano}}, \bibnamefont{and}
  \bibinfo{author}{\bibfnamefont{J.}~\bibnamefont{Reppy}},
  \bibinfo{journal}{Physical Review Letters.} \textbf{\bibinfo{volume}{51}},
  \bibinfo{pages}{666} (\bibinfo{year}{1983}).

\bibitem[{\citenamefont{Reppy}(1984)}]{reppy19844He}
\bibinfo{author}{\bibfnamefont{J.}~\bibnamefont{Reppy}},
  \bibinfo{journal}{Physica B+C.} \textbf{\bibinfo{volume}{126}},
  \bibinfo{pages}{335} (\bibinfo{year}{1984}).

\bibitem[{\citenamefont{Weichman}(2008)}]{weichman2008dirty}
\bibinfo{author}{\bibfnamefont{P.}~\bibnamefont{Weichman}},
  \bibinfo{journal}{Arxiv Preprint arXiv:0810.3263.}  (\bibinfo{year}{2008}).

\bibitem[{\citenamefont{Fisher et~al.}(1989)\citenamefont{Fisher, Weichman,
  Grinstein, and Fisher}}]{fisher1989boson}
\bibinfo{author}{\bibfnamefont{M.}~\bibnamefont{Fisher}},
  \bibinfo{author}{\bibfnamefont{P.}~\bibnamefont{Weichman}},
  \bibinfo{author}{\bibfnamefont{G.}~\bibnamefont{Grinstein}},
  \bibnamefont{and} \bibinfo{author}{\bibfnamefont{D.}~\bibnamefont{Fisher}},
  \bibinfo{journal}{Physical Review B.} \textbf{\bibinfo{volume}{40}},
  \bibinfo{pages}{546} (\bibinfo{year}{1989}).

\bibitem[{\citenamefont{Giamarchi and Schulz}(1988)}]{giamarchi1988anderson}
\bibinfo{author}{\bibfnamefont{T.}~\bibnamefont{Giamarchi}} \bibnamefont{and}
  \bibinfo{author}{\bibfnamefont{H.}~\bibnamefont{Schulz}},
  \bibinfo{journal}{Physical Review B.} \textbf{\bibinfo{volume}{37}},
  \bibinfo{pages}{325} (\bibinfo{year}{1988}).

\bibitem[{\citenamefont{Pollet et~al.}(2009)\citenamefont{Pollet, ProkofÕev,
  Svistunov, and Troyer}}]{pollet2009absence}
\bibinfo{author}{\bibfnamefont{L.}~\bibnamefont{Pollet}},
  \bibinfo{author}{\bibfnamefont{N.}~\bibnamefont{ProkofÕev}},
  \bibinfo{author}{\bibfnamefont{B.}~\bibnamefont{Svistunov}},
  \bibnamefont{and} \bibinfo{author}{\bibfnamefont{M.}~\bibnamefont{Troyer}},
  \bibinfo{journal}{Physical Review Letters.} \textbf{\bibinfo{volume}{103}},
  \bibinfo{pages}{140402} (\bibinfo{year}{2009}).

\bibitem[{\citenamefont{Greiner et~al.}(2002)\citenamefont{Greiner, Mandel,
  Esslinger, H\"{a}nsch, and Bloch}}]{greiner2002quantum}
\bibinfo{author}{\bibfnamefont{M.}~\bibnamefont{Greiner}},
  \bibinfo{author}{\bibfnamefont{O.}~\bibnamefont{Mandel}},
  \bibinfo{author}{\bibfnamefont{T.}~\bibnamefont{Esslinger}},
  \bibinfo{author}{\bibfnamefont{T.}~\bibnamefont{H\"{a}nsch}},
  \bibnamefont{and} \bibinfo{author}{\bibfnamefont{I.}~\bibnamefont{Bloch}},
  \bibinfo{journal}{Nature.} \textbf{\bibinfo{volume}{415}},
  \bibinfo{pages}{39} (\bibinfo{year}{2002}).

\bibitem[{\citenamefont{Schulte et~al.}(2005)\citenamefont{Schulte,
  Drenkelforth, Kruse, Ertmer, Arlt, Sacha, Zakrzewski, and
  Lewenstein}}]{schulte2005routes}
\bibinfo{author}{\bibfnamefont{T.}~\bibnamefont{Schulte}},
  \bibinfo{author}{\bibfnamefont{S.}~\bibnamefont{Drenkelforth}},
  \bibinfo{author}{\bibfnamefont{J.}~\bibnamefont{Kruse}},
  \bibinfo{author}{\bibfnamefont{W.}~\bibnamefont{Ertmer}},
  \bibinfo{author}{\bibfnamefont{J.}~\bibnamefont{Arlt}},
  \bibinfo{author}{\bibfnamefont{K.}~\bibnamefont{Sacha}},
  \bibinfo{author}{\bibfnamefont{J.}~\bibnamefont{Zakrzewski}},
  \bibnamefont{and}
  \bibinfo{author}{\bibfnamefont{M.}~\bibnamefont{Lewenstein}},
  \bibinfo{journal}{Physical Review Letters.} \textbf{\bibinfo{volume}{95}},
  \bibinfo{pages}{170411} (\bibinfo{year}{2005}).

\bibitem[{\citenamefont{Fallani et~al.}(2007)\citenamefont{Fallani, Lye,
  Guarrera, Fort, and Inguscio}}]{fallani2007ultracold}
\bibinfo{author}{\bibfnamefont{L.}~\bibnamefont{Fallani}},
  \bibinfo{author}{\bibfnamefont{J.}~\bibnamefont{Lye}},
  \bibinfo{author}{\bibfnamefont{V.}~\bibnamefont{Guarrera}},
  \bibinfo{author}{\bibfnamefont{C.}~\bibnamefont{Fort}}, \bibnamefont{and}
  \bibinfo{author}{\bibfnamefont{M.}~\bibnamefont{Inguscio}},
  \bibinfo{journal}{Physical Review Letters.} \textbf{\bibinfo{volume}{98}},
  \bibinfo{pages}{130404} (\bibinfo{year}{2007}).

\bibitem[{\citenamefont{Billy et~al.}(2008)\citenamefont{Billy, Josse, Zuo,
  Bernard, Hambrecht, Lugan, Cl{\'e}ment, Sanchez-Palencia, Bouyer, and
  Aspect}}]{billy2008direct}
\bibinfo{author}{\bibfnamefont{J.}~\bibnamefont{Billy}},
  \bibinfo{author}{\bibfnamefont{V.}~\bibnamefont{Josse}},
  \bibinfo{author}{\bibfnamefont{Z.}~\bibnamefont{Zuo}},
  \bibinfo{author}{\bibfnamefont{A.}~\bibnamefont{Bernard}},
  \bibinfo{author}{\bibfnamefont{B.}~\bibnamefont{Hambrecht}},
  \bibinfo{author}{\bibfnamefont{P.}~\bibnamefont{Lugan}},
  \bibinfo{author}{\bibfnamefont{D.}~\bibnamefont{Cl{\'e}ment}},
  \bibinfo{author}{\bibfnamefont{L.}~\bibnamefont{Sanchez-Palencia}},
  \bibinfo{author}{\bibfnamefont{P.}~\bibnamefont{Bouyer}}, \bibnamefont{and}
  \bibinfo{author}{\bibfnamefont{A.}~\bibnamefont{Aspect}},
  \bibinfo{journal}{Nature} \textbf{\bibinfo{volume}{453}},
  \bibinfo{pages}{891} (\bibinfo{year}{2008}).

\bibitem[{\citenamefont{White et~al.}(2009)\citenamefont{White, Pasienski,
  McKay, Zhou, Ceperley, and DeMarco}}]{white2009strongly}
\bibinfo{author}{\bibfnamefont{M.}~\bibnamefont{White}},
  \bibinfo{author}{\bibfnamefont{M.}~\bibnamefont{Pasienski}},
  \bibinfo{author}{\bibfnamefont{D.}~\bibnamefont{McKay}},
  \bibinfo{author}{\bibfnamefont{S.}~\bibnamefont{Zhou}},
  \bibinfo{author}{\bibfnamefont{D.}~\bibnamefont{Ceperley}}, \bibnamefont{and}
  \bibinfo{author}{\bibfnamefont{B.}~\bibnamefont{DeMarco}},
  \bibinfo{journal}{Physical Review Letters.} \textbf{\bibinfo{volume}{102}},
  \bibinfo{pages}{55301} (\bibinfo{year}{2009}).

\bibitem[{\citenamefont{Giamarchi et~al.}(2001)\citenamefont{Giamarchi,
  Le~Doussal, and Orignac}}]{giamarchi2001competition}
\bibinfo{author}{\bibfnamefont{T.}~\bibnamefont{Giamarchi}},
  \bibinfo{author}{\bibfnamefont{P.}~\bibnamefont{Le~Doussal}},
  \bibnamefont{and} \bibinfo{author}{\bibfnamefont{E.}~\bibnamefont{Orignac}},
  \bibinfo{journal}{Physical Review B.} \textbf{\bibinfo{volume}{64}},
  \bibinfo{pages}{245119} (\bibinfo{year}{2001}).

\bibitem[{\citenamefont{Altman et~al.}(2004)\citenamefont{Altman, Kafri,
  Polkovnikov, and Refael}}]{altman2004pha}
\bibinfo{author}{\bibfnamefont{E.}~\bibnamefont{Altman}},
  \bibinfo{author}{\bibfnamefont{Y.}~\bibnamefont{Kafri}},
  \bibinfo{author}{\bibfnamefont{A.}~\bibnamefont{Polkovnikov}},
  \bibnamefont{and} \bibinfo{author}{\bibfnamefont{G.}~\bibnamefont{Refael}},
  \bibinfo{journal}{Physical Review Letters.} \textbf{\bibinfo{volume}{93}},
  \bibinfo{pages}{150402} (\bibinfo{year}{2004}).

\bibitem[{\citenamefont{Altman et~al.}(2010)\citenamefont{Altman, Kafri,
  Polkovnikov, and Refael}}]{altman2010superfluid}
\bibinfo{author}{\bibfnamefont{E.}~\bibnamefont{Altman}},
  \bibinfo{author}{\bibfnamefont{Y.}~\bibnamefont{Kafri}},
  \bibinfo{author}{\bibfnamefont{A.}~\bibnamefont{Polkovnikov}},
  \bibnamefont{and} \bibinfo{author}{\bibfnamefont{G.}~\bibnamefont{Refael}},
  \bibinfo{journal}{Physical Review B.} \textbf{\bibinfo{volume}{81}},
  \bibinfo{pages}{174528} (\bibinfo{year}{2010}).

\bibitem[{\citenamefont{Weichman and
  Mukhopadhyay}(2008)}]{weichman2008particle}
\bibinfo{author}{\bibfnamefont{P.}~\bibnamefont{Weichman}} \bibnamefont{and}
  \bibinfo{author}{\bibfnamefont{R.}~\bibnamefont{Mukhopadhyay}},
  \bibinfo{journal}{Physical Review B.} \textbf{\bibinfo{volume}{77}},
  \bibinfo{pages}{214516} (\bibinfo{year}{2008}).

\bibitem[{\citenamefont{Dasgupta and Ma}(1980)}]{dasgupta1980low}
\bibinfo{author}{\bibfnamefont{C.}~\bibnamefont{Dasgupta}} \bibnamefont{and}
  \bibinfo{author}{\bibfnamefont{S.}~\bibnamefont{Ma}},
  \bibinfo{journal}{Physical Review B.} \textbf{\bibinfo{volume}{22}},
  \bibinfo{pages}{1305} (\bibinfo{year}{1980}).

\bibitem[{\citenamefont{Bhatt and Lee}(1982)}]{bhatt1982scaling}
\bibinfo{author}{\bibfnamefont{R.}~\bibnamefont{Bhatt}} \bibnamefont{and}
  \bibinfo{author}{\bibfnamefont{P.}~\bibnamefont{Lee}},
  \bibinfo{journal}{Physical Review Letters.} \textbf{\bibinfo{volume}{48}},
  \bibinfo{pages}{344} (\bibinfo{year}{1982}).

\bibitem[{\citenamefont{Fisher}(1994)}]{fisher1994random}
\bibinfo{author}{\bibfnamefont{D.}~\bibnamefont{Fisher}},
  \bibinfo{journal}{Physical Review B.} \textbf{\bibinfo{volume}{50}},
  \bibinfo{pages}{3799} (\bibinfo{year}{1994}).

\bibitem[{\citenamefont{Fisher}(1999)}]{fisher1999phase}
\bibinfo{author}{\bibfnamefont{D.}~\bibnamefont{Fisher}},
  \bibinfo{journal}{Physica A: Statistical Mechanics and its Applications.}
  \textbf{\bibinfo{volume}{263}}, \bibinfo{pages}{222} (\bibinfo{year}{1999}).

\bibitem[{\citenamefont{Vosk and Altman}(2011)}]{vosk2011superfluid}
\bibinfo{author}{\bibfnamefont{R.}~\bibnamefont{Vosk}} \bibnamefont{and}
  \bibinfo{author}{\bibfnamefont{E.}~\bibnamefont{Altman}},
  \bibinfo{journal}{Arxiv Preprint arXiv:1104.2063.}  (\bibinfo{year}{2011}).

\bibitem[{\citenamefont{Bruder et~al.}(2005)\citenamefont{Bruder, Fazio, and
  Sch\"{o}n}}]{bruder2005bose}
\bibinfo{author}{\bibfnamefont{C.}~\bibnamefont{Bruder}},
  \bibinfo{author}{\bibfnamefont{R.}~\bibnamefont{Fazio}}, \bibnamefont{and}
  \bibinfo{author}{\bibfnamefont{G.}~\bibnamefont{Sch\"{o}n}},
  \bibinfo{journal}{Annalen der Physik.} \textbf{\bibinfo{volume}{14}},
  \bibinfo{pages}{566} (\bibinfo{year}{2005}).

\bibitem[{\citenamefont{Teichmann et~al.}(2009)\citenamefont{Teichmann,
  Hinrichs, Holthaus, and Eckardt}}]{teichmann2009bose}
\bibinfo{author}{\bibfnamefont{N.}~\bibnamefont{Teichmann}},
  \bibinfo{author}{\bibfnamefont{D.}~\bibnamefont{Hinrichs}},
  \bibinfo{author}{\bibfnamefont{M.}~\bibnamefont{Holthaus}}, \bibnamefont{and}
  \bibinfo{author}{\bibfnamefont{A.}~\bibnamefont{Eckardt}},
  \bibinfo{journal}{Physical Review B.} \textbf{\bibinfo{volume}{79}},
  \bibinfo{pages}{100503} (\bibinfo{year}{2009}).

\bibitem[{\citenamefont{Gottlob and Hasenbusch}(1993)}]{gottlob1993critical}
\bibinfo{author}{\bibfnamefont{A.}~\bibnamefont{Gottlob}} \bibnamefont{and}
  \bibinfo{author}{\bibfnamefont{M.}~\bibnamefont{Hasenbusch}},
  \bibinfo{journal}{Physica A: Statistical Mechanics and its Applications}
  \textbf{\bibinfo{volume}{201}}, \bibinfo{pages}{593} (\bibinfo{year}{1993}).

\bibitem[{\citenamefont{Harris}(1974)}]{harris1974effect}
\bibinfo{author}{\bibfnamefont{A.}~\bibnamefont{Harris}},
  \bibinfo{journal}{Journal of Physics C: Solid State Physics.}
  \textbf{\bibinfo{volume}{7}}, \bibinfo{pages}{1671} (\bibinfo{year}{1974}).

\bibitem[{\citenamefont{Vojta}(2010)}]{vojta2010quantum}
\bibinfo{author}{\bibfnamefont{T.}~\bibnamefont{Vojta}},
  \bibinfo{journal}{Journal of Low Temperature Physics.}
  \textbf{\bibinfo{volume}{161}}, \bibinfo{pages}{299} (\bibinfo{year}{2010}).

\bibitem[{\citenamefont{Kov{\'a}cs and
  Igl{\'o}i}(2010)}]{kovacs2010renormalization}
\bibinfo{author}{\bibfnamefont{I.}~\bibnamefont{Kov{\'a}cs}} \bibnamefont{and}
  \bibinfo{author}{\bibfnamefont{F.}~\bibnamefont{Igl{\'o}i}},
  \bibinfo{journal}{Physical Review B.} \textbf{\bibinfo{volume}{82}},
  \bibinfo{pages}{054437} (\bibinfo{year}{2010}).

\bibitem[{\citenamefont{Motrunich et~al.}(2000)\citenamefont{Motrunich, Mau,
  Huse, and Fisher}}]{motrunich2000infinite}
\bibinfo{author}{\bibfnamefont{O.}~\bibnamefont{Motrunich}},
  \bibinfo{author}{\bibfnamefont{S.}~\bibnamefont{Mau}},
  \bibinfo{author}{\bibfnamefont{D.}~\bibnamefont{Huse}}, \bibnamefont{and}
  \bibinfo{author}{\bibfnamefont{D.}~\bibnamefont{Fisher}},
  \bibinfo{journal}{Physical Review B.} \textbf{\bibinfo{volume}{61}},
  \bibinfo{pages}{1160} (\bibinfo{year}{2000}).

\bibitem[{\citenamefont{Stauffer and Aharony}(1994)}]{dietrich1994introduction}
\bibinfo{author}{\bibfnamefont{D.}~\bibnamefont{Stauffer}} \bibnamefont{and}
  \bibinfo{author}{\bibfnamefont{A.}~\bibnamefont{Aharony}},
  \emph{\bibinfo{title}{Introduction to Percolation Theory.}}
  (\bibinfo{publisher}{Taylor \& Francis, London}, \bibinfo{year}{1994}).

\bibitem[{\citenamefont{Bissbort and
  Hofstetter}(2009)}]{bissbort2009stochastic}
\bibinfo{author}{\bibfnamefont{U.}~\bibnamefont{Bissbort}} \bibnamefont{and}
  \bibinfo{author}{\bibfnamefont{W.}~\bibnamefont{Hofstetter}},
  \bibinfo{journal}{EPL.} \textbf{\bibinfo{volume}{86}}, \bibinfo{pages}{50007}
  (\bibinfo{year}{2009}).

\bibitem[{\citenamefont{Kr\"{u}ger et~al.}(2010)\citenamefont{Kr\"{u}ger, Hong,
  and Phillips}}]{kruger2010two}
\bibinfo{author}{\bibfnamefont{F.}~\bibnamefont{Kr\"{u}ger}},
  \bibinfo{author}{\bibfnamefont{S.}~\bibnamefont{Hong}}, \bibnamefont{and}
  \bibinfo{author}{\bibfnamefont{P.}~\bibnamefont{Phillips}},
  \bibinfo{journal}{Arxiv Preprint arXiv:1006.2395.}  (\bibinfo{year}{2010}).

\bibitem[{\citenamefont{Huse}(2011)}]{huse2011pc}
\bibinfo{author}{\bibfnamefont{D.}~\bibnamefont{Huse}},
  \bibinfo{journal}{Private communication.}  (\bibinfo{year}{2011}).

\bibitem[{\citenamefont{ProkofÕev and Svistunov}(2004)}]{prokof2004superfluid}
\bibinfo{author}{\bibfnamefont{N.}~\bibnamefont{ProkofÕev}} \bibnamefont{and}
  \bibinfo{author}{\bibfnamefont{B.}~\bibnamefont{Svistunov}},
  \bibinfo{journal}{Physical Review Letters.} \textbf{\bibinfo{volume}{92}},
  \bibinfo{pages}{15703} (\bibinfo{year}{2004}).

\bibitem[{\citenamefont{Roscilde and Haas}(2007)}]{roscilde2007mott}
\bibinfo{author}{\bibfnamefont{T.}~\bibnamefont{Roscilde}} \bibnamefont{and}
  \bibinfo{author}{\bibfnamefont{S.}~\bibnamefont{Haas}},
  \bibinfo{journal}{Physical Review Letters.} \textbf{\bibinfo{volume}{99}},
  \bibinfo{pages}{47205} (\bibinfo{year}{2007}).

\bibitem[{\citenamefont{Sengupta and Haas}(2007)}]{sengupta2007quantum}
\bibinfo{author}{\bibfnamefont{P.}~\bibnamefont{Sengupta}} \bibnamefont{and}
  \bibinfo{author}{\bibfnamefont{S.}~\bibnamefont{Haas}},
  \bibinfo{journal}{Physical Review Letters.} \textbf{\bibinfo{volume}{99}},
  \bibinfo{pages}{50403} (\bibinfo{year}{2007}).

\bibitem[{\citenamefont{Fernandes and Schmalian}(2011)}]{fernandes2011complex}
\bibinfo{author}{\bibfnamefont{R.}~\bibnamefont{Fernandes}} \bibnamefont{and}
  \bibinfo{author}{\bibfnamefont{J.}~\bibnamefont{Schmalian}},
  \bibinfo{journal}{Physical Review Letters.} \textbf{\bibinfo{volume}{106}},
  \bibinfo{pages}{67004} (\bibinfo{year}{2011}).

\bibitem[{\citenamefont{Bray-Ali et~al.}(2006)\citenamefont{Bray-Ali, Moore,
  Senthil, and Vishwanath}}]{bray2006ordering}
\bibinfo{author}{\bibfnamefont{N.}~\bibnamefont{Bray-Ali}},
  \bibinfo{author}{\bibfnamefont{J.}~\bibnamefont{Moore}},
  \bibinfo{author}{\bibfnamefont{T.}~\bibnamefont{Senthil}}, \bibnamefont{and}
  \bibinfo{author}{\bibfnamefont{A.}~\bibnamefont{Vishwanath}},
  \bibinfo{journal}{Physical Review B} \textbf{\bibinfo{volume}{73}},
  \bibinfo{pages}{064417} (\bibinfo{year}{2006}).

\bibitem[{\citenamefont{Wang and Sandvik}(2006)}]{wang2006low}
\bibinfo{author}{\bibfnamefont{L.}~\bibnamefont{Wang}} \bibnamefont{and}
  \bibinfo{author}{\bibfnamefont{A.}~\bibnamefont{Sandvik}},
  \bibinfo{journal}{Physical Review Letters.} \textbf{\bibinfo{volume}{97}},
  \bibinfo{pages}{117204} (\bibinfo{year}{2006}).

\bibitem[{\citenamefont{Wang and Sandvik}(2010)}]{wang2010low}
\bibinfo{author}{\bibfnamefont{L.}~\bibnamefont{Wang}} \bibnamefont{and}
  \bibinfo{author}{\bibfnamefont{A.}~\bibnamefont{Sandvik}},
  \bibinfo{journal}{Physical Review B} \textbf{\bibinfo{volume}{81}},
  \bibinfo{pages}{054417} (\bibinfo{year}{2010}).

\bibitem[{\citenamefont{Sandvik}(2006)}]{sandvik2006quantum}
\bibinfo{author}{\bibfnamefont{A.}~\bibnamefont{Sandvik}},
  \bibinfo{journal}{Physical Review Letters.} \textbf{\bibinfo{volume}{96}},
  \bibinfo{pages}{207201} (\bibinfo{year}{2006}).

\bibitem[{\citenamefont{Moore}(2011)}]{moore2011pc}
\bibinfo{author}{\bibfnamefont{J.}~\bibnamefont{Moore}},
  \bibinfo{journal}{Private communication.}  (\bibinfo{year}{2011}).

\bibitem[{\citenamefont{Bouadim et~al.}(2010)\citenamefont{Bouadim, Loh,
  Randeria, and Trivedi}}]{bouadim2010single}
\bibinfo{author}{\bibfnamefont{K.}~\bibnamefont{Bouadim}},
  \bibinfo{author}{\bibfnamefont{Y.}~\bibnamefont{Loh}},
  \bibinfo{author}{\bibfnamefont{M.}~\bibnamefont{Randeria}}, \bibnamefont{and}
  \bibinfo{author}{\bibfnamefont{N.}~\bibnamefont{Trivedi}},
  \bibinfo{journal}{Arxiv preprint arXiv:1011.3275.}  (\bibinfo{year}{2010}).

\bibitem[{\citenamefont{Crane et~al.}(2007)\citenamefont{Crane, Armitage,
  Johansson, Sambandamurthy, Shahar, and Gr\"{u}ner}}]{crane2007survival}
\bibinfo{author}{\bibfnamefont{R.}~\bibnamefont{Crane}},
  \bibinfo{author}{\bibfnamefont{N.}~\bibnamefont{Armitage}},
  \bibinfo{author}{\bibfnamefont{A.}~\bibnamefont{Johansson}},
  \bibinfo{author}{\bibfnamefont{G.}~\bibnamefont{Sambandamurthy}},
  \bibinfo{author}{\bibfnamefont{D.}~\bibnamefont{Shahar}}, \bibnamefont{and}
  \bibinfo{author}{\bibfnamefont{G.}~\bibnamefont{Gr\"{u}ner}},
  \bibinfo{journal}{Physical Review B.} \textbf{\bibinfo{volume}{75}},
  \bibinfo{pages}{184530} (\bibinfo{year}{2007}).

\bibitem[{\citenamefont{Frydman et~al.}(2002)\citenamefont{Frydman, Naaman, and
  Dynes}}]{frydman2002universal}
\bibinfo{author}{\bibfnamefont{A.}~\bibnamefont{Frydman}},
  \bibinfo{author}{\bibfnamefont{O.}~\bibnamefont{Naaman}}, \bibnamefont{and}
  \bibinfo{author}{\bibfnamefont{R.}~\bibnamefont{Dynes}},
  \bibinfo{journal}{Physical Review B.} \textbf{\bibinfo{volume}{66}},
  \bibinfo{pages}{052509} (\bibinfo{year}{2002}).

\bibitem[{\citenamefont{Sherman et~al.}(2011)\citenamefont{Sherman, Kopnov,
  Shahar, and Frydman}}]{sherman2011superconducting}
\bibinfo{author}{\bibfnamefont{D.}~\bibnamefont{Sherman}},
  \bibinfo{author}{\bibfnamefont{G.}~\bibnamefont{Kopnov}},
  \bibinfo{author}{\bibfnamefont{D.}~\bibnamefont{Shahar}}, \bibnamefont{and}
  \bibinfo{author}{\bibfnamefont{A.}~\bibnamefont{Frydman}},
  \bibinfo{journal}{Arxiv preprint arXiv:1112.0776.}  (\bibinfo{year}{2011}).

\bibitem[{\citenamefont{Allain et~al.}(2011)\citenamefont{Allain, Han, and
  Bouchiat}}]{allain2011gate}
\bibinfo{author}{\bibfnamefont{A.}~\bibnamefont{Allain}},
  \bibinfo{author}{\bibfnamefont{Z.}~\bibnamefont{Han}}, \bibnamefont{and}
  \bibinfo{author}{\bibfnamefont{V.}~\bibnamefont{Bouchiat}},
  \bibinfo{journal}{Arxiv preprint arXiv:1109.6910.}  (\bibinfo{year}{2011}).

\bibitem[{\citenamefont{Yu et~al.}(2011)\citenamefont{Yu, Yin, Sullivan, Xia,
  Huan, Paduan-Filho, Oliveira, Haas, Steppke, Miclea et~al.}}]{yu2011bose}
\bibinfo{author}{\bibfnamefont{R.}~\bibnamefont{Yu}},
  \bibinfo{author}{\bibfnamefont{L.}~\bibnamefont{Yin}},
  \bibinfo{author}{\bibfnamefont{N.}~\bibnamefont{Sullivan}},
  \bibinfo{author}{\bibfnamefont{J.}~\bibnamefont{Xia}},
  \bibinfo{author}{\bibfnamefont{C.}~\bibnamefont{Huan}},
  \bibinfo{author}{\bibfnamefont{A.}~\bibnamefont{Paduan-Filho}},
  \bibinfo{author}{\bibfnamefont{N.}~\bibnamefont{Oliveira}},
  \bibinfo{author}{\bibfnamefont{S.}~\bibnamefont{Haas}},
  \bibinfo{author}{\bibfnamefont{A.}~\bibnamefont{Steppke}},
  \bibinfo{author}{\bibfnamefont{C.}~\bibnamefont{Miclea}},
  \bibnamefont{et~al.}, \bibinfo{journal}{Arxiv Preprint arXiv:1109.4403.}
  (\bibinfo{year}{2011}).

\bibitem[{\citenamefont{Wang et~al.}(2011)\citenamefont{Wang, Guo, and
  Sandvik}}]{wang2011incompressible}
\bibinfo{author}{\bibfnamefont{Y.}~\bibnamefont{Wang}},
  \bibinfo{author}{\bibfnamefont{W.}~\bibnamefont{Guo}}, \bibnamefont{and}
  \bibinfo{author}{\bibfnamefont{A.}~\bibnamefont{Sandvik}},
  \bibinfo{journal}{Arxiv preprint arXiv:1110.3213.}  (\bibinfo{year}{2011}).

\end{thebibliography}

\end{document}